\documentclass[11pt,letterpaper]{article}

\usepackage{latexsym,multirow}
\usepackage{amssymb,amsmath, bm, mathtools}
\usepackage{graphicx}
\usepackage[color,all,import,arrow]{xy}
\usepackage{enumerate}
\usepackage{setspace}

\usepackage{natbib}
\usepackage[pdftex, bookmarksopen=true, bookmarksnumbered=true,
pdfstartview=FitH, breaklinks=true, urlbordercolor={0 1 0}, citebordercolor={0 0 1}]{hyperref}
\usepackage{colortbl}
\usepackage{subfigure}

\usepackage{dcolumn}
\newcolumntype{.}{D{.}{.}{-1}}
\newcolumntype{d}[1]{D{.}{.}{#1}}
\usepackage{theorem}
\theoremstyle{plain}
\theoremheaderfont{\scshape}
\newtheorem{assumption}{Assumption}

\newtheorem{corollary}{Corollary}

\newtheorem{theorem}{Theorem}
\newtheorem{example}{Example}
\newtheorem{remark}{Remark}

\newtheorem{lemma}{Lemma}
\newtheorem{estimator}{Estimator}

\newcommand{\ind}{\mbox{$\perp\!\!\!\perp$}}

\providecommand{\norm}[1]{\lVert#1\rVert}

\usepackage{rotating}

\usepackage[compact]{titlesec}

\allowdisplaybreaks

\newcommand\spacingset[1]{\renewcommand{\baselinestretch}%
{#1}\small\normalsize}

\newcommand{\blind}{0}

\newcommand*{\QEDB}{\hfill\ensuremath{\square}}

\newcommand{\logit}{\text{logit}}

\newcommand{\sumn}{\sum_{i=1}^n}

\def\score{\mathfrak{s}}
\def\P{\mathbb{P}}
\def\d{\textup{d}}
\def\F{\mathbb{F}}
\def\C{\mathbb{C}}
\def\H{\mathcal{H}}

\newcommand{\E}{\mathbb{E}}

\def\T{\tiny\textsc{t}}

\begin{document} 

\newcommand{\tit}{Principal Stratification with Continuous Post-Treatment Variables: Nonparametric Identification and Semiparametric Estimation}
%
%
\spacingset{1.25}

\if0\blind

{\title{\bf\tit}

\author{Sizhu Lu\thanks{Department of Statistics, University of California, Berkeley, CA 94720, USA. Email: \href{mailto:sizhu_lu@berkeley.edu}{sizhu\_lu@berkeley.edu}}
  \and
     Zhichao Jiang\thanks{School of Mathematics,
      Sun Yat-sen University, Guangzhou,  Guangdong 510275, China. Email:
      \href{mailto:jiangzhch7@mail.sysu.edu.cn}{jiangzhch7@mail.sysu.edu.cn} }
    \and Peng Ding\thanks{Department of Statistics, University of California, Berkeley, CA 94720, USA. Email: \href{mailto:pengdingpku@berkeley.edu}{pengdingpku@berkeley.edu}} 
}

\date{
\today
}

\maketitle

}\fi

\if1\blind
\title{\bf \tit}

\maketitle
\fi

\pdfbookmark[1]{Title Page}{Title Page}

\thispagestyle{empty}
\setcounter{page}{0}
         
\begin{abstract}
Post-treatment variables often complicate causal inference. They appear in many scientific problems, including noncompliance, truncation by death, mediation, and surrogate endpoint evaluation. Principal stratification is a strategy to address these challenges by adjusting for the potential values of the post-treatment variables, defined as the principal strata. It allows for characterizing treatment effect heterogeneity across principal strata and unveiling the mechanism of the treatment's impact on the outcome related to post-treatment variables. However, the existing literature has primarily focused on binary post-treatment variables, leaving the case with continuous post-treatment variables largely unexplored. This gap persists due to the complexity of infinitely many principal strata, which present challenges to both the identification and estimation of causal effects. We fill this gap by providing nonparametric identification and semiparametric estimation theory for principal stratification with continuous post-treatment variables. We propose to use working models to approximate the underlying causal effect surfaces and derive the efficient influence functions of the corresponding model parameters. Based on the theory, we construct doubly robust estimators and implement them in an R package.

\end{abstract}


\clearpage
\spacingset{1.5}

\section{Introduction}
\label{sec::introduction}
Principal stratification is an approach to adjusting for post-treatment variables in causal inference. 
Because the post-treatment variable is affected by the treatment, the adjustment cannot be simply achieved by conditioning on its observed value.
Principal stratification, first proposed by \citet{frangakis2002principal}, addresses this issue by conditioning on the principal strata defined by the joint potential values of the post-treatment variable. The causal effects within principal strata, often referred to as the principal stratification average causal effects (PCEs), delineate causal mechanisms through the post-treatment variable while maintaining valid interpretations as subgroup causal effects. Over the past two decades, 
principal stratification has been successfully applied in a wide range of problems, including noncompliance \citep{angrist1996identification,frumento2012evaluating,mealli2013using},
truncation by death \citep{rubin2006causal,ding2011identifiability,wang2017causal,yang2016using},
missing data \citep{frangakis1999addressing,mattei2014identification},
mediation \citep{rubin2004direct,gallop2009mediation,elliott2010bayesian,mattei2011augmented,daniels2012bayesian,zigler2012estimating},
and surrogate evaluation \citep{frangakis2002principal,gilbert2008evaluating,huang2011comparing,li2010bayesian,jiang2016principal}.

The identification of PCEs presents challenges due to the fundamental problem of causal inference: only one potential value of the post-treatment variable is observed. Theoretical investigations into the identification of PCEs have mostly focused on binary or discrete post-treatment variables \citep[e.g.,][]{cheng2006bounds,imai2008sharp,ding2011identifiability,mattei2011augmented,mealli2013using,mealli2016identification,jiang2016principal, yang2016using,forastiere2016identification,jiang2020multiply}.
\citet{jiang2021identification} provide formal identification results for PCEs with a general post-treatment variable. However, their theory requires the existence of additional variables satisfying certain conditional independence assumptions, which can be difficult to meet in practice. In contrast to the lack of theory for continuous post-treatment variables, real-world applications frequently involve continuous post-treatment variables. For example, \citet{efron1991compliance}, \citet{jin2008principal}, and \citet{bartolucci2011modeling} consider a noncompliance problem in which the post-treatment variable is the proportion of the drug or placebo taken and takes continuous values ranging from zero to one. \citet{schwartz2011bayesian} study the causal mechanism of the effect of physical activity on cardiovascular disease mediated by a continuous post-treatment body mass index.

Dichotomizing the post-treatment variable is commonly employed in practice \citep[e.g.,][]{sjolander2009sensitivity,jiang2020multiply}.
The dichotomization approach, however, not only discards important information for understanding causal mechanisms but also poses challenges to valid inference due to the arbitrary choice of thresholds. Without dichotomization, the existing literature mostly employs modeling approaches to offer practical solutions while avoiding formal discussion of identifiability. For instance, \citet{conlon2014surrogacy} and \citet{jin2008principal} assume strong parametric models for the potential values of both the outcome and the post-treatment variable and specify informative priors for the parameters.  \citet{bartolucci2011modeling}  adopt similar outcome models but propose the use of copula to relax restrictions on potential values of the post-treatment variable.  \citet{schwartz2011bayesian} further relax the restrictions on the post-treatment variable under a Dirichlet process mixture model within a semiparametric Bayesian framework. Subsequently, many more complex models have been proposed for the application of principal stratification in various scenarios with continuous post-treatment variables \citep[e.g.,][]{zigler2012estimating,comment2019survivor,kim2019bayesian}, all of which rely heavily on modeling assumptions. However, estimates from these models may be unstable when its identifiability is not assured.
Therefore, there is a need for methods capable of handling continuous post-treatment variables without relying on strong modeling assumptions.

We develop a framework for principal stratification with a continuous post-treatment variable. In Section~\ref{sec::assumptions}, we introduce the notation and two identification assumptions. In Section~\ref{sec::identification}, we establish the nonparametric identification of PCEs under the principal ignorability assumption, which is commonly used in practice \citep[e.g.,][]{egleston2009estimation,jo2009use,jo2011use,stuart2015assessing,feller2017principal,mattei2023assessing,cheng2023multiply} and has been utilized in \citet{jiang2020multiply} to establish the identification of PCEs for a binary post-treatment variable.  
We also use a copula function to obtain 
the joint distribution of potential values of the post-treatment variable from their marginal distributions. It can be viewed as a generalization of the monotonicity assumption for a binary post-treatment variable \citep{jiang2021identification} and has also been popular in various settings \citep[e.g.,][]{roy2008principal,bartolucci2011modeling,daniels2012bayesian,conlon2014surrogacy, yang2018using}. 

The continuous post-treatment variable generates infinitely many principal strata, making the estimation of PCEs a difficult task. We propose to project the PCEs onto pre-specified working models to approximate the underlying complex causal effect surfaces. The projection parameter corresponds to the model-based parameter when the working model is correctly specified, but remains a well-defined causal parameter even if the working model is not correct. This approach has been adopted in other contexts of causal inference \citep{neugebauer2007nonparametric,kennedy2019robust,ye2023instrumented}. Based on the identification formulas, we propose two estimators that depend on different components of the observed data distribution.

The existence of multiple estimators motivates the derivation of a more efficient estimator.
In Section~\ref{sec::est}, we derive the efficient influence function (EIF) of the PCEs to motivate another estimator. In Section~\ref{sec::asymptotics}, we study the asymptotic properties of the proposed estimator based on EIF. The estimator is doubly robust in that it is consistent if either the treatment probability or the outcome mean model is correctly specified, contingent on the correct specification of the principal density model, and is semiparametrically efficient if all models are correctly specified with requirements on their convergence rates. In Section~\ref{sec::simulation}, we evaluate the finite sample performance of various estimators through simulations. In Section~\ref{sec::application}, we apply the estimators to analyze a randomized experiment in the youth labor market in Uganda. In Section~\ref{sec::discussion}, we conclude and discuss some extensions.

We use the following notation. Let $\norm{r}_{2} = \{\int r(v)^{2}\d\P(v)\}^{1/2}$ denote the $L_{2}(\P)$ norm where $\P(\cdot)$ denotes the distribution of the observed data. Let $p(\cdot)$ denote the probability density function of continuous random variables. 
Write $b_n=O_\P(a_n)$ if $b_n/a_n$ is bounded in probability and $b_n=o_\P(a_n)$ if $b_n/a_n$ converges to $0$ in probability.
Let $1(\cdot)$ be the indicator function. 
Let $\dot{f}(\eta)=\partial f(\eta) / \partial \eta$ denote the gradient of function $f$ and 
$\ddot{f}(\eta)=\partial^{2} f(\eta) / \partial\eta \partial\eta^{\T}$ denote the Hessian matrix of $f$ with respect to $\eta$. 
Let $\dot c_u(u,v)$ and $\dot c_v(u,v)$ denote the partial derivative of a bivariate function $c(u,v)$ with respect to $u$ and $v$, respectively.

\section{Setup and assumptions}
\label{sec::assumptions}
\subsection{Setup}
We generalize the principal stratification method in \citet{frangakis2002principal} to the case when the post-treatment variable is continuous. 
For unit $i=1,\ldots,n$, let $Z_i$ be the treatment, $S_i$ the post-treatment variable, $Y_i$ the outcome, and $X_i$ a set of pre-treatment covariates. Let $Y_{iz}$ and $S_{iz}$ denote the potential values of the outcome and the post-treatment variable if the treatment were set to $Z_i=z$, respectively. Assume all variables are identically and independently sampled from a super population and drop the subscript $i$ for simplicity.

We revisit four illustrative examples of real-world applications with a continuous post-treatment variable.
\begin{example}
    \citet{efron1991compliance} and \citet{jin2008principal} study the treatment effect of cholestyramine on lowering cholesterol levels. Let $Z$ denote the binary treatment where patients with $Z=1$ if the unit were assigned active pills of the drug, and $Z=0$ if the unit were assigned placebo pills. Let $Y$ denote the cholesterol reduction. Partial noncompliance arises in this trial because the patients may only take a proportion of the assigned dose. The post-treatment variable $S$ represents the proportion of the assigned dose taken, ranging from zero to one. To measure the causal effect of the drug that is actually taken, it is crucial to adjust for $S$ in the analysis.
\end{example}

\begin{example}
    \citet{gilbert2008evaluating} and \citet{gilbert2015surrogate} study the surrogate endpoint evaluation in randomized controlled trials, using an HIV vaccine trial and a herpes zoster vaccine trial, respectively. The binary treatment variable $Z$ is the randomized assignment of the vaccine or placebo. The post-treatment variable $S$ is the potential surrogate endpoint, which is some vaccine-induced immune responses, e.g., the 50\% neutralization titers for HIV gp120 in the HIV trial and the varicella zoster virus antibody titer in the herpes zoster trial. The outcome $Y$ is the infection of the disease. \citet{gilbert2008evaluating} and \citet{gilbert2015surrogate} propose using principal stratification criteria to evaluate how well the biomarkers act as an approximate surrogate for the endpoint. 
\end{example}

\begin{example}
    \citet{schwartz2011bayesian} study the effect of physical activity on cardiovascular disease and how the effect varies across levels of body mass index (BMI). In this study, the binary treatment variable $Z$ denotes physical activity with $Z=1$ for high-level exercisers and $Z=0$ for low-level exercisers. The outcome $Y$ is the indicator of cardiovascular disease. The scientific question of interest concerns the heterogeneity of the effect of physical activity on cardiovascular disease across different BMI level subgroups, denoted by a continuous post-treatment variable $S$.To address this,  \citet{schwartz2011bayesian} employ principal stratification to adjust for $S$ using a Bayesian semiparametric model.
\end{example}

\begin{example}
\label{exmp:yin2022}
    \citet{yin2022learning} conduct a randomized experiment in China to learn the effect of credit supply on consumption behavior. They randomly assign credit card users to treatment and control groups, denoted as $Z=1$ and $Z=0$, respectively. Users in the treatment group are offered an increase in their available credit, while users in the control group maintain their current credit limits. The outcome of interest $Y$ is the change in their average monthly consumption following the treatment. To assess how the effect of credit increase varies across users with different expectations in their future outcome, we adjust for the expected changes in monthly income 6 months after the experiment, defined as a continuous variable $S$. We analyze these data in Section~\ref{sec::application}.
\end{example}

The principal stratum $U$ is defined by the joint potential values of $S$, i.e., $U=(S_{1},S_{0})$. When $S$ is continuous, the number of principal strata is infinity. For ease of notation, we use $U=s_{1}s_{0}$ to denote the principal stratum $(S_1 = s_{1},S_0 = s_{0})$.
Our goal is to identify and estimate the PCE surface as a function of $s_{1}$ and $s_{0}$
\begin{eqnarray*}
\tau\left(s_{1},s_{0}\right) &=& \E\left(Y_{1}-Y_{0}\mid U=s_{1}s_{0}\right)
\end{eqnarray*}
and the two corresponding average potential outcome surfaces
\begin{eqnarray*}
m_z(s_1,s_0)& = &  \E\left(Y_{z}\mid U=s_{1}s_{0}\right)   \quad \text{for $z = 0,1$.}
\end{eqnarray*}

\subsection{Assumptions}
In this subsection, we discuss two crucial identification assumptions. We consider the scenario in which treatment ignorability holds for both the outcome and the post-treatment variable.
\begin{assumption}[Treatment ignorability]
\label{assump:treatment_ignorability}
$Z \ind (S_{0},S_{1},Y_{0},Y_{1})\mid X$. 
\end{assumption}
Assumption~\ref{assump:treatment_ignorability} is a standard identification assumption for the average causal effect.
It holds in randomized experiments by design. However, it is untestable in observational studies and requires the collection of all confounders between the treatment and the outcome, as well as those between the treatment and the post-treatment variable \citep{rosenbaum1983central}. 
Under Assumption~\ref{assump:treatment_ignorability}, we can identify the average causal effect of the treatment on the outcome, $\E(Y_1-Y_0)$, and that on the post-treatment variable, $\E(S_1-S_0)$. However, the PCEs are generally not identifiable since we cannot observe $S_1$ and $S_0$ simultaneously.

When $S$ is binary, the monotonicity assumption, i.e., $S_1\geq S_0$, is sufficient for the identification of the distribution of $U$ \citep[e.g.,][]{angrist1996identification, jiang2020multiply}, which serves as the basis for identifying the PCEs. With a continuous $S$, the distribution of $U$ becomes a density function and is not identifiable even with monotonicity. 

Under Assumption~\ref{assump:treatment_ignorability}, we can identify only the marginal distribution of $S_{z}$ given $X$ by $p(S_z\mid X) = p(S\mid Z=z,X)$ for $z=0,1$.
Therefore, we propose to use a copula function $\C_{\rho}(\cdot,\cdot)$ to establish a mapping between the joint distribution and the marginal distributions of $S_{z}$:
\begin{eqnarray*}
\P(S_1\leq s_1, S_0\leq s_0\mid X) \ = \ \C_{\rho}\{\F_1(s_1, X),\F_0(s_0, X)\},
\end{eqnarray*}
where the parameter $\rho$ indexes the association between $S_{1}$ and $S_{0}$ given $X$ (see \citealp{jiang2021identification, roy2008principal, bartolucci2011modeling, yang2018using, sun2023principal} for similar copula approaches), and $\F_z(s_z, X)=\P(S_{z}\leq s_z\mid X)$ denotes the marginal cumulative distribution function of $S_z$ given $X$ for $z=0,1$. 

 The literature takes different perspectives on the parameter $\rho$,  which governs the association in the copula function and thereby influences the identification of the distribution of $U$. For instance,
   \citet{efron1991compliance} assume that $S_1$ and $S_0$ for each unit are at the same percentile and impute the counterfactual from the marginal distributions of $S_z$, effectively specifying $\rho$ as a known constant. In contrast,  \citet{bartolucci2011modeling}  treat $\rho$ as a parameter and propose a parametric model to identify and estimate it, while \citet{sun2023principal} leverage repeated measurements of $S$ for identification and estimation. However, these approaches rely on strong parametric models or additional information. 

To avoid additional assumptions, we treat the copula function as known with a pre-specified sensitivity parameter $\rho$, directing our attention toward scenarios where the joint density of $(S_1,S_0)$ is identifiable. We will use $\C(\cdot,\cdot)$ to denote the copula function and omit the subscript $\rho$ for ease of notation.
Given a pre-specified value of $\rho$, we can identify the density of $U$ given $X$ by
\begin{eqnarray}
    p\left(U=s_{1}s_{0}\mid X\right) &=& \frac{\partial^2 \P(S_1\leq s_1, S_0\leq s_0\mid X)}{\partial s_1 \partial s_0} \notag \\
    &=& \frac{\partial^2 \C(\F_1(s_1, X),\F_0(s_0, X))}{\partial s_1 \partial s_0} \notag \\
    &=& c(\F_1(s_1, X),\F_0(s_0, X))p_1(s_1,X)p_0(s_0,X), \label{equ::principal_density}
\end{eqnarray}
where $c(u,v)=\partial^2{\C(u,v)}/\partial u\partial v$ denotes the copula density function that equals the second order cross derivative of the copula function $\C(\cdot, \cdot)$, and $p_z(s_z,X)=p(S=s_z\mid Z=z,X)$ is the density function of $S$ given $Z=z$ and $X$ at $S=s_z$ for $z=0,1$.
For a continuous $S$, $p(U=s_{1}s_{0}\mid X)$ is the density function of $U$ given $X$. We define $e(s_{1},s_{0},X)=p(U=s_{1}s_{0}\mid X) $ and refer to it as
the {\it principal density}. With a binary $S$, it reduces to the principal score in \citet{follmann2000effect,hill2002differential,joffe2007defining,ding2017principal,feller2017principal,jiang2020multiply}. 

With the known copula function and association parameter $\rho$, there is a one-to-one mapping between $e(s_1,s_0,X)$ and $\{p_1(s_1,X),p_0(s_0,X)\}$, and thus we will refer to both as principal density.
By averaging the principal density over $X$, we can obtain the marginal density of $U$:
\begin{eqnarray*}
    e\left(s_{1},s_{0}\right) &=& p\left(U=s_{1}s_{0}\right)\ =\ \E\left\{ e\left(s_{1},s_{0},X\right)\right\}.
\end{eqnarray*}

Next, we generalize the principal ignorability assumption with a binary post-treatment variable \citep{jo2009use,jo2011use,stuart2015assessing,ding2017principal,jiang2020multiply} to the following assumption with a continuous post-treatment variable.
\begin{assumption}[Principal ignorability]
\label{assump:principal_ignorability}
$\E(Y_{1}\mid U=s_{1}s_{0},X) = \E(Y_{1}\mid U=s_{1}s^{\prime}_{0},X)$ for any $s_{0}, s^{\prime}_{0}$ and $\E(Y_{0}\mid U=s_{1}s_{0},X) = \E(Y_{0}\mid U=s^{\prime}_{1}s_{0},X)$ for any $s_{1}, s^{\prime}_{1}$.
\end{assumption}

Assumption~\ref{assump:principal_ignorability} requires that conditional on the covariates, the expectation of the potential outcome under treatment, $Y_{1}$, does not vary across principal strata with the same value of $S_{1}$, and similarly, the expectation of the potential outcome under control, $Y_{0}$, does not vary across principal strata with the same value of $S_{0}$. A stronger version of the principal ignorability assumes conditional independence $Y_1 \ind S_0 \mid S_1, X \textup{ and } Y_0 \ind S_1 \mid S_0, X$. Assumption~\ref{assump:principal_ignorability} is weaker as it only constrains the expected values of potential outcomes rather than the entire distribution, conditioned on the covariates. Under Assumption~\ref{assump:treatment_ignorability}, Assumption~\ref{assump:principal_ignorability} is equivalent to
\begin{eqnarray}
\E(Y_{1}\mid U=s_{1}s_{0}, Z=1, S=s_{1},X) &=& \E(Y_{1}\mid U=s_{1}s^{\prime}_{0}, Z=1, S=s_{1},X), \label{equ:pi-1}\\
\E(Y_{0}\mid U=s_{1}s_{0}, Z=0, S=s_{0},X) &=& \E(Y_{0}\mid U=s^{\prime}_{1}s_{0}, Z=0, S=s_{0},X). \label{equ:pi-2}\
\end{eqnarray}
The observed stratum $(Z=1,S=s_{1})$ is a mixture of the principal strata $U=s_{1}s_{0}$ for all possible values of $s_{0}$. Therefore,~\eqref{equ:pi-1} means that within the observed
stratum $(Z=1,S=s_{1})$, the average potential outcome under treatment does not vary across the principal strata conditional on the covariates. So the conditional
expectations in~\eqref{equ:pi-1} simplify to the outcome mean $\mu_1(s_1,X)=\E(Y\mid Z=1,S=s_{1},X)$.
Similar argument applies to the observed stratum $(Z=0,S=s_{0})$; the conditional expectations in \eqref{equ:pi-2} simplify to the outcome mean $\mu_0(s_0,X)=\E(Y\mid Z=0,S=s_{0},X)$.

\section{Nonparametric identification and projection parameters}
\label{sec::identification}

\subsection{Identification of PCEs}
In this subsection, we discuss the nonparametric identification of the PCE surfaces. From~\eqref{equ:pi-1}~and~\eqref{equ:pi-2}, we can identify the surfaces of the average potential outcomes by averaging the outcome means $\mu_z(s_z,X)$ over the distribution of $X$ given $U$.
This leads to the identification formulas in the following theorem.

\begin{theorem}
\label{thm:identification}
Suppose that $e(s_{1},s_{0},X)$ is identifiable and Assumptions~\ref{assump:treatment_ignorability} and \ref{assump:principal_ignorability} hold. For any $(s_{1},s_{0})$ such that $e(s_{1},s_{0})>0$, we have
\begin{eqnarray}
\E\left(Y_{z}\mid U=s_{1}s_{0}\right) &=& \E\left\{ \frac{e(s_{1},s_{0},X)}{e(s_{1},s_{0})}\mu_{z}\left(s_{z},X\right)\right\}\quad \text{for $z=0,1$.}
\label{equ:pd+om+equ1}
\end{eqnarray}

\end{theorem}
Theorem~\ref{thm:identification} expresses the distribution of $X$ given $U$ as a function of the principal density. Thus,
the formulas rely on both the principal density and the outcome mean, extending those in \citet{jiang2020multiply} based on the principal score and the outcome models. Note that we do not need  $e(s_1,s_0)>0$ for all $s_1,s_0$ as we focus only on the strata with $e(s_1,s_0)>0$.

The existence of alternative identification formulas in \citet{jiang2020multiply} motivates us to explore other identification formulas with a continuous $S$. The following theorem presents the identification formula based on the treatment probability and principal density. 
\begin{theorem}
\label{thm:non_regular}
Suppose that Assumptions~\ref{assump:treatment_ignorability} and \ref{assump:principal_ignorability} hold, and $0<\pi(x)=\P(Z=1\mid X=x)<1$ for all $x$ in the support of $X$.
Assume the potential outcome surface $m_{z}(s_{1},s_{0})$, the principal density $e(s_{1},s_{0},X)$ and $p_{z}\left(s_{z},X\right)$, and the marginal density $e(s_{1},s_{0})$ are continuous in $s_{1}$ and $s_{0}$ for $z=0,1$. For any $(s_{1},s_{0})$ such that $e(s_{1},s_{0})>0$, we have
\begin{eqnarray}
 \nonumber   \E\left( Y_{1}\mid U=s_{1}s_{0}\right) &=& \lim_{\epsilon\rightarrow0}\E\left\{ Y_{1}\mid S_{1}\in B_{\epsilon}\left(s_{1}\right),S_{0}\in B_{\epsilon}\left(s_{0}\right)\right\}  \\
 \nonumber	&=& \lim_{\epsilon\rightarrow0}\E\left[ \frac{\P\left\{ S_{1}\in B_{\epsilon}\left(s_{1}\right),S_{0}\in B_{\epsilon}\left(s_{0}\right)\mid X\right\} }{\P\left\{ S_{1}\in B_{\epsilon}\left(s_{1}\right),S_{0}\in B_{\epsilon}\left(s_{0}\right)\right\} }\frac{1\left\{ S_{1}\in B_{\epsilon}\left(s_{1}\right)\right\} }{\P\left\{S_{1}\in B_{\epsilon}\left(s_{1}\right)\mid X\right\}}\frac{ZY}{\pi\left(X\right)}\right] \\
\label{equ:identification-2}	&=& \lim_{\epsilon\rightarrow0}\E\left\{ \frac{e\left(s_{1},s_{0},X\right)}{e\left(s_{1},s_{0}\right)}\frac{1\left(s_{1}-\epsilon\leq S<s_{1}+\epsilon\right)}{2\epsilon \cdot p_{1}\left(s_{1},X\right)}\frac{ZY}{\pi\left(X\right)}\right\}, 
\end{eqnarray}
where $B_{\epsilon}(s_{z})=\{s\in\mathbb{R}\mid s_{z} - \epsilon \leq s < s_{z} + \epsilon\}$ is the $\epsilon$-ball centered at $s_{z}$ for $z=0,1$.
\end{theorem}
We can obtain a similar identification formula for $\E(Y_{0}\mid U=s_{1}s_{0})$. Similar to the identification of the classic average treatment effect, the identification here requires the overlap condition that $0<\pi(x)<1$. Equation~\eqref{equ:identification-2} expresses the average potential outcome under treatment in principal stratum $U=s_1s_0$ as the limit of a weighted average of the outcome for the treated units. The weights have two main components: $1/\pi(X)$ corresponds to the treatment probability, while  $ e(s_{1},s_{0},X) /\{e(s_1,s_0)p_{1}(s_{1},X)\}$ corresponds to the principal density. The indicator function $Z 1(s_{1}-\epsilon\leq S<s_{1}+\epsilon)/2\epsilon $ limits the identification formula to units with $S$ in the neighborhood of $s_1$, indicating that $\E( Y_{1}\mid U=s_{1}s_{0})$ may not be regular because the parameter is not Frechet differentiable \citep[for technical details see ][]{bickel1993efficient}. 
Similar argument applies to $\E( Y_{0}\mid U=s_{1}s_{0})$. Therefore, the estimation of $\tau(s_1,s_0)$ based on Theorem~\ref{thm:non_regular} requires using nonparametric local methods such as kernel methods.

Unlike  \citet{jiang2020multiply}, it is difficult to obtain an identification formula based only on the treatment probability and outcome model with a continuous $S$. Achieving this requires replacing the principal density in the identification formula of Theorem~\ref{thm:identification} with the treatment probability.
However, the principal density $p(U=s_1 s_0\mid X)=c(\F_1(s_1,X),\F_0(s_0,X))p(S_{1}=s_{1}\mid X)p(S_{0}=s_{0}\mid X)$ is no longer a linear function of the marginal distributions of $S_z$ as in the case of binary $S$. Instead, it depends on the copula function $c(\cdot,\cdot)$, which generally takes a complicated form. Therefore, the principal density plays a crucial role in the identification formulas of the PCE surface when the post-treatment variable $S$ is continuous. We will further demonstrate this fact later in Section~\ref{sec::est}.

\subsection{Projection parameters}
Using the identification formulas in Theorems~\ref{thm:identification} and~\ref{thm:non_regular}, we can construct nonparametric estimators for the average potential outcome surfaces $m_1(s_1,s_0)$ and $m_0(s_1,s_0)$. However, they are local parameters that are not pathwise differentiable. Therefore, their efficient influence functions with finite variances do not exist \citep{bickel1993efficient}, leading to substantial finite-sample variability in their estimates. In addition, because the average potential outcome surfaces are two-dimensional functions, their interpretation in practical settings may pose challenges.

To improve precision and interpretation, we project the mean surface $m_{z}(s_{1},s_{0})$ onto a pre-specified working model $f_{z}(s_{1},s_{0};\eta_z)$, where $\eta_z\in\mathbb{R}^{q}$ is a finite-dimensional parameter \citep{neugebauer2007nonparametric,kennedy2019robust,ye2023instrumented}. In particular, we define our parameters of interest as 
\begin{eqnarray}
    \eta_{1} &=& \arg\min_{\eta\in\mathbb{R}^{q}}\E\left[w_{1}\left(S_{1},S_{0}\right)\left\{m_{1}\left(S_{1},S_{0}\right)-f_{1}\left(S_{1},S_{0};\eta\right)\right\}^{2}\right], \label{equ:opt_m1} \\
    \eta_{0} &=& \arg\min_{\eta\in\mathbb{R}^{q}}\E\left[w_{0}\left(S_{1},S_{0}\right)\left\{m_{0}\left(S_{1},S_{0}\right)-f_{0}\left(S_{1},S_{0};\eta\right)\right\}^{2}\right], \label{equ:opt_m0}
\end{eqnarray}
where $w_{1}(s_{1},s_{0})$ and $w_{0}(s_{1},s_{0})$ are some user-specified weight functions. The formulation in~\eqref{equ:opt_m1}~and~\eqref{equ:opt_m0} does not require a correct specification of the working model, and the pre-specified weights and the working models could be different for the two mean surfaces. We can view the projection coefficient $\eta_z$ as the solution to the nonlinear weighted least squares of the true population parameters $m_z(S_1, S_0)$ for $z=0,1$. Constant weight functions, where $w_1(s_{1},s_{0}) = w_0(s_{1},s_{0}) = 1$, lead to uniform weighting across the entire support of $(S_1,S_0)$. 
In contrast, if domain knowledge indicates that particular values of $(s_1,s_0)$ are more important, we can employ non-constant weight functions to assign greater weights to them accordingly.
From~\eqref{equ:opt_m1}~and~\eqref{equ:opt_m0}, $\eta_{z}$ is the parameter that minimizes the discrepancy between the true mean surface $m_{z}$ and the working model $f_{z}$, with the discrepancy defined by the choice of the weighting function $w_{z}$.

The identification of the projected parameters $\eta_1$ and $\eta_0$ follows directly from the identification of the average potential outcome surfaces in Theorems~\ref{thm:identification} and~\ref{thm:non_regular} if the optimization problems in~\eqref{equ:opt_m1} and~\eqref{equ:opt_m0} have unique solutions, which depends on 
the functional form of $f_z$ and the support of $e(s_1,s_0)$.

\begin{remark}
We can also  project the PCE surface $\tau\left(s_{1},s_{0}\right)$ using a similar approach:
\begin{equation}
\label{equ:opt}
\eta_{\tau}=\arg\min_{\eta\in\mathbb{R}^{q}}\E\left[ w_{\tau}\left(S_{1},S_{0}\right)\left\{\tau\left(S_{1},S_{0}\right)-f_{\tau}\left(S_{1},S_{0};\eta\right)\right\}^{2}\right] .
\end{equation}
The quantities $\eta_{\tau}$ and $\eta_1 - \eta_0$ are closely related under some specific choices of $f_1$, $f_0$, and $f_{\tau}$, but are different in general. An advantage of the separately projected parameters is that we can also consider nonlinear causal estimands, for instance, $m_1(S_1,S_0)/m_0(S_1,S_0)$, and approximate it by the ratio between two projected mean surfaces $f_1(S_1,S_0;\eta_1) / f_0(S_1,S_0;\eta_0)$. We focus on the separately projected parameters $\eta_1$ and $\eta_0$ defined in~\eqref{equ:opt_m1} and~\eqref{equ:opt_m0} in the main text and relegate the result for $\eta_{\tau}$ and the discussion on the special case when these two ways of projection are equivalent to the Supplementary Material. 
\end{remark}

\subsection{Estimation of the projection parameters}

Next, we introduce two estimators of the projected parameters corresponding to the two identification formulas in Theorems~\ref{thm:identification} and~\ref{thm:non_regular}. 
These estimators rely on different components of the observed data distribution. For simplicity, we use $\hat \pi$, $\hat e$, and $\hat \mu$ to denote the estimated treatment probability, 
principal density, and outcome mean, respectively, suppressing their dependence on the covariates $X$.
As established earlier, we only need to estimate the marginal distribution $p_z(s_z, X)$ for $z=0,1$ and apply the copula function to obtain the principal density estimator. Throughout this subsection, we assume the optimization problems in~\eqref{equ:opt_m1} and~\eqref{equ:opt_m0} have unique solutions.

We consider the estimation of $\eta_z$ for $z=0,1$. From~\eqref{equ:opt_m1}~and~\eqref{equ:opt_m0}, we can write the projected quantity $\eta_z$ as
\begin{eqnarray}
    \eta_{z} &=& \arg\min_{\eta}\E\left[ w_z\left(S_{1},S_{0}\right)\left\{m_{z}\left(S_{1},S_{0}\right)-f_z\left(S_{1},S_{0};\eta\right)\right\}^{2}\right] \notag \\
    &=& \arg \min_{\eta} \iint w_z\left(s_{1}, s_{0}\right) \left\{m_{z}\left( s_{1},s_{0} \right)-f_z\left(s_{1}, s_{0} ; \eta\right)\right\}^{2}e\left(s_{1},s_{0}\right) \d s_{1}\d s_{0} \label{equ:objective_eta1}
\end{eqnarray}
for $z=0,1$.
Taking the first order derivative with respect to $\eta$ on the objective function in~\eqref{equ:objective_eta1}, we have
\begin{equation}
    \iint w_z\left(s_{1}, s_{0}\right) \dot{f}_z\left(s_{1},s_{0} ; \eta_z\right)e\left(s_{1},s_{0}\right) \left\{m_{z}\left(s_{1}, s_{0}\right)-f_z\left(s_{1}, s_{0} ; \eta_z\right)\right\} \d s_{1}\d s_{0}=0.
    \label{equ:foc_m1}
\end{equation}
Equation \eqref{equ:foc_m1} motivates an estimating equation based on Theorem~\ref{thm:identification}.

\begin{lemma}[Estimating equation based on principal density and outcome mean models]
\label{lemma:pd+om_ee}
Suppose that Assumptions~\ref{assump:treatment_ignorability} and \ref{assump:principal_ignorability} hold. For $z=0,1$, define
\begin{eqnarray}
    && D_{z,\textup{pd+om}}\left(Y,S,Z,X;\eta, e, \mu_{z}\right) \notag \\
    &=& \iint w_z\left(s_{1}, s_{0}\right) \dot{f}_z\left(s_{1},s_{0} ; \eta \right) e\left(s_{1}, s_{0}, X\right) \left\{\mu_{z}\left(s_{z},X\right)-f_z\left(s_{1}, s_{0} ; \eta\right)\right\} \d s_{1}\d s_{0}. \label{equ:pd+om_ee}
\end{eqnarray}
We have 
\begin{eqnarray}
    \E\left\{D_{z,\textup{pd+om}}\left(Y,S,Z,X;\eta_z, e, \mu_{z}\right)\right\}&=& 0 \label{equ:pd+om_ee2}
\end{eqnarray}
and thus $D_{z,\textup{pd+om}}$ is a valid estimating equation for $\eta_z$ for $z=0,1$.
\end{lemma}
Theorem~\ref{thm:identification} indicates that
\begin{eqnarray}
m_{z}\left(s_{1}, s_{0}\right)-f_z\left(s_{1}, s_{0}; \eta\right) &=& \E\left[\frac{e\left(s_{1}, s_{0}, X\right)}{e\left(s_{1},s_{0}\right)} \left\{\mu_{z}\left(s_{z},X\right)-f_z\left(s_{1}, s_{0} ; \eta\right)\right\}\right]. \label{eqn::mf_to_muf}
\end{eqnarray}
Plugging~\eqref{eqn::mf_to_muf} into~\eqref{equ:foc_m1} then yields the validity of the estimating equation in Lemma~\ref{lemma:pd+om_ee}.
Equation~\eqref{equ:pd+om_ee} motivates the following two-step estimator.

\begin{estimator}[Estimator based on principal density and outcome mean models] We provide a two-step estimator $\hat{\eta}_{z,\textup{pd+om}}$ based on the principal density and outcome mean model for $z=0,1$.
    \begin{itemize}
    \item Step 1: Estimate the principal density $\hat{e}=(\hat p_1,\hat p_0)$ and outcome mean $\hat{\mu}_{z}$.
    \item Step 2: Obtain the estimator $\hat{\eta}_{z,\textup{pd+om}}$ by plugging $\hat{e}$ and $\hat{\mu}_{z}$ into~\eqref{equ:pd+om_ee} and solving for the empirical analog of the estimating equation~\eqref{equ:pd+om_ee2}
    \begin{equation*}
        \sum_{i=1}^{n}D_{z,\textup{pd+om}}\left(Y_{i},S_{i},Z_{i},X_{i}; \hat{\eta}_{z,\textup{pd+om}},\hat{e},\hat{\mu}_{z} \right) = 0.
    \end{equation*}
\end{itemize}
\end{estimator}
The consistency and  root-$n$ convergence rate of $\hat{\eta}_{z,\textup{pd+om}}$ to $\eta_z$ require the consistent estimation and root-$n$ convergence rate of the principal density and outcome mean models.

Next, we construct estimating equations based on the identification formula in Theorem~\ref{thm:non_regular}.
\begin{lemma}[Estimating equation based on treatment probability and principal density]
\label{lemma:weighting_ee}
Suppose that Assumptions~\ref{assump:treatment_ignorability} and \ref{assump:principal_ignorability} hold. Define
\begin{eqnarray}
    && D_{1,\textup{tp+pd}}\left(Y,S,Z,X; \eta,\pi,e\right) \notag \\
    &=& \int  w_1\left(S, s_{0}\right) \dot{f}_1\left(S,s_{0} ; \eta\right) \frac{e\left(S, s_{0}, X\right)}{p_{1}(S,X)} \frac{Z}{\pi(X)}\left\{Y-f_1\left(S, s_{0} ; \eta\right)\right\} \d s_{0} \label{equ:weighting_ee1}
\end{eqnarray}
and
\begin{eqnarray}
    && D_{0,\textup{tp+pd}}\left(Y,S,Z,X; \eta,\pi,e\right) \notag \\
    &=& \int w_0\left(s_1, S\right) \dot{f}_0\left(s_1,S ; \eta\right)  \frac{e\left(s_1, S, X\right)}{p_{0}(S,X)} \frac{1-Z}{1-\pi(X)}\left\{Y-f_0\left(s_1,S ; \eta\right)\right\} \d s_{1}. \label{equ:weighting_ee2}
\end{eqnarray}
We have 
\begin{eqnarray}
    \E\left\{D_{z,\textup{tp+pd}}\left(Y,S,Z,X; \eta_z,\pi,e\right)\right\}&=&0
    \label{equ:weighting_ee3}
\end{eqnarray}
and thus  $D_{z,\textup{tp+pd}}$ is a valid estimating equation for $\eta_z$ for $z=0,1$.
\end{lemma}
The weights based on principal density and treatment probability in~\eqref{equ:weighting_ee1}~and~\eqref{equ:weighting_ee2} have a similar form as those in Theorem~\ref{thm:non_regular}. 

These equations depend on the treatment probability and principal density while agnostic of the outcome means, motivating the following two-step weighting estimator that does not require modeling the outcome.

\begin{estimator}[Estimator based on treatment probability and principal density] 
We provide a two-step estimator $\hat{\eta}_{z,\textup{tp+pd}}$ based on the treatment probability and principal density for $z=0,1$.
    \begin{itemize}
    \item Step 1: Estimate the treatment probability $\hat{\pi}$ and principal density $\hat{e}=(\hat{p}_{1},\hat{p}_{0})$.
    \item Step 2: Obtain the estimator $\hat{\eta}_{z,\textup{tp+pd}}$ by plugging  $\hat{\pi}$ and $\hat{e}$ into~\eqref{equ:weighting_ee1}~and~\eqref{equ:weighting_ee2} and solving for the empirical analog of the estimating equation~\eqref{equ:weighting_ee3}
    \begin{equation*}
        \sumn D_{z,\textup{tp+pd}}\left(Y_{i},S_{i},Z_{i},X_{i}; \hat{\eta}_{z,\textup{tp+pd}},\hat{\pi},\hat{e} \right) = 0.
    \end{equation*}
    \end{itemize}
\end{estimator}
The consistency and root-$n$ convergence rate of $\hat{\eta}_{z,\textup{tp+pd}}$ to $\eta_z$ require the consistent estimation and root-$n$ convergence rate of the treatment probability and principal density models. 

The variance estimators of both  $\hat{\eta}_{z,\textup{pd+om}}$ and $\hat{\eta}_{z,\textup{tp+pd}}$ can be obtained using nonparametric bootstrap. 
We can also use classic variance estimators constructed from the estimating equations for Z-estimators if the corresponding models are correctly specified. 
Because estimating equations~\eqref{equ:weighting_ee1}~and~\eqref{equ:weighting_ee2} involve  the inverse of the treatment probability, $\hat{\eta}_{z,\textup{tp+pd}}$ becomes unstable if the estimated treatment probability is close to zero or one. In this case, $\hat{\eta}_{z,\textup{pd+om}}$ is comparatively more stable, though it relies on the extrapolation using the outcome model. Interestingly, the inverse of the marginal distribution $p_z(S,X)$ in~\eqref{equ:weighting_ee1}~and~\eqref{equ:weighting_ee2}  does not lead to a similar problem because the ratios $e(S,s_0,X) /p_1(S,X)$ and $e(s_1,S,X) /p_0(S,X)$ are always bounded in $[0,1]$.
The computation of both  $\hat{\eta}_{z,\textup{pd+om}}$ and $\hat{\eta}_{z,\textup{pd+om}}$ requires numerical integration. 
Due to the additional layer of integration in~\eqref{equ:pd+om_ee} compared to~\eqref{equ:weighting_ee1}~and~\eqref{equ:weighting_ee2}, $\hat{\eta}_{z,\textup{pd+om}}$ is generally more computationally intensive than $\hat{\eta}_{z,\textup{tp+pd}}$.

\section{Semiparametric estimation based on EIF}
\label{sec::est}

The availability of multiple estimators for $\eta_z$ serves as a catalyst for devising the most efficient estimator.
In this section, we first derive the EIF for $\eta_z$ and then construct the estimator based on the EIF, then provide some representative examples of working models. 

\subsection{EIF and semiparametric estimation}
For ease of notation, we only present the result for $\eta_1$; the results for $\eta_0$ and $\eta_\tau$ can be obtained similarly and are provided in the Supplementary Material.

We introduce additional notation. Define 
\begin{eqnarray*}
\nu_1(s_1,s_0,X)&=&  w_1\left(s_{1},s_{0}\right)\dot{f}_1\left(s_{1},s_{0};\eta_{1}\right)e\left(s_{1},s_{0},X\right)\left\{\mu_{1}\left(s_{1},X\right) - f_1\left(s_{1},s_{0};\eta_{1}\right)\right\},\\
r_u(s_1,s_0,S,X)&=& 1- \frac{\dot{c}_{u}(s_1,s_0,X)}{c(s_1,s_0,X)}\left\{1(S\leq s_1)-\F_1(s_1,X)\right\},\\
r_v(s_1,s_0,S,X)&=& 1- \frac{\dot{c}_{v}(s_1,s_0,X)}{c(s_1,s_0,X)}\left\{1(S\leq s_0)-\F_0(s_0,X)\right\}.
\end{eqnarray*}
The term $\nu_1(s_1,s_0,X)$ is the integrand of the estimating equation in~\eqref{equ:pd+om_ee}, and $r_u(s_1,s_0,S,X)$ and $r_v(s_1,s_0,S,X)$ are quantities related to $S_1$ and $S_0$. 
The term $ \dot{c}_{u}(s_1,s_0,X)/c(s_1,s_0,X)$ is equal to $\partial \log c(u,v)/\partial u$ evaluated at $\F_1(s_1,X)$ and $\F_0(s_0,X)$, which can be viewed as the score function of the principal density with respect to $u$.  Similar interpretation applies to $ \dot{c}_{v}(s_1,s_0,X)/c(s_1,s_0,X)$.
We can show that 
\begin{eqnarray*}
\E\{r_u(s_1,s_0,S,X) \mid Z=1,X\}\ =\ \E\{r_v(s_1,s_0,S,X) \mid Z=0,X\} \ = \ 1.
\end{eqnarray*}
The following theorem presents the EIF for $\eta_1$.
\begin{theorem}[EIF for $\eta_1$]
\label{thm:eif_m1}
Under the nonparametric model, the EIF for $\eta_{1}$ is
\begin{equation}
    \varphi_{1}\left(V\right) = H_{1}^{-1} \left\{\ell_{1}\left(X\right)+\ell_{2}\left(S,Z,X\right)+\ell_{3}\left(Y,S,Z,X\right)\right\},
    \label{equ:eif_m1}
\end{equation}
where  
\begin{eqnarray*}
H_{1} &=& \E\left(w_1(S_1,S_0)\left[\dot{f}_1\left(S_{1},S_{0};\eta_{1}\right)\dot{f}_1\left(S_{1},S_{0};\eta_{1}\right)^{\T} - \left\{m_1(S_1,S_0) - f_1(S_1,S_0;\eta_1)\right\}\ddot{f}_1(S_1,S_0;\eta_1) \right] \right)
\end{eqnarray*}
and
\begin{eqnarray*}
    \ell_{1}\left(X\right) &=& \iint  \nu_1(s_1,s_0,X) \d s_{1}\d s_{0},\\
    \ell_{2}\left(S,Z,X\right) &=& \frac{Z}{\pi\left(X\right)}\left\{\frac{\int \nu_1(S,s_0,X)\d s_0}{p_1(S,X)} - \iint \nu_1(s_1,s_0,X) r_u(s_1,s_0,S,X) \d s_1 \d s_0\right\}\\
    &&+ \frac{1-Z}{1-\pi\left(X\right)}\left\{\frac{\int \nu_1(s_1,S,X)\d s_1}{p_0(S,X)} - \iint \nu_1(s_1,s_0,X) r_v(s_1,s_0,S,X) \d s_1 \d s_0\right\},\\
    \ell_{3}\left(Y,S,Z,X\right) &=& \frac{Z}{\pi\left(X\right)}\frac{ \int w_1\left(S,s_{0}\right)\dot{f}_1\left(S,s_{0};\eta_{1}\right)e\left(S,s_{0},X\right)\d s_{0}}{p_{1}\left(S,X\right)}\left\{Y-\mu_{1}\left(S,X\right)\right\}.
\end{eqnarray*}
\end{theorem}

We suppress the dependence of $H_1$, $\ell_1$, $\ell_2$, and $\ell_3$ on $\eta_1$ for simplicity.
We show in the proof of Theorem~\ref{thm:eif_m1} that $2H_1$ is equal to the Hessian matrix of the objective function in~\eqref{equ:opt_m1} evaluated at $\eta_1$.
The three terms $\ell_1(X)$, $\ell_2(S,Z,X)$, and $\ell_3(Y,S,Z,X)$ correspond to the three levels of perturbation on different components of the observed conditional distributions.
The function $\ell_1(X)$  is the same as the estimating equation in~\eqref{equ:pd+om_ee}, while $\ell_3(Y,S,Z,X)$ has a similar form to the estimating equation in~\eqref{equ:weighting_ee1} by replacing the residual $Y-f_1(S,s_0;\eta)$ inside the integral with $Y-\mu_1(S,X)$.
The function $\ell_2(S,Z,X)$ takes a more complex form, partly due to the association between $S_1$ and $S_0$ specified in the copula function, which hinders clear explanation. In the special case when $S_1$ and $S_0$ are independent conditional on $X$, $c(u,v)=1$, $\dot c_u(u,v) = \dot c_v(u,v) = 0$, and thus $r_u(s_1,s_0,S,X)=r_v(s_1,s_0,S,X)=1$. 
The function  $\ell_2(S,Z,X)$ then simplifies to
\begin{eqnarray*}
 \ell_2(S,Z,X)&=& \frac{Z}{\pi\left(X\right)}\left[ u_1(S,X) - \E\{u_1(S,X)\mid X,Z=1\}  \right] \\
 &&+ \frac{1-Z}{1-\pi\left(X\right)} \left[u_1(S,X) - \E\{u_1(S,X)\mid X,Z=0\} \right], 
\end{eqnarray*}
where $u_1(s_z,X)=\int \nu_1(s_1,s_0,X)\d s_{1-z}/p_z(s_z,X)$ for $z=0,1$. 
Thus,  $\ell_2(S,Z,X)$ consists of the inverse probability weighting of the residuals of the ratio $u_1(s_z,X)$ in this special case.
Importantly, all the terms in the influence function involve the principal density, which further demonstrates the difficulty of avoiding it in identification.

The absence of the term corresponding to the perturbation of $\P(Z=1\mid X)$  implies that knowing the treatment probability would not help improve the estimation of our causal quantities.
A similar phenomenon appears in the estimation of the classic average treatment effect \citep{hahn1998role} and has been studied in \citet{hitomi2008puzzling}.

The EIF motivates the following estimating equation for $\eta_1$.
\begin{lemma}[Estimating equation based on EIF]
\label{lemma:eif_ee}
    Suppose that Assumptions~\ref{assump:treatment_ignorability} and~\ref{assump:principal_ignorability} hold. Define
    \begin{equation}
    \label{equ:eif_ee}
    D_{1,\textup{eif}}\left(Y,S,Z,X; \eta_1, \pi, e, \mu_{1}\right) = \ell_{1}\left(X\right)+\ell_{2}\left(S,Z,X\right)+\ell_{3}\left(Y,S,Z,X\right).
    \end{equation}
    We have 
    \begin{eqnarray}
    \E\left\{D_{1,\textup{eif}}\left(Y,S,Z,X; \eta_1, \pi, e, \mu_{1}\right)\right\} &=& 0.
    \label{equ:eif_ee2}
    \end{eqnarray}

\end{lemma}
This follows directly from the zero property of the EIF. Lemma~\ref{lemma:eif_ee} leads to the following estimation strategy.
\begin{estimator}[Estimator based on EIF]
We provide a two-step estimator $\hat{\eta}_{1,\textup{eif}}$ motivated by the form of EIF. It is based on all three nuisance parameters: the treatment probability, principal density, and outcome mean.
    \begin{itemize}
    \item Step 1: Estimate the treatment probability $\hat \pi$,  principal density $\hat e=(\hat p_1, \hat p_0)$, and outcome mean $\hat \mu_1$.
    \item Step 2: Obtain the estimator $\hat{\eta}_{1,\textup{eif}}$ by plugging the estimated values of the nuisance parameters $(\hat \pi, \hat e, \hat \mu_1)$ into~\eqref{equ:eif_ee} and solving for the empirical analog of the estimating equation~\eqref{equ:eif_ee2}
    \begin{equation}
        \sumn \hat{D}_{1,\textup{eif}}\left(Y_{i},S_{i},Z_{i},X_{i}; \hat{\eta}_{1,\textup{eif}}, \hat{\pi}, \hat{e}, \hat{\mu}_{1}\right) = 0.
        \label{equ:z_estimator_ee}
    \end{equation}
    \end{itemize}
We can similarly construct a two-step estimator $\hat{\eta}_{0,\textup{eif}}$ based on the EIF for $\eta_0$, details of which are relegated to the Supplementary Material.
\end{estimator}

\subsection{Representative working models}
Next, we provide more details for the EIF and the estimator under some representative working models. 
\begin{example}[Working model is constant]
\label{exmp:constant}
Suppose that the weight function and the working model are both constant, i.e., $w_1(s_{1},s_{0})=1$ and $f_1\left(s_{1},s_{0};\eta \right)=\eta$ for all $s_1,s_0$ on the support of $S_1,S_0$. The parameter of interest is simplified to
\begin{eqnarray*}
    \eta_{1} & =&  \arg\min_{\eta}\E \left\{m_{1}\left(S_{1},S_{0}\right)-\eta \right\}^{2} \ =\  \E \left\{m_{1}\left(S_{1},S_{0}\right)\right\} \ = \ \E(Y_{1}).
\end{eqnarray*}
From Theorem~\ref{thm:eif_m1}, we have $H_1=1$ and
\begin{eqnarray*}
\ell_{1}\left(X\right) &=& \E(Y\mid Z=1,X)-\eta_1,\\
\ell_{2}\left(S,Z,X\right) &=& \frac{Z}{\pi\left(X\right)} \{\mu_{1}\left(S,X\right) -\E(Y\mid Z=1,X)\}, \\
\ell_{3}\left(Y,S,Z,X\right) &=&  \frac{Z}{\pi\left(X\right)} \left\{Y-\mu_{1}\left(S,X\right)\right\}.
\end{eqnarray*}
Therefore,  the EIF for $\eta_{1}$ is
\begin{equation*}
    \varphi_{1}\left(V\right)= \frac{Z}{\pi\left(X\right)}\left\{ Y-\E\left(Y\mid Z=1,X\right)\right\} +\E\left(Y\mid Z=1,X\right)-\eta_{1},
\end{equation*}
which is the EIF for the average potential outcome under treatment in observational studies \citep{hahn1998role}. Based on $\varphi_1(V)$, we can derive the canonical doubly robust estimator for $E(Y_1)$ \citep{bang2005doubly}.
\end{example}
Example \ref{exmp:constant} shows that when we project the PCE surface onto a constant, the parameter of interest becomes the average potential outcome $\E(Y_1)$ and the EIF reduces to that for $\E(Y_1)$ in the classic setting. This example acts as a sanity check of Theorem~\ref{thm:eif_m1}.

In the following example, we explore the use of a linear working model, which strikes a balance between reasonable approximation and clear interpretation.
\begin{example}[Working model is linear in $\eta$]
\label{exmp:linear_eta}
Suppose that the working model $f_1(s_{1},s_{0};\eta )$ is linear in $\eta$, i.e., $f_1(s_{1},s_{0};\eta )=\eta^{\T}g_1(s_{1},s_{0})$ for some known pre-specified function $g_1$ but unknown $\eta$. 
We introduce the notation 
\begin{eqnarray*}
G_1(s_1,s_0,X; h) &=& w_1(s_1,s_0)e(s_1,s_0,X)g_1(s_1,s_0)h(s_1,s_0,X)^{\T},
\end{eqnarray*}
where $h(s_1,s_0,X)$ can take various forms below as a scalar function or a vector function whose dimension is compatible with $g_1(s_1,s_0)$. We write it as $G_1(s_1,s_0,X;1)$ when $h(s_1,s_0,X)=1$ is a constant function. 
The EIF for $\eta_1$ is 
\begin{equation*}
    \varphi_{1}\left(V\right)=H_{1}^{-1}\left\{A_1(Y,S,Z,X) -A_2(Y,S,Z,X) \eta_1\right\},
\end{equation*}
where 
\begin{eqnarray*}
    H_{1} &=& \iint w_1\left(s_{1},s_{0}\right)g_1\left(s_{1},s_{0}\right)g_1\left(s_{1},s_{0}\right)^{\T}e\left(s_{1},s_{0}\right)\d s_{1}\d s_{0}
\end{eqnarray*}
and
\begin{eqnarray*}
    && A_1(Y,S,Z,X) \\
    &=& \frac{Z}{\pi\left(X\right)}\left[\frac{\int G_1(S,s_0,X;\mu_1)\d s_0}{p_1(S,X)} - \iint G_1(s_1,s_0,X;\mu_1) r_u(s_1,s_0,S,X) \d s_1 \d s_0\right]\\
    &&+ \frac{1-Z}{1-\pi\left(X\right)}\left[\frac{\int G_1(s_1,S,X;\mu_1)\d s_1}{p_0(S,X)}  - \iint G_1(s_1,s_0,X;\mu_1)  r_v(s_1,s_0,S,X) \d s_1 \d s_0\right]\\
    &&+ \iint G_1(s_1,s_0,X; \mu_1)\d s_1 \d s_0 + \frac{Z}{\pi\left(X\right)}\frac{ \int G_1(S,s_0,X;1)\d s_{0}}{p_{1}\left(S,X\right)}\left\{Y-\mu_{1}\left(S,X\right)\right\},\\
    && A_2(Y,S,Z,X) \\
    &=&\frac{Z}{\pi\left(X\right)}\left[\frac{\int G_1(S,s_0,X;g_1)\d s_0}{p_1(S,X)} - \iint G_1(s_1,s_0,X;g_1) r_u(s_1,s_0,S,X) \d s_1 \d s_0\right]\\
    &&+ \frac{1-Z}{1-\pi\left(X\right)}\left[\frac{\int G_1(s_1,S,X;g_1)\d s_1}{p_0(S,X)}  - \iint G_1(s_1,s_0,X;g_1)  r_v(s_1,s_0,S,X) \d s_1 \d s_0\right]\\
    &&+ \iint G_1(s_1,s_0,X; g_1)\d s_1 \d s_0.
\end{eqnarray*}

We can therefore construct an estimator for $\eta_1$ by solving for Equation~\eqref{equ:z_estimator_ee}. Define the following estimated values
\begin{eqnarray*}
\hat G_1(s_1,s_0,X_i; h)&=& w_1(s_1,s_0)g_1(s_1,s_0)\hat e(s_1,s_0,X_i)h(s_1,s_0,X_i)^{\T},\\
\hat r_u(s_1,s_0,S_i,X_i)&=& 1- \frac{\hat{\dot{c}}_{u}(s_1,s_0,X_i)}{\hat c(s_1,s_0,X_i)}\left\{1(S_i\leq s_1)-\hat \F_1(s_1,X_i)\right\},\\
\hat r_v(s_1,s_0,S_i,X_i)&=& 1- \frac{\hat{\dot{c}}_{v}(s_1,s_0,X_i)}{\hat c(s_1,s_0,X_i)}\left\{1(S_i\leq s_0)-\hat \F_0(s_0,X_i)\right\},
\end{eqnarray*}
where $\hat{\dot{c}}_{*}(s_1,s_0,X_i)/\hat c(s_1,s_0,X_i)$ is the value of  $\dot{c}_{*}(s_1,s_0,X_i)/c(s_1,s_0,X_i)$  for $*=u,v$ evaluated at the estimated distribution functions $\hat \F_1(s_1,X_i)$ and $\hat \F_0(s_0,X_i)$. Then, the estimator is 
\begin{eqnarray*}
\hat{\eta}_{1,\textup{eif}} &=& \frac{ \sumn \hat A_{i2}}{ \sumn \hat A_{i1}},
\end{eqnarray*}
 where
\begin{eqnarray*}
    \hat A_{i1} &=& \frac{Z_i}{\hat \pi\left(X_i\right)}\left[\frac{\int \hat G_1(S_i,s_0,X_i;\hat \mu_1)\d s_0}{\hat p_1(S_i,X_i)} - \iint \hat G_1(s_1,s_0,X_i;\hat \mu_1) \hat r_u(s_1,s_0,S_i,X_i) \d s_1 \d s_0\right]\\
    &&+ \frac{1-Z_i}{1-\hat \pi\left(X_i\right)}\left[\frac{\int \hat G_1(s_1,S_i,X_i;\hat \mu_1)\d s_1}{\hat p_0(S_i,X_i)}  - \iint \hat G_1(s_1,s_0,X_i;\hat \mu_1)  \hat r_v(s_1,s_0,S_i,X_i) \d s_1 \d s_0\right]\\
    &&+ \iint \hat G_1(s_1,s_0,X_i;\hat \mu_1)\d s_1 \d s_0 + \frac{Z_i}{\hat \pi\left(X_i\right)}\frac{ \int \hat G_1(S_i,s_0,X_i;1)\d s_{0}}{\hat p_{1}\left(S_i,X_i\right)}\left\{Y_i-\hat \mu_{1}\left(S_i,X_i\right)\right\}
\end{eqnarray*}
and
\begin{eqnarray*}
    \hat A_{i2} &=& \frac{Z_i}{\hat \pi\left(X_i\right)}\left[\frac{\int \hat G_1(S_i,s_0,X_i;g_1)\d s_0}{\hat p_1(S_i,X_i)} - \iint \hat G_1(s_1,s_0,X_i;g_1) \hat r_u(s_1,s_0,S_i,X_i) \d s_1 \d s_0\right]\\
    &&+ \frac{1-Z_i}{1-\hat \pi\left(X_i\right)}\left[\frac{\int \hat G_1(s_1,S_i,X_i;g_1)\d s_1 }{\hat p_0(S_i,X_i)} - \iint \hat G_1(s_1,s_0,X_i;g_1)  \hat r_v(s_1,s_0,S_i,X_i) \d s_1 \d s_0\right]\\
    &&+ \iint \hat G_1(s_1,s_0,X_i; g_1)\d s_1 \d s_0
\end{eqnarray*}
are the fitted values of $A_1$ and $A_2$.
\end{example}
Under the linear working model, the EIF exhibits linearity with respect to $\eta_1$. This linearity allows closed-form solutions to the system of linear equations, significantly reducing computational complexity in estimation. We implement our proposed estimation procedure based on this closed-form solution in an R package.
The function $g_1(s_{1},s_{0})$ can be tailored to suit various real-world applications. We discuss some specific choices of $g_1(s_1,s_0)$, such as polynomial functions of $(s_1,s_0)$, and provide the corresponding closed-form EIFs in Section~\ref{sec::more_examples_linear_form} in the Supplementary Material.

When the working model is nonlinear in $\eta_1$, the empirical estimating equation~\eqref{equ:z_estimator_ee} often involves nonlinear functions in $\eta_1$.
Consequently, estimating  $\eta_1$ requires the use of numerical methods such as Newton's method,  posing computational challenges.

\section{Asymptotic theory and statistical inference}
\label{sec::asymptotics}
In this section, we derive the asymptotic properties of the estimators based on the EIF. We show it is generally doubly robust and exhibits a parametric convergence rate under certain regularity conditions even when nuisance functions are estimated at slower rates. Similar to Section~\ref{sec::est}, we only give the result for $\hat{\eta}_{1,\textup{eif}}$.

We introduce additional notation to study the estimator under model misspecification. Let $\bar{\pi}, \bar{e}, \bar{\mu}_{1}$ denote the probability limit of the estimated treatment probability $\hat \pi$, principal density $\hat e$, and outcome mean $\hat \mu_1$, respectively. Therefore, $\bar{\pi}=\pi$ indicates that the treatment probability model is correctly specified,  $\bar{e}=e$ indicates that the principal density model is correctly specified, and $\bar{\mu}_1=\mu_1$ indicates that the outcome mean model is correctly specified. 

We first introduce the following two regularity conditions on the estimated nuisance parameters and the estimating equation.  
\begin{assumption}[Donsker]
Let $\eta$ and $\xi$ denote the parameter of interest and the nuisance parameters, respectively. The function $v=(y,s,z,x)\mapsto D_{1,\textup{eif}}(v; \tilde{\eta}, \tilde{\xi})$ indexed by $(\tilde{\eta}, \tilde{\xi})$ satisfies
\begin{enumerate}
    \item[(a)] the class of functions $\{ D_{1,\textup{eif}}(\cdot; \tilde{\eta}, \tilde{\xi}):  \norm{\tilde{\eta} - \eta} < \delta, d(\tilde{\xi}, \bar{\xi}) < \delta \}$ is Donsker for some $\delta>0$, and 
    \item[(b)] $\E\{\norm{D_{1,\textup{eif}}(\cdot; \tilde{\eta},\tilde{\xi}) - D_{1,\textup{eif}}(\cdot; \eta, \bar{\xi} )}^{2} \}  \rightarrow 0$ as $(\tilde{\eta}, \tilde{\xi}) \rightarrow (\eta, \bar{\xi})$.
\end{enumerate}
\label{assump:donsker}
\end{assumption}
\begin{assumption}[Differentiablility]
The mapping $\tilde{\eta} \mapsto \E\{D_{1,\textup{eif}}(V; \tilde{\eta}, \tilde{\xi})\}$ is differentiable at $\eta$ uniformly in $\tilde{\xi}$ in a neighborhood of $\bar{\xi}$ with nonsingular derivative matrices $C(\eta, \tilde{\xi})\rightarrow C(\eta, \bar{\xi}) \equiv \partial \E\{D_{1,\textup{eif}}(V;\tilde{\eta},\bar{\xi})\} / \partial \tilde{\eta} |_{\tilde{\eta}=\eta}$.
\label{assump:differentiablility}
\end{assumption}

Assumption~\ref{assump:donsker} restricts the model complexity of the nuisance parameters. It is a standard regularity condition of Z-estimators \citep{van2000asymptotic}. Moreover, we could use sample-splitting and cross-fitting methods to estimate the model parameters to relax the Donsker conditions \citep{chernozhukov2018double}. Assumption~\ref{assump:differentiablility} is another regularity condition that validates the use of the Delta method when we derive the asymptotic distribution of $\hat \eta_{1,\textup{eif}}$.

The following theorem gives a general result for the asymptotic distribution of $\hat{\eta}_{1,\textup{eif}}$. 

\begin{theorem}[Consistency] 
\label{thm:consistency}
Suppose that the probability limit of the estimated nuisance parameters satisfies that either $(\bar{\pi}, \bar{e})=(\pi, e)$ or $(\bar{e}, \bar{\mu}_{1})=(e, \mu_{1})$,  $0<\pi(x)<1$ for all $x$ in the support of $X$, and the weight function $w_1(s_1,s_0)$ and the working model $f_1(s_1,s_0)$ are bounded. 
\begin{enumerate}[(a)]
\item Under Assumptions~\ref{assump:treatment_ignorability} and~\ref{assump:principal_ignorability}, $\hat{\eta}_{1,\textup{eif}}$ is a consistent estimator of $\eta_1$.
\item under Assumptions~\ref{assump:treatment_ignorability}~to~\ref{assump:differentiablility},
\begin{equation*}
    \norm{\hat{\eta}_{1,\textup{eif}} - \eta_1}_2 = O_{\P}\left\{ n^{-1/2} + \textup{Rem} \right\},
\end{equation*}
where
\begin{eqnarray*}
    \textup{Rem} &=& \norm{\hat{\pi}-\pi}_{2}\norm{\hat{\mu}_{1}-\mu_{1}}_{2}+\norm{\hat{\pi}-\pi}_{2}\left(\norm{\hat{p}_{1}-p_{1}}_{2}+\norm{\hat{p}_{0}-p_{0}}_{2}\right)\\
    &&+ \norm{\hat{\mu}_{1}-\mu_{1}}_{2}\left(\norm{\hat{p}_{1}-p_{1}}_{2}+\norm{\hat{p}_{0}-p_{0}}_{2}\right)+ \norm{\hat{p}_{1}-p_{1}}_{2}\norm{\hat{p}_{0}-p_{0}}_{2}.
\end{eqnarray*}
\end{enumerate}
\end{theorem}
Theorem~\ref{thm:consistency} shows that the asymptotic bias of $\hat{\eta}_{1,\textup{eif}}$ is determined by a function of the asymptotic biases of the nuisance parameters. While $\norm{\hat{\pi}-\pi}_{2}\norm{\hat{\mu}_{1}-\mu_{1}}_{2}$ converges to zero if either the treatment probability or the outcome mean model is correctly specified, the other terms rely heavily on the principal density model. In particular, the last term $ \norm{\hat{p}_{1}-p_{1}}_{2}\norm{\hat{p}_{0}-p_{0}}_{2}$ converges to zero only if the principal density model is correctly specified. As a result, the consistency of the estimator $\hat{\eta}_{1,\textup{eif}}$ for $\eta_1$ is contingent on a correct specification of the principal density model, and either the treatment probability or outcome mean model, leading to the double robustness of $\hat{\eta}_{1,\textup{eif}}$. This contrasts with the triply robust property demonstrated in \citet{jiang2020multiply}, which shows that the estimator can be consistent even in the absence of a correct principal score model, provided the treatment probability and outcome mean models are correct. The result echos the discussion on the crucial role of the principal density in identifying the PCE surface with a continuous $S$.

Theorem~\ref{thm:consistency} also implies the result on root-$n$ consistency of $\hat{\eta}_{1,\textup{eif}}$, provided the following assumption on the convergence rates of the 
treatment probability, principal density, and the outcome mean.
\begin{assumption}[Convergence rates of nuisance parameters]
The convergence rates of the estimators of the nuisance parameters satisfy
\begin{enumerate}
    \item[(a)] $\norm{\hat{\pi} - \pi}_{2} \norm{\hat{p}_{z} - p_{z}}_{2} = o_{\P}(n^{-1/2})$ for $z=0,1$,
    \item[(b)] $\norm{\hat{p}_{z} - p_{z}}_{2} \norm{\hat{\mu}_{1} - \mu_{1}}_{2} = o_{\P}(n^{-1/2})$ for $z=0,1$,
    \item[(c)] $\norm{\hat{\pi} - \pi}_{2} \norm{\hat{\mu}_{1} - \mu_{1}}_{2} = o_{\P}(n^{-1/2})$, and
    \item[(d)] $\norm{\hat{p}_{z}-p_{z}}_{2} = o_{\P}(n^{-1/4})$ for $z=0,1$.
\end{enumerate}
\label{assump:nuisance_rate}
\end{assumption}

Assumption~\ref{assump:nuisance_rate} essentially requires that the models on treatment probability, principal density, and outcome mean are all correctly specified. Moreover, the conditions impose more stringent requirements on the convergence rates of the models. If domain knowledge enables us to correctly specify parametric models to estimate the treatment probability and the outcome model, parametric estimation guarantees a root-$n$ rate of nuisance parameters $\pi$ and $\mu_z$, thus Assumptions \ref{assump:nuisance_rate}(a)--\ref{assump:nuisance_rate}(c) are achieved. However, parametric models might be restrictive in observational studies with multi-dimensional covariates. We can instead use flexible machine learning models for estimation and still achieve the required rates \citep{chernozhukov2018double}. For the conditional density estimation of $p_z$, we can use nonparametric density estimation such as kernel density estimation when the covariates $X$ is low-dimensional. With larger dimensions of the covariates, we need to impose more restrictions on the principal density to achieve the required convergence rate of $\hat{p}_z$ in Assumption~\ref{assump:nuisance_rate}(d).

\begin{corollary}[Root-$n$ consistency]
\label{cor:double_robustness}
Under Assumptions~\ref{assump:treatment_ignorability}--\ref{assump:nuisance_rate}, we have $\norm{\hat{\eta}_{1,\textup{eif}} - \eta_1}_2 = O_{\P}( n^{-1/2}  )$.
\end{corollary}

The next theorem shows the asymptotic Normality and semiparametric efficiency of $\hat{\eta}_{1,\textup{eif}}$. 
\begin{theorem}[Asymptotic distribution]
Under Assumptions~\ref{assump:treatment_ignorability}--\ref{assump:nuisance_rate}, we have
\begin{equation*}
\sqrt{n}\left(\hat{\eta}_{1,\textup{eif}}-\eta_1 \right) \overset{\textup{d}}{\rightarrow}\mathcal{N}\left(0,\E\left\{\varphi_{1}\left(V;\eta_1,\xi\right)\varphi_{1}\left(V;\eta_1,\xi\right)^{\T}\right\}\right)
\end{equation*}
where $\varphi_{1}(V;\eta_1,\xi)$ is the EIF for $\eta_1$ as in Theorem \ref{thm:eif_m1}. That is,  $\hat{\eta}_{1,\textup{eif}}$ achieves the semiparametric efficiency bound.
\label{thm:asymptotic_distribution}
\end{theorem}

If Assumption \ref{assump:nuisance_rate} holds, we can construct a consistent estimator of the asymptotic variance using the sample analog of the semiparametric efficiency bound, i.e.,
$
n^{-1}\sumn \varphi_1(V;\hat{\eta}_{1,\textup{eif}}, \hat{\xi}) \varphi_1(V;\hat{\eta}_{1,\textup{eif}}, \hat{\xi})^{\T}.
$
This, however, requires all the nuisance models to be correctly specified.
Under certain smoothness conditions on the nuisance models, we can estimate the asymptotic variance of $\hat{\eta}_{1,\textup{eif}}$ using nonparametric bootstrap even if Assumption \ref{assump:nuisance_rate} fails \citep{kennedy2019robust}. If neither Assumption \ref{assump:nuisance_rate} holds nor the requirements for valid nonparametric bootstrap are satisfied, the statistical inference needs more careful procedures. 
 This is beyond the scope of this paper and we leave it to future work. 

\section{Simulation}
\label{sec::simulation}
In this section, we evaluate the finite sample performance of various estimators for $\eta_1$ defined in Equation~\eqref{equ:opt_m1}. We employ a linear projection working model $f_1(s_1,s_0;\eta_1)=\beta_1^1 s_1 + \beta_1^0 s_0 + \alpha_1$ anduse the uniform weighting function $w_1(s_1, s_0)=1$ for any $s_1$ and $s_0$. 

Generate the covariates $X=(X_1,X_2)^{\T}\in \mathbb{R}^2$ from two independent standard Normal distributions and the treatment by $Z\mid X \sim \textup{Bernoulli}\{\pi(X)\}$. Given a pre-specified value of $\rho$, generate the potential values of post-treatment variable by $S_1=0.6X_1 + 0.8\{\rho^{1/2}\epsilon_\rho + (1-\rho)^{1/2}\epsilon_{1}\}$ and $S_0=0.6X_2 + 0.8\{\rho^{1/2}\epsilon_\rho + (1-\rho)^{1/2}\epsilon_{0}\}$, where $\epsilon_\rho$, $\epsilon_{1}$, and $\epsilon_{0}$ are independently from standard normal distribution.
Generate the potential outcomes by $Y_z\mid S_1,S_0,X\sim \mathcal{N}(\mu_z(S_1,S_0,X), 1)$ for $z=0,1$. The observed post-treatment variable and outcome are then determined by $S=ZS_1+(1-Z)S_0$ and $Y=ZY_1+(1-Z)Y_0$, respectively. We consider three different data-generating regimes.  In the first regime, $\pi(X)=0.5$, $\mu_1(S_1,S_0,X)=S_1+X_1$ and $\mu_0(S_1,S_0,X)=S_0+X_2$. In the second regime, $\pi(X)=\logit\{(\tilde{X}_1 + \tilde{X}_2)/2\}$, where $\tilde{X}_j=\{(X_j+0.25)^2-1\}/\sqrt{2}$ for $j=1,2$, and $\mu_z(S_1,S_0,X)$ is the same as in the first regime. In the third regime, $\pi(X)$ is the same as in the first regime,  and $\mu_1(S_1,S_0,X)=S_1+2\sqrt{2}\tilde{X}_1$ and $\mu_0(S_1,S_0,X)=S_0+2\sqrt{2}\tilde{X}_2$.  

We consider three estimators: the EIF based estimator $\hat \eta_{1,\textup{eif}}$, and the estimators based on the estimating equations~\eqref{equ:pd+om_ee}~and~\eqref{equ:weighting_ee1}, denoted by $\hat{\eta}_{1,\textup{pd+om}}$ and $\hat{\eta}_{1,\textup{tp+pd}}$, respectively. We estimate the treatment probability by logistic regressions with linear predictors $(X_1,X_2)$, and outcome means by linear regressions with predictors $(S,X_1,X_2)$ fitted separately in the treatment and control groups. Therefore, the treatment probability model is misspecified in the second regime and the outcome mean model is misspecified in the third regime. We estimate the principal density using the nonparametric kernel conditional density estimation and Gaussian copula with the correctly specified correlation coefficient $\rho$. We take the numerical integration when computing the single and double integrals.

For each data generating regime, we generate a random sample of size $n$ and compute the values of three estimators. We compare their performance by repeating the process over 500 Monte Carlo samples and compute the bias and the standard deviation (SD) of the three estimators. To avoid repetition, we only report the results for $\beta_1^1$, which is the projection coefficient of $s_1$ in the working model $f_1(s_1,s_0;\eta_1)=\beta_1^1 s_1 + \beta_1^0 s_0 + \alpha_1$. To explore how the finite-sample performance varies across different sample sizes $n$, we choose four sample sizes $n=200,500,1000,2000$. We generate data following the aforementioned processes with various values of the correlation coefficient $\rho=0, 0.2, 0.5$. 

Tables~\ref{tab::simu_res_rho0}~to~\ref{tab::simu_res_rho0.5} show the biases and the root mean squared errors (RMSEs) of the three estimators across various values of the sensitivity parameter $\rho=0, 0.2, 0.5$ under each data generating regime. The performance is similar across different $\rho$ values. The EIF estimator $\hat{\beta}^1_{1,\textup{eif}}$ is consistent for the true value if either the treatment probability or the outcome mean model is correctly specified. The weighting estimator $\hat{\beta}^1_{1,\textup{tp+pd}}$ performs well when the treatment probability model is correctly specified and is inconsistent otherwise. The finite-sample performance of $\hat{\beta}^1_{1,\textup{pd+om}}$ is not satisfying under all three regimes. There exists a large finite-sample bias even when all models are correctly specified, although the bias decreases as the sample size increases. Overall, we recommend using the EIF estimator as it is generally robust to the misspecification of either the treatment probability or the outcome model.

\begin{table}[h]
\doublespacing
\centering
\caption{Biases and RMSEs of the three estimators $\hat\beta_{1,\textup{eif}}^1$, $\hat\beta_{1,\textup{tp+pd}}^1$, and $\hat\beta_{1,\textup{pd+om}}^1$ with $\rho = 0$. In Regime 1, both the treatment probability and outcome mean models are correct; In Regime 2, the treatment probability model is misspecified; In Regime 3, the outcome mean models are misspecified. The principal density model is correct in all three regimes.}
\begin{small}

\begin{tabular}{lllcclcclcc}
\\
\hline
\hline
& & &\multicolumn{2}{c}{$\hat\beta_{1,\textup{eif}}^1$}& &\multicolumn{2}{c}{$\hat\beta_{1,\textup{tp+pd}}^1$}&   &\multicolumn{2}{c}{$\hat\beta_{1,\textup{pd+om}}^1$}\\
\cline{4-5} \cline{7-8} \cline{10-11}
& & &Bias&  RMSE&   &Bias&  RMSE&   &Bias&  RMSE\\
\hline
Regime 1 & $n = 200$ &  & 0.003 & 0.150 &  & 0.010 & 0.138 &  & -0.147 & 0.199\\
 & $n = 500$ &  & -0.004 & 0.099 &  & -0.003 & 0.092 &  & -0.126 & 0.155 \\
 & $n = 1000$ &  & 0.004 & 0.067 &  & 0.005 & 0.064 &  & -0.096 & 0.115\\
 & $n = 2000$ &  & 0.001 & 0.046 &  & 0.003 & 0.045 &  & -0.081 & 0.092\\
 \hline
Regime 2 & $n = 200$ &  & 0.008 & 0.156 &  & 0.099 & 0.165 &  & -0.154 & 0.208\\
 & $n = 500$ &  & 0.003 & 0.100 &  & 0.096 & 0.130 &  & -0.122 & 0.151\\
 & $n = 1000$ &  & 0.008 & 0.069 &  & 0.104 & 0.122 &  & -0.095 & 0.114\\
 & $n = 2000$ &  & 0.006 & 0.051 &  & 0.102 & 0.111 &  & -0.078 & 0.090\\
 \hline
Regime 3 & $n = 200$ &  & 0.024 & 0.464 &  & 0.027 & 0.440 &  & -0.131 & 0.422\\
 & $n = 500$ &  & -0.012 & 0.316 &  & -0.011 & 0.302 &  & -0.133 & 0.318\\
 & $n = 1000$ &  & 0.016 & 0.213 &  & 0.018 & 0.207 &  & -0.082 & 0.215\\
 & $n = 2000$ &  & 0.014 & 0.148 &  & 0.015 & 0.145 &  & -0.068 & 0.162\\
\hline
\end{tabular}
\end{small}
\label{tab::simu_res_rho0}
\end{table}

\begin{table}[h]
\doublespacing
\centering
\caption{Biases and RMSEs of the three estimators $\hat\beta_{1,\textup{eif}}^1$, $\hat\beta_{1,\textup{tp+pd}}^1$, and $\hat\beta_{1,\textup{pd+om}}^1$ with $\rho = 0.2$. In Regime 1, both the treatment probability and outcome mean models are correct; In Regime 2, the treatment probability model is misspecified; In Regime 3, the outcome mean models are misspecified. The principal density model is correct in all three regimes.}
\begin{small}

\begin{tabular}{lllcclcclcc}
\\
\hline
\hline
& & &\multicolumn{2}{c}{$\hat\beta_{1,\textup{eif}}^1$}& &\multicolumn{2}{c}{$\hat\beta_{1,\textup{tp+pd}}^1$}&   &\multicolumn{2}{c}{$\hat\beta_{1,\textup{pd+om}}^1$}\\
\cline{4-5} \cline{7-8} \cline{10-11}
& & &Bias&  RMSE&   &Bias&  RMSE&   &Bias&  RMSE\\
\hline
Regime 1 & $n = 200$ &  & -0.026 & 0.161 &  & -0.042 & 0.154 &  & -0.184 & 0.233\\
 & $n = 500$ &  & -0.008 & 0.097 &  & -0.023 & 0.094 &  & -0.132 & 0.160\\
 & $n = 1000$ &  & -0.002 & 0.066 &  & -0.015 & 0.067 &  & -0.102 & 0.120\\
 & $n = 2000$ &  & -0.003 & 0.046 &  & -0.014 & 0.048 &  & -0.086 & 0.097\\
 \hline
Regime 2 & $n = 200$ &  & -0.005 & 0.156 &  & 0.059 & 0.149 &  & -0.170 & 0.220\\
 & $n = 500$ &  & -0.009 & 0.100 &  & 0.064 & 0.111 &  & -0.134 & 0.162\\
 & $n = 1000$ &  & 0.002 & 0.068 &  & 0.079 & 0.102 &  & -0.100 & 0.120\\
 & $n = 2000$ &  & 0.002 & 0.049 &  & 0.078 & 0.091 &  & -0.083 & 0.095\\
 \hline
Regime 3 & $n = 200$ &  & 0.030 & 0.490 &  & 0.011 & 0.464 &  & -0.140 & 0.439\\
 & $n = 500$ &  & -0.009 & 0.325 &  & -0.024 & 0.311 &  & -0.140 & 0.327\\
 & $n = 1000$ &  & 0.018 & 0.215 &  & 0.005 & 0.209 &  & -0.087 & 0.217\\
 & $n = 2000$ &  & 0.015 & 0.151 &  & 0.005 & 0.147 &  & -0.073 & 0.166\\
\hline
\end{tabular}
\end{small}
\label{tab::simu_res_rho0.2}
\end{table}

\begin{table}[h]
\doublespacing
\centering
\caption{Biases and RMSEs of the three estimators $\hat\beta_{1,\textup{eif}}^1$, $\hat\beta_{1,\textup{tp+pd}}^1$, and $\hat\beta_{1,\textup{pd+om}}^1$ with $\rho = 0.5$. In Regime 1, both the treatment probability and outcome mean models are correct; In Regime 2, the treatment probability model is misspecified; In Regime 3, the outcome mean models are misspecified. The principal density model is correct in all three regimes.}
\begin{small}

\begin{tabular}{lllcclcclcc}
\\
\hline
\hline
& & &\multicolumn{2}{c}{$\hat\beta_{1,\textup{eif}}^1$}& &\multicolumn{2}{c}{$\hat\beta_{1,\textup{tp+pd}}^1$}&   &\multicolumn{2}{c}{$\hat\beta_{1,\textup{pd+om}}^1$}\\
\cline{4-5} \cline{7-8} \cline{10-11}
& & &Bias&  RMSE&   &Bias&  RMSE&   &Bias&  RMSE\\
\hline
Regime 1 & $n = 200$ &  & -0.055 & 0.169 &  & -0.101 & 0.185 &  & -0.193 & 0.244\\
 & $n = 500$ &  & -0.031 & 0.103 &  & -0.064 & 0.115 &  & -0.139 & 0.167\\
 & $n = 1000$ &  & -0.015 & 0.070 &  & -0.049 & 0.084 &  & -0.104 & 0.123\\
 & $n = 2000$ &  & -0.010 & 0.050 &  & -0.036 & 0.061 &  & -0.081 & 0.094\\
 \hline
Regime 2 & $n = 200$ &  & -0.048 & 0.174 &  & -0.017 & 0.151 &  & -0.188 & 0.241\\
 & $n = 500$ &  & -0.033 & 0.110 &  & 0.008 & 0.098 &  & -0.142 & 0.172\\
 & $n = 1000$ &  & -0.013 & 0.071 &  & 0.034 & 0.076 &  & -0.102 & 0.123\\
 & $n = 2000$ &  & -0.009 & 0.050 &  & 0.041 & 0.063 &  & -0.083 & 0.095\\
 \hline
Regime 3 & $n = 200$ &  & 0.025 & 0.549 &  & -0.023 & 0.517 &  & -0.155 & 0.496\\
 & $n = 500$ &  & -0.005 & 0.356 &  & -0.043 & 0.344 &  & -0.141 & 0.355\\
 & $n = 1000$ &  & 0.017 & 0.244 &  & -0.013 & 0.239 &  & -0.085 & 0.237\\
 & $n = 2000$ &  & 0.014 & 0.168 &  & -0.010 & 0.164 &  & -0.074 & 0.180\\
\hline
\end{tabular}
\end{small}
\label{tab::simu_res_rho0.5}
\end{table}

\section{Application}
\label{sec::application}
In this section, we revisit Example~\ref{exmp:yin2022} and analyze the data from a randomized experiment in China \citep{yin2022learning}. The study aims to understand the causal effect of offered credit limit increase on consumer income expectation and economic decision-making. The randomized controlled trial is conducted in collaboration with a large commercial bank in China. In the trial, bank customers are randomly assigned to two groups: customers in the treatment group, denoted by $Z=1$, are immediately offered an increase in their available credit, while customers in the control group, denoted by $Z=0$, are not qualified for the credit limit increase until 6 months after the trial. The outcome of interest $Y$ represents customers' consumption of non-durable goods in dollars 6 months after the treatment. The difference-in-means estimator of the average treatment effect $\E\{Y(1)-Y(0)\}$ is 678.487 with a $95\%$ confidence interval (592.157, 764.818), which indicates that the treatment leads to an increase in future consumption. 

In addition to the treatment and the outcome, the survey also collected data on customers' expectations for their income in 6 months in both a pre-experiment survey right before the treatment and a post-experiment survey after the survey. 
We define the post-treatment variable of interest $S$ as the difference in customers' income expectations between the pre- and post-treatment surveys. People's expectations change with time, and the treatment has a significantly positive effect on people's expected future income on average. The difference-in-means estimator of $\E\{S(1)-S(0)\}$ is 749.590 with a $95\%$ confidence interval (635.246, 863.934). Based on these data, we aim to understand how the increase in credit limit affects consumption behavior through changes in income expectations. We explore how the PCE surface $\tau(s_1,s_0)$ varies with different values of $s_1$ and $s_0$, which could provide insights into the causal mechanism of the treatment effect. 

We specify our working models $f_1$ and $f_0$ to be linear in $s_1$ and $s_0$ with an intercept, i.e., $f_z(s_1, s_0;\eta_z) = \beta_z^1 s_1 + \beta_z^0 s_0 + \alpha_z $ for $z=0,1$, and thus the estimated PCE surface has a functional form of 
\begin{eqnarray*}
    f_{\tau}(s_1, s_0;\eta_{\tau}) & = & \beta_{\tau}^1 s_1 + \beta_{\tau}^0 s_0 + \alpha_{\tau},
\end{eqnarray*}
where $\beta_{\tau}^1 = \beta_{1}^{1} - \beta_{0}^{1}$, $\beta_{\tau}^0 = \beta_{1}^{0} - \beta_{0}^{0}$, and $\alpha_{\tau} = \alpha_1 - \alpha_0$. Under the linear working models $f_1$ and $f_0$, we have $\tau(s_1,s_0)-\tau(s'_1,s_0)=  \beta_{\tau}^1 (s_1-s'_1)$ for all $s_0$. Therefore,  $\beta_\tau^1$ quantifies the extent to which the effect of credit limit increase on the consumption changes with varying potential values of income expectation under treatment. Similarly, $\beta_\tau^0$ measures the changes in the treatment effect with varying potential values of income expectation under control.  Moreover,  when  $s_1=s_0=s$, $\tau(s_1,s_0)= (\beta_{\tau}^1+ \beta_{\tau}^0 ) s + \alpha_{\tau} $, indicating the magnitude of the treatment effect not attributed to the change in expected future income.

\begin{figure}[h]
    \centering
    \includegraphics[width=\textwidth]{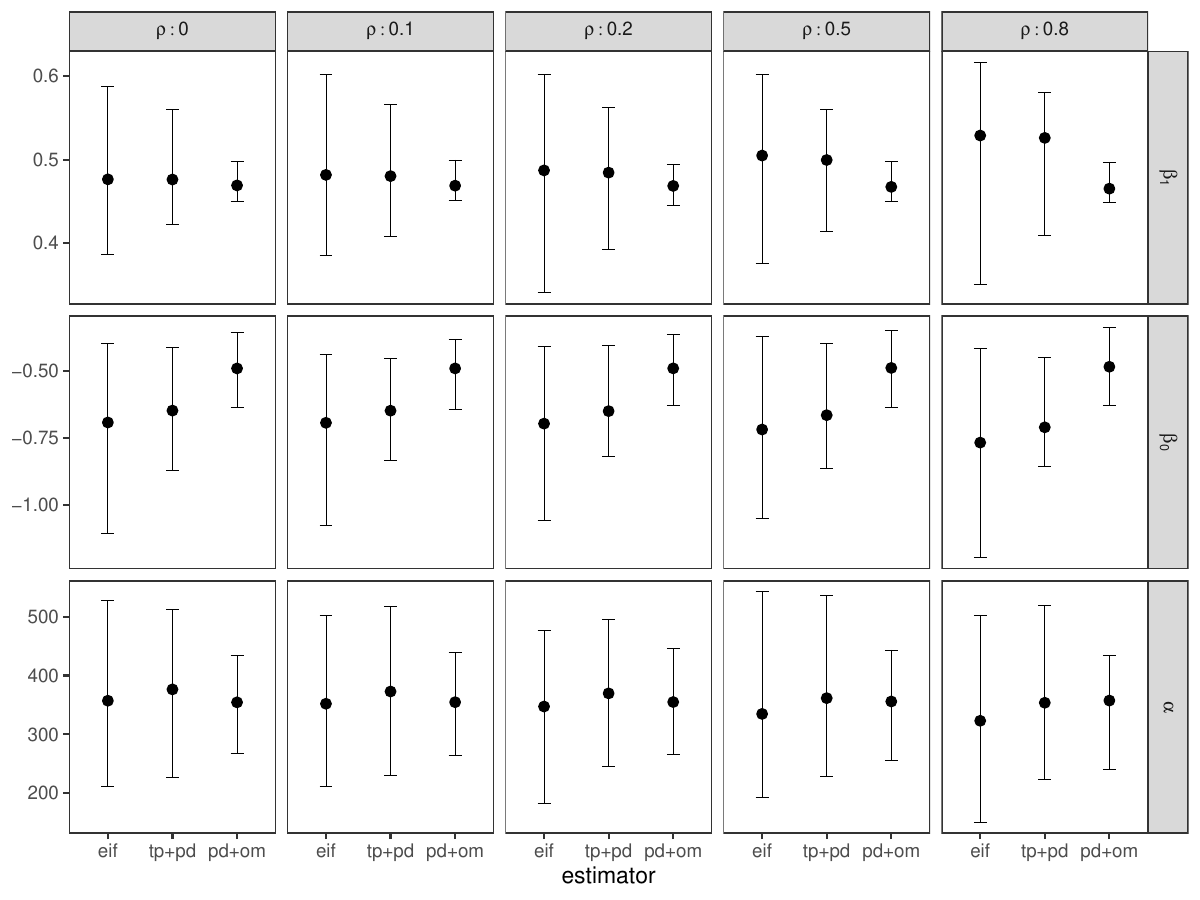}
    \caption{Estimated PCE surface of the credit limit increase on the future consumption. The rows correspond to the three parameters in the working model and the columns correspond to different values of $\rho$. Each subplot shows the point estimates and the corresponding 95\% confidence intervals for the three estimation strategies.}
    \label{fig:app_results}
\end{figure}

In this application, Assumption~\ref{assump:treatment_ignorability} is guaranteed by the design of the original randomized controlled trial. For Assumption~\ref{assump:principal_ignorability}, we conditional on the gender, age, annual income, and credit score.
We use logistic regression for the treatment probability model $\pi(X)$, kernel density estimation for the conditional densities $p_z(s_z, X)$, and linear regression for the outcome models $\mu_z(s_z,X)$, for $z=0,1$. To get the estimated principal density, we specify the Gaussian copula with different values of correlation coefficient $\rho$. We compute point estimators of the coefficients of the EIF estimator and the two basic estimators, and construct the 95\% confidence intervals using the nonparametric bootstrap. The results of $(\hat{\beta}_{\tau,*}^1, \hat{\beta}_{\tau,*}^0, \hat{\alpha}_{\tau,*}) \ (* = \textup{eif},\textup{tp+pd},\textup{pd+om})$ are summarized in Figure~\ref{fig:app_results} across various values of the sensitivity parameter $\rho$. The point estimators do not vary much across different values of $\rho$ and estimation strategies. By nature of the randomized controlled trials, the treatment is randomly assigned thus the treatment probability model is guaranteed to have a correct specification, leading to similar point estimators of the weighting estimator and the EIF estimator. 
Overall, the estimated PCE surface exhibits an increasing trend in $S_1$ and a decreasing trend in $S_0$, suggesting that the credit limit increase offer positively affects customers' consumption behavior with heterogeneity across principal strata. For example, consider the EIF estimated PCE surface when $\rho = 0$. The estimated functional form is $f_\tau(s_1,s_0) = 0.476 s_1 - 0.693 s_0 + 357.067$.  This suggests that as customers update their expected future income, the treatment effect on consumption tends to increase. When $s_1 = s_0 = s$, the direct effect of the credit limit increase offer is estimated as $-0.217 s + 357.067$, which is positive for most observed customers (77.4 percentile of observed $S$). Notably, customers with higher expected future income tend to experience a smaller direct effect.   

\section{Discussion}
\label{sec::discussion}
This paper studies the PCEs involving continuous post-treatment variables, extending binary post-treatment variables methodologies. 
We identify the PCEs by generalizing the principal ignorability assumption and introducing a copula model for the joint potential values of the post-treatment variable.
Because the PCEs are two-dimensional functions in the continuous case, we project these functions onto a pre-specified working model to enhance estimation precision and interpretation. 
We derive various estimators for the projection parameters using different identification formulas and the EIFs for the parameters. The proposed EIF estimator is doubly robust and attains the semiparametric efficiency bound contingent on the performance of estimated nuisance parameters.

The identification of PCEs relies on the crucial principal ignorability assumption, which is untestable using observed data. This necessitates sensitivity analysis to assess the robustness of conclusions to the violation of principal ignorability. We introduce the following sensitivity parameters: 
\begin{eqnarray}
    \varepsilon_1(s_0,X) &=& \frac{\E(Y_1\mid U=s_1s_0, X)}{\E(Y_1\mid S_1=s_1, X)} \quad \textup{and} \quad
    \varepsilon_0(s_1,X) \ =\  \frac{\E(Y_0\mid U=s_1s_0, X)}{\E(Y_0\mid S_0=s_0, X)}.
\label{equ:sensitivity}
\end{eqnarray}
Values of these two parameters characterize the extent of principal ignorability violation; both $ \varepsilon_1(s_0,X)$ and $ \varepsilon_1(s_0,X)$ become $1$ when principal ignorability holds.
This sensitivity analysis setup generalizes that for a binary post-treatment variable in \citet{jiang2020multiply}.  Under~\eqref{equ:sensitivity}, we can obtain results for identification, estimation, and semiparametric theory for the projection parameters in parallel to those under principal ignorability. We leave this to future work.

We outline several potential extensions of our framework in the appendix. First, we can extend the framework to settings with multiple continuous post-treatment variables, in which $S$ is a random vector. Extending the principal ignorability to encompass vector $S$, we can derive analogous identification formulas for the projection parameters. However, this extension poses challenges in specifying the copula model and complicates the derivation of the EIF. Second, we can consider the conditional PCEs, defined as     $ \E(Y_1-Y_0\mid U=s_1s_0,W=w)$, where $W$ is a subset of the observed covariates $X$. These conditional PCEs are identifiable under the same set of assumptions, and we can similarly define projection parameters using working models. We leave the comprehensive investigation of these extensions to future work.

\pdfbookmark[1]{References}{References}
\spacingset{1.45}
\bibliographystyle{Chicago}
\bibliography{ps_continuous_post_treatment}

\newpage

\appendix

\setcounter{equation}{0}
\setcounter{figure}{0}
\setcounter{theorem}{0}
\setcounter{lemma}{0}
\setcounter{section}{0}
\setcounter{corollary}{0}
\setcounter{example}{0}
\renewcommand {\theequation} {S\arabic{equation}}
\renewcommand {\thefigure} {S\arabic{figure}}
\renewcommand {\thetheorem} {S\arabic{theorem}}
\renewcommand {\thelemma} {S\arabic{lemma}}
\renewcommand {\thesection} {S\arabic{section}}
\renewcommand {\thecorollary} {S\arabic{corollary}}
\renewcommand {\theexample} {S\arabic{example}}

\begin{center}
  \LARGE {\bf Supplementary Material}
\end{center}

Section~\ref{app::extension} extends the framework to two scenarios for multiple post-treatment variables and conditional PCEs.

Section~\ref{app::proof} provides proofs of the theorems, lemmas, and examples. 

Section~\ref{app::example} discusses more examples of the functional forms of working models and provides the forms of EIFs. 

Section~\ref{app::eif-eta0} gives the EIF for $\eta_0$ with a linear working model example. 

Section~\ref{app::eif-tau} discusses the alternative projection parameter of interest where we directly project the PCE surface to a working model. We provide EIF and additional results on the relationship between the two projection methods.

\section{Extensions}
\label{app::extension}
\subsection{Multiple post-treatment variables}
\label{sec::multiple_s}
We can extend the framework to settings with multiple continuous post-treatment variables. For simplicity, we focus the case with two continuous post-treatment variables. Suppose the two post-treatment variables, denoted by $S$ and $T$, are both continuous with potential outcomes $(S_1,S_0)$ and $(T_1,T_0)$. Their joint value determines the principal strata $U=s_1s_0t_1t_0$. Again, there are infinitely many principal strata when both post-treatment variables are continuous, and the PCE surface can be defined as a function of $s_1$, $s_0$, $t_1$, and $t_0$:
\[
\tau(s_1,s_0,t_1,t_0)=\E(Y_1-Y_0\mid U=s_1s_0t_1t_0).
\]
We generalize the principal ignorability assumption  as 
$\E(Y_{1}\mid U=s_{1}s_{0}t_1t_0,X) = \E(Y_{1}\mid U=s_{1}s^{\prime}_{0}t_1t^\prime_0,X)$ for any $s_{0}, s^{\prime}_{0},t_0$, and $t^\prime_0$, and $\E(Y_{0}\mid U=s_{1}s_{0}t_1t_0,X) = \E(Y_{0}\mid U=s^{\prime}_{1}s_{0}t^\prime_1 t_0,X)$ for any $s_{1}, s^{\prime}_{1},t_1$, and $t^\prime$.
Under treatment ignorability, the principal ignorability assumption leads to the identification of the mean surfaces:
\begin{eqnarray*}
\E(Y_1\mid U=s_1s_0t_1t_0, X) &=& \E(Y_1\mid S=s_1, T=t_1, Z=1, X) \ = \ \E(Y\mid S=s_1, T=t_1, Z=1, X),\\
\E(Y_0\mid U=s_1s_0t_1t_0, X) &=& \E(Y_0\mid S=s_0, T=t_0, Z=0, X) \ = \ \E(Y\mid S=s_0, T=t_0, Z=0, X).
\end{eqnarray*}
Similar to the case with a single post-treatment variable, the identification of PCE surface requires that of the principal density, denoted by $e(s_1,s_0,t_1,t_0,X)=p(U=s_1s_0t_1t_0\mid X)$.
Because the marginal distributions $p(S_1=s_1,T_1=t_1\mid X)$ and $p(S_0=s_0,T_0=t_0\mid X)$ are identifiable under treatment ignorability, we can specify a copula model to obtain the principal density from these marginal distributions. However, this copula approach becomes challenging with large $K$. Let $\mu_1(s_1,t_1,X)=\E(Y\mid S=s_1, T=t_1, Z=1, X)$ and  $\mu_0(s_0,t_0,X)=\E(Y\mid S=s_0, T=t_0, Z=0, X)$ denote the conditional outcome means. We can obtain the following identification formula for the  average potential outcome surface:
\[
\E(Y_z\mid U=s_1s_0t_1t_0) = \E\left\{ \frac{e(s_1,s_0,t_1,t_0,X)}{e(s_1,s_0,t_1,t_0)} \mu_z(s_z,t_z,X)  \right\},
\]
where $e(s_1,s_0,t_1,t_0)=\E\{e(s_1,s_0,t_1,t_0,X)\}$.
We can also generalize the projection approach with multiple post-treatment variables and derive the EIF for the projection parameters. However, the derivation of the EIF becomes considerably more complex due to the increased number of models to perturb when computing the derivatives.

\subsection{Conditional PCE}
In addition to the marginal PCEs, we can also consider the conditional PCE surface:
\begin{eqnarray*}
    \tau(s_1,s_0,w) &=& \E(Y_1-Y_0\mid U=s_1s_0,W=w),
\end{eqnarray*}
where $W$ is a subset of the observed covariates $X$. Let  $m_z(s_1,s_0,w)=\E(Y_z\mid U=s_1s_0,W=w)$ denote the conditional potential outcome surfaces for $z=0,1$. Under Assumptions~\ref{assump:treatment_ignorability} and~\ref{assump:principal_ignorability}, we have the following identification formula for $m_1(s_1,s_0,w)$:
\begin{eqnarray*}
    m_1(s_1,s_0,w) &=& \E\left\{\E(Y_1\mid U=s_1s_0,W=w,X)\mid U=s_1s_0,W=w\right\} \\
    &=& \E\left\{\frac{e(s_1,s_0,X\mid W=w)}{e(s_1,s_0\mid W=w)}\mu_1(s_1,X)\mid W=w\right\},
\end{eqnarray*}
where $e(s_1,s_0,X\mid W=w)=p(U=s_1s_0\mid X, W=w)$ is the conditional principal density on the subset $W=w$, and $e(s_1,s_0\mid W=w)=\E\{e(s_1,s_0,X\mid W=w)\mid W=w\}$ is marginalized over $X\mid W=w$. The identification of $m_0(s_1,s_0,w)$ can be achieved similarly. Employing the same projection approach, we define our causal parameter of interest as the projection parameter
\begin{eqnarray*}
    \eta_z &=& \arg\min_{\eta} \E\left[w_z(S_1,S_0,W)\left\{m_z(S_1,S_0,W)-f_z(S_1,S_0,W;\eta)\right\}^2\right],
\end{eqnarray*}
where the pre-specified weights $w_z(s_1,s_0,w)$ and working models $f_z(s_1,s_0,w)$ are functions of $w$ as well.
We can then develop the estimation strategies and derive the efficiency theory for $\eta_z$ following a similar approach in the main text.

In the special case when $W=X$, $m_1(s_1,s_0,x)=\E(Y_1\mid U=s_1s_0,X=x)=\E(Y_1\mid S_1=s_1,X=x)$ is not a function of $s_0$. Similarly, $m_0(s_1,s_0,x)$ is not a function of $s_1$. We can simplify the notation to $m_z(s_z,x)$ for $z=0,1$, and specify the working models as functions solely of $(s_z,x)$. 

\section{Proofs}
\label{app::proof}
In this section, we provide proof of all the results in the main text. We first introduce the following lemma and will use it to simplify the proofs.

\begin{lemma}\label{lem:weight} Let $X$ and $Y$ be two random variables
with densities $f_{1}(x)$ and $f_{2}(y)$. For any function $h(\cdot)$
with $\E\{h(X)\}<\infty$, we have 
\begin{eqnarray*}
\E\{h(X)\}=\E\left\{ \frac{f_{1}(Y)}{f_{2}(Y)}h(Y)\right\} .
\end{eqnarray*}
\end{lemma}
The proof is straightforward and thus omitted.

\subsection{Proof of Theorem \ref{thm:identification}}
We have 
\begin{align*} 
\E\left(Y_{1}\mid X,U=s_{1}s_{0}\right) & = \E\left(Y_{1}\mid X,U=s_{1}s_{0}\text{ or }s_{1}s_{0}^{\prime},\forall s_{0}^{\prime}\right)\\
& = \E\left(Y_{1}\mid X,Z=1,U=s_{1}s_{0}\text{ or }s_{1}s_{0}^{\prime},\forall s_{0}^{\prime}\right) \\
& = \E\left(Y_1\mid X,Z=1,S_1=s_{1}\right)\\
& = \mu_{1}\left(s_{1},X\right),
\end{align*}
where the first equality follows from principal ignorability and the third equality follows from treatment ignorability.
Therefore, 
\begin{align}
\notag \E\left(Y_{1}\mid U=s_{1}s_{0}\right) & = \E\left\{ \E\left(Y_{1}\mid X,U=s_{1}s_{0}\right)\mid U=s_{1}s_{0}\right\} \\
\notag & = \E\left\{ \mu_{1}\left(s_{1},X\right)\mid U=s_{1}s_{0}\right\} \\
\notag & = \E\left\{ \frac{p\left(X\mid U=s_{1}s_{0}\right)}{p\left(X\right)}\mu_{1}\left(s_{1},X\right)\right\}  \\
\label{eqn::lambda-mu}& = \E\left\{ \frac{e\left(s_{1},s_{0},X\right)}{e\left(s_{1},s_{0}\right)}\mu_{1}\left(s_{1},X\right)\right\},
\end{align}
where the first equality follows from the law of iterated expectations and the third equality follows from Lemma~\ref{lem:weight}.
The identification formulas of $\E\left(Y_{0}\mid U=s_{1}s_{0}\right)$ and $\tau(s_1,s_0)$ can be obtained  similarly. \QEDB

\subsection{Proof of Theorem \ref{thm:non_regular}}
The derivation in~\eqref{equ:identification-2} follows directly from the law of iterated expectations. 

\subsection{Proof of Lemma \ref{lemma:pd+om_ee}}
\allowdisplaybreaks
From Theorem \ref{thm:identification}, we have
\begin{align}
m_{1}\left(s_{1}, s_{0}\right)-f_1\left(s_{1}, s_{0} ; \eta_1\right) &=\E\left\{Y_{1} - f_1\left(s_{1}, s_{0} ; \eta_1\right)\mid U=s_{1} s_{0}\right\} \notag \\
&=\E\left[\frac{e\left(s_{1}, s_{0}, X\right)}{e\left(s_{1},s_{0}\right)} \left\{\mu_{1}\left(s_{1},X\right)-f_1\left(s_{1}, s_{0} ; \eta_1\right)\right\}\right] \label{equ:pd+om_ee_id_lemma_proof}.
\end{align}
Plugging~\eqref{equ:pd+om_ee_id_lemma_proof} into \eqref{equ:foc_m1} yields $\E\left\{D_{1,\textup{pd+om}}\left(Y,S,Z,X;\eta_1, e, \mu_{1}\right)\right\}=0$. Similarly, we can obtain $\E\left\{D_{0,\textup{pd+om}}\left(Y,S,Z,X;\eta_0, e, \mu_{0}\right)\right\}=0$. \QEDB

\subsection{Proof of Lemma \ref{lemma:weighting_ee}}
\allowdisplaybreaks
We only prove the result for $\eta_1$. Consider the expectation of the term within the integral in the estimating equation. We have
\begin{eqnarray*} 
&& \E\left[ w_1\left(S, s_{0}\right) \dot{f}_1\left(S,s_{0} ; \eta_1\right)\frac{e\left(S, s_{0}, X\right)}{p_{1}(S,X)}\frac{Z}{\pi(X)}\left\{Y-f_1\left(S,s_{0};\eta_1\right)\right\}\right] \\
&= & \E\left[ \E\left\{w_1\left(S, s_{0}\right) \dot{f}_1\left(S,s_{0} ; \eta_1\right)e\left(S, s_{0}, X\right)\frac{Y-f_1\left(S,s_{0};\eta_1\right)}{p_{1}(S,X)}\mid Z=1,X\right\}\right] \\
&= & \E\left[ \E\left\{w_1\left(S, s_{0}\right) \dot{f}_1\left(S,s_{0} ; \eta_1\right)e\left(S, s_{0}, X\right)\frac{\mu_1(S,X)-f_1\left(S,s_{0};\eta_1\right)}{p_{1}(S,X)}\mid Z=1,X\right\}\right]  \\
&= & \E\left[ \int \left\{w_1\left(s_1, s_{0}\right) \dot{f}_1\left(s_1,s_{0} ; \eta_1\right)e\left(s_1, s_{0}, X\right)\frac{\mu_1(s_1,X)-f_1\left(s_1,s_{0};\eta_1\right)}{p_{1}(s_1,X)}\right\}p(S=s_1\mid Z=1,X) \d s_1\right] \\
&=&  \E\left[ \int w_1\left(s_1, s_{0}\right) \dot{f}_1\left(s_1,s_{0} ; \eta_1\right)e\left(s_1, s_{0}, X\right)\left\{\mu_1(s_1,X)-f_1\left(s_1,s_{0};\eta_1\right) \right\} \d s_1\right]  \\
&=&  \int \E\left[ w_1\left(s_1, s_{0}\right) \dot{f}_1\left(s_1,s_{0} ; \eta_1\right)e\left(s_1, s_{0}, X\right)\left\{\mu_1(s_1,X)-f_1\left(s_1,s_{0};\eta_1\right) \right\}\right] \d s_1, 
\end{eqnarray*}
where the first two equalities follow from the law of iterated expectation. Therefore,
\begin{eqnarray*}
&& \E\left\{ D_{1,\textup{tp+pd}}\left(Y,S,Z,X;\eta_1,\pi,e\right)\right\} \\
&=& \int \E\left[ w_1\left(S, s_{0}\right) \dot{f}_1\left(S,s_{0} ; \eta_1\right)e\left(S, s_{0}, X\right)\frac{1}{p_{1}(S,X)}\frac{Z}{\pi(X)}\left\{Y-f_1\left(S,s_{0};\eta_1\right)\right\}\right] \d s_{0}\\
&=& \E \left[ \iint w_1\left(s_{1}, s_{0}\right) \dot{f}_1\left(s_{1},s_{0};\eta_1\right)e\left(s_{1},s_{0},X\right)\left\{\mu_{1}\left(s_{1},X\right)-f_1\left(s_{1},s_{0};\eta_1\right)\right\} \d s_{1}\d s_{0}\right]\\
&=& 0,
\end{eqnarray*}
where the last equality follows from Lemma~\ref{lemma:pd+om_ee}. Similarly, we can obtain $\E\left\{D_{0,\textup{tp+pd}}\left(Y,S,Z,X;\eta_0, e, \mu_{0}\right)\right\}=0$. \QEDB

\subsection{Proof of Theorem \ref{thm:eif_m1}}
\subsubsection{Nonparametric model setup}
We follow the semiparametric theory in \citet{bickel1993efficient} to derive the EIF for our projection parameter $\eta_{1}$. We denote the vector of all observed variables $V=(Y,S,Z,X)$ and factorize the likelihood as 
\[
p(V)=p(X)p(Z\mid X)p(S\mid Z,X)p(Y\mid S,Z,X).
\]
To derive the EIF for $\eta_{1}$, we consider
a one-dimensional parametric submodel $p_{\theta}(V)$ which contains
the true model $p(V)$ at $\theta=0$. We use $\theta$ in the subscript
to denote the quantity with respect to the submodel, e.g., $e_{\theta}\left(s_{1},s_{0}\right)$ is the value of $e\left(s_{1},s_{0}\right)$ with respect to the submodel. We use dot to denote the partial derivative with respect to $\theta$, e.g., $\dot{e}_{\theta}\left(s_{1},s_{0}\right) = \partial e_{\theta}\left(s_{1},s_{0}\right) / \partial \theta$, and use $\score_{\theta}(\cdot)$ to denote the score function with respect to the submodel. The decomposition of the score function under the submodel is 
\[
\score_{\theta}(V)=\score_{\theta}(X)+\score_{\theta}(Z\mid X)+\score_{\theta}(S\mid Z,X)+\score_{\theta}(Y\mid S,Z,X),
\]
where $\score_{\theta}(X)=\partial\log p_{\theta}(X)/\partial\theta$,
$\score_{\theta}(Z\mid X)=\partial\log p_{\theta}(Z\mid X)/\partial\theta$,
$\score_{\theta}(S\mid Z,X)=\partial\log p_{\theta}(S\mid Z,X)/\partial\theta$,
and $\score_{\theta}(Y\mid S,Z,X)=\partial\log p_{\theta}(Y\mid S,Z,X)/\partial\theta$
are the score functions corresponding to the four components of the
likelihood function. We write $\left.\score_{\theta}(\cdot)\right|_{\theta=0}$ as $\score(\cdot)$, which is the score function evaluated at the true model $\theta=0$ under the one-dimensional submodel. The tangent space 
\[
\Lambda=\H_{1}\oplus \H_{2}\oplus \H_{3}\oplus \H_{4}
\]
is the direct sum of 
\begin{align*}
\H_{1} & =\{h(X):\mathbb{E}\{h(X)\}=0\},\\
\H_{2} & =\{h(Z,X):\mathbb{E}\{h(Z,X)\mid X\}=0\},\\
\H_{3} & =\{h(S,Z,X):\mathbb{E}\{h(S,Z,X)\mid Z,X\}=0\},\\
\H_{4} & =\{h(Y,Z,S,X):\mathbb{E}\{h(Y,Z,S,X)\mid Z,S,X\}=0\},
\end{align*}
where $\H_{1}$, $\H_{2}$, $\H_{3}$, and $\H_{4}$ are orthogonal to
each other. Let $\varphi_{1}\left(V\right)$ denote the EIF for $\eta_{1}$.
It must satisfy the equation 
\[
\left.\dot{\eta}_{1,\theta}\right|_{\theta=0}:=\left.\frac{\partial\eta_{1,\theta}}{\partial\theta}\right|_{\theta=0}=\mathbb{E}\left\{ \varphi_{1}(V)s(V)\right\} .
\]
To find such $\varphi_{1}(V)$, we need to calculate the derivative $\left.\dot{\eta}_{1,\theta}\right|_{\theta=0}$. We first state several lemmas to simplify the proof.

\subsubsection{Lemmas}
\begin{lemma}
For $e\left(s_{1},s_{0},X\right)$ and  $e\left(s_{1},s_{0}\right)$, we have
\begin{eqnarray*}
    && \left.\dot{e}_{\theta}(s_1,s_0,X)\right|_{\theta=0} \\
    &=& e(s_1, s_0, X)\left\{ \score(S=s_1\mid Z=1,X) + \score(S=s_0\mid Z=0,X) \right\} \\
    &&+ \dot{c}_{u}(\F_1(s_1, X),\F_0(s_0, X)) \E\left\{1(S\leq s_1)\score(S\mid Z=1,X)\mid Z=1,X \right\} p_{1}(s_1,X)p_{0}(s_0,X) \\
    &&+ \dot{c}_{v}(\F_1(s_1, X),\F_0(s_0, X)) \E\left\{1(S\leq s_0)\score(S\mid Z=0,X)\mid Z=0,X \right\} p_{1}(s_1,X)p_{0}(s_0,X)
\end{eqnarray*}
and
\begin{eqnarray*}
 && \left.\dot{e}_{\theta}\left(s_{1},s_{0}\right)\right|_{\theta=0} \\
 &=& \E\left[ e\left(s_{1},s_{0},X\right)\left\{\score\left(X\right) + \score(S=s_1\mid Z=1,X) + \score(S=s_0\mid Z=0,X) \right\} \right] \\
 &&+ \E\left[\dot{c}_{u}(\F_1(s_1, X),\F_0(s_0, X)) \E\left\{1(S\leq s_1)\score(S\mid Z=1,X)\mid Z=1,X \right\} p_{1}(s_1,X)p_{0}(s_0,X) \right]\\
 &&+ \E\left[\dot{c}_{v}(\F_1(s_1, X),\F_0(s_0, X)) \E\left\{1(S\leq s_0)\score(S\mid Z=0,X)\mid Z=0,X \right\} p_{1}(s_1,X)p_{0}(s_0,X) \right].
\end{eqnarray*}

\label{lemma:e}
\end{lemma}
\noindent {\it Proof of Lemma \ref{lemma:e}.}
By the identification of the principal density in \eqref{equ::principal_density}, we have the Gateaux derivative
\begin{eqnarray}
    \left.\dot{e}_{\theta}(s_1,s_0,X)\right|_{\theta=0} &=& \left. \frac{\partial}{\partial \theta}e_{\theta}(s_1,s_0,X)\right|_{\theta=0} \notag \\
    &=& \left. \frac{\partial}{\partial \theta}\left\{c(\F_{1,\theta}(s_1, X), \F_{0,\theta}(s_0, X))p_{1,\theta}(s_1,X)p_{0,\theta}(s_0,X)\right\}\right|_{\theta=0} \notag \\
    &=& c(\F_1(s_1, X),\F_0(s_0, X))  \left.\dot{p}_{1,\theta}(s_1,X)\right|_{\theta=0}p_{0}(s_0,X) \label{equ::lemma_joint_density_dot_1} \\
    &&+ c(\F_1(s_1, X),\F_0(s_0, X)) p_{1}(s_1,X)\left.\dot{p}_{0,\theta}(s_0,X)\right|_{\theta=0} \label{equ::lemma_joint_density_dot_2} \\
    &&+ \left. \frac{\partial}{\partial \theta}\left\{c(\F_{1,\theta}(s_1, X), \F_{0,\theta}(s_0, X))\right\}\right|_{\theta=0} p_{1}(s_1,X)p_{0}(s_0,X). \label{equ::lemma_joint_density_dot_3}
\end{eqnarray}
We compute the three terms in \eqref{equ::lemma_joint_density_dot_1}--\eqref{equ::lemma_joint_density_dot_3} separately. First, for \eqref{equ::lemma_joint_density_dot_1}, we have
\begin{eqnarray*}
    && c(\F_1(s_1, X),\F_0(s_0, X))  \left.\dot{p}_{1,\theta}(s_1,X)\right|_{\theta=0}p_{0}(s_0,X) \\
    &=& c(\F_1(s_1, X),\F_0(s_0, X)) p_{1}(s_1,X)p_{0}(s_0,X)\score(S=s_1\mid Z=1,X) \\
    &=& e(s_1, s_0, X) \score(S=s_1\mid Z=1,X).
\end{eqnarray*}
Similarly, the second term \eqref{equ::lemma_joint_density_dot_2} is equal to $e(s_1, s_0, X) \score(S=s_0\mid Z=0,X)$.

The derivative in \eqref{equ::lemma_joint_density_dot_3} is
\begin{eqnarray*}
    && \left. \frac{\partial}{\partial \theta}\left\{c(\F_{1,\theta}(s_1, X),\F_{0,\theta}(s_0, X))\right\}\right|_{\theta=0} \\
    &=& \dot{c}_{u}(\F_1(s_1, X),\F_0(s_0, X)) \left. \frac{\partial}{\partial \theta}\F_{1,\theta}(s_1, X)\right|_{\theta=0} + \dot{c}_{v}(\F_1(s_1, X),\F_0(s_0, X)) \left. \frac{\partial}{\partial \theta}\F_{0,\theta}(s_0, X)\right|_{\theta=0} \\
    &=& \dot{c}_{u}(\F_1(s_1, X),\F_0(s_0, X)) \E\left\{1(S\leq s_1)\score(S\mid Z=1,X)\mid Z=1,X \right\} \\
    &&+ \dot{c}_{v}(\F_1(s_1, X),\F_0(s_0, X)) \E\left\{1(S\leq s_0)\score(S\mid Z=0,X)\mid Z=0,X \right\},
\end{eqnarray*}
where the last equality follows from the fact that for $z=0,1$,
\begin{eqnarray*}
    \left. \frac{\partial}{\partial \theta}\F_{z,\theta}(s_z, X)\right|_{\theta=0}
    &=& \left. \frac{\partial}{\partial \theta}\E_{\theta}\{1(S\leq s_z)\mid Z=z, X\}\right|_{\theta=0} \\
    &=& \E\left\{1(S\leq s_z)\score(S\mid Z=z,X)\mid Z=z,X \right\}.
\end{eqnarray*}
Combining all three terms, we have
\begin{eqnarray*}
    && \left.\dot{e}_{\theta}(s_1,s_0,X)\right|_{\theta=0} \\
    &=& e(s_1, s_0, X)\left\{ \score(S=s_1\mid Z=1,X) + \score(S=s_0\mid Z=0,X) \right\} \\
    && + \dot{c}_{u}(\F_1(s_1, X),\F_0(s_0, X)) \E\left\{1(S\leq s_1)\score(S\mid Z=1,X)\mid Z=1,X \right\} p_{1}(s_1,X)p_{0}(s_0,X) \\
    && + \dot{c}_{v}(\F_1(s_1, X),\F_0(s_0, X)) \E\left\{1(S\leq s_0)\score(S\mid Z=0,X)\mid Z=0,X \right\} p_{1}(s_1,X)p_{0}(s_0,X).
\end{eqnarray*}
For  $e(s_1,s_0)$, we have
\begin{eqnarray*}
 && \left.\dot{e}_{\theta}\left(s_{1},s_{0}\right)\right|_{\theta=0} \\
 &=&\left.\frac{\partial}{\partial\theta}\E_{\theta}\left\{ e_{\theta}\left(s_{1},s_{0},X\right)\right\} \right|_{\theta=0}\\
 &=& \left.\E_\theta\left\{ e_\theta\left(s_{1},s_{0},X\right)\score_\theta\left(X\right)\right\}\right|_{\theta=0} + \E\left\{ \left.\dot{e}_{\theta}\left(s_{1},s_{0},X\right)\right|_{\theta=0}\right\} \\
 &=& \E\left[ e\left(s_{1},s_{0},X\right)\left\{\score\left(X\right) + \score(S=s_1\mid Z=1,X) + \score(S=s_0\mid Z=0,X) \right\} \right] \\
 &&+ \E\left[\dot{c}_{u}(\F_1(s_1, X),\F_0(s_0, X)) \E\left\{1(S\leq s_1)\score(S\mid Z=1,X)\mid Z=1,X \right\} p_{1}(s_1,X)p_{0}(s_0,X) \right]\\
 &&+ \E\left[\dot{c}_{v}(\F_1(s_1, X),\F_0(s_0, X)) \E\left\{1(S\leq s_0)\score(S\mid Z=0,X)\mid Z=0,X \right\} p_{1}(s_1,X)p_{0}(s_0,X) \right].
\end{eqnarray*}
\QEDB

\begin{lemma}
For $\mu_{z}\left(s_{z},X\right)$, we have
\begin{align*}
\left.\dot{\mu}_{1,\theta}\left(s_{1},X\right)\right|_{\theta=0} & 
=\E\left[ \left\{Y-\mu_{1}\left(s_{1},X\right)\right\}\score\left(Y\mid S=s_{1},Z=1,X\right)\mid S=s_{1},Z=1,X\right],\\
\left.\dot{\mu}_{0,\theta}\left(s_{0},X\right)\right|_{\theta=0} & 
=\E\left[ \left\{Y-\mu_{0}\left(s_{0},X\right)\right\}\score\left(Y\mid S=s_{0},Z=0,X\right)\mid S=s_{0},Z=0,X\right] .
\end{align*}
\label{lemma:mu_X}
\end{lemma}
\noindent {\it Proof of Lemma \ref{lemma:mu_X}.}
We have
\begin{align*}
\left.\dot{\mu}_{1,\theta}\left(s_{1},X\right)\right|_{\theta=0} & =\left.\frac{\partial}{\partial\theta}\E_{\theta}\left( Y\mid S=s_{1},Z=1,X\right) \right|_{\theta=0}\\
 & =\E\left\{ Y\score\left(Y\mid S=s_{1},Z=1,X\right)\mid S=s_{1},Z=1,X\right\} \\
 & =\E\left[ \left\{Y-\mu_{1}\left(s_{1},X\right)\right\}\score\left(Y\mid S=s_{1},Z=1,X\right)\mid S=s_{1},Z=1,X\right] ,
\end{align*}
where the last equality follows from $\E\left\{\mu_{1}\left(s_{1},X\right)\score\left(Y\mid S=s_{1},Z=1,X\right)\mid S=s_{1},Z=1,X\right\}=0$. The derivation of $\left.\dot{\mu}_{0,\theta}\left(s_{0},X\right)\right|_{\theta=0}$ is omitted due to similarity. \QEDB

\begin{lemma}
\label{lemma:lambda}
Denote $\lambda_{1}\left(s_{1},s_{0}\right)=\E\left\{e\left(s_{1},s_{0},X\right)\mu_{1}\left(s_{1},X\right)\right\}$, $\lambda_{0}\left(s_{1},s_{0}\right)=\E\left\{e\left(s_{1},s_{0},X\right)\mu_{0}\left(s_{0},X\right)\right\}$, and $\lambda \left(s_{1},s_{0}\right) = \lambda_{1}\left(s_{1},s_{0}\right) - \lambda_{0}\left(s_{1},s_{0}\right)$. We have
\begin{eqnarray*}
\left.\dot{\lambda}_{\theta}\left(s_{1},s_{0}\right)\right|_{\theta=0} 
&=& \left.\dot{\lambda}_{1,\theta}\left(s_{1},s_{0}\right)\right|_{\theta=0} - \left.\dot{\lambda}_{0,\theta}\left(s_{1},s_{0}\right)\right|_{\theta=0},
\end{eqnarray*}
where
\begin{eqnarray*}
\left.\dot{\lambda}_{z,\theta}\left(s_{1},s_{0}\right)\right|_{\theta=0}
&=& \E\left\{ e\left(s_{1},s_{0},X\right)\mu_{z}\left(s_{z},X\right)\score\left(X\right)\right\} 
+\E\left\{ \left.\dot{e}_{\theta}\left(s_{1},s_{0},X\right)\right|_{\theta=0}\mu_{z}\left(s_{z},X\right)\right\} \\
&&+ \E\left\{ e\left(s_{1},s_{0},X\right)\left.\dot{\mu}_{z,\theta}\left(s_{z},X\right)\right|_{\theta=0}\right\}.
\end{eqnarray*}
\end{lemma}
\noindent {\it Proof of Lemma \ref{lemma:lambda}.}
By the chain rule, we have
\begin{eqnarray*}
\left.\dot{\lambda}_{z,\theta}\left(s_{1},s_{0}\right)\right|_{\theta=0}
&=&\left.\frac{\partial}{\partial\theta}\E_{\theta}\left\{ e_{\theta}\left(s_{1},s_{0},X\right)\mu_{1,\theta}\left(s_{1},X\right)\right\} \right|_{\theta=0}\\
&=&\E\left\{ e\left(s_{1},s_{0},X\right)\mu_{z}\left(s_{z},X\right)\score\left(X\right)\right\} +\E\left\{ \left.\dot{e}_{\theta}\left(s_{1},s_{0},X\right)\right|_{\theta=0}\mu_{z}\left(s_{z},X\right)\right\} \\
&&+ \E\left\{ e\left(s_{1},s_{0},X\right)\left.\dot{\mu}_{z,\theta}\left(s_{z},X\right)\right|_{\theta=0}\right\}.
\end{eqnarray*}
\QEDB

\subsubsection{EIF for $\eta_1$}
We derive the EIF for $\eta_1$. The EIFs for $\eta_0$ and $\eta_\tau$ can be obtained similarly. 
 Because Equation~\eqref{equ:foc_m1} holds for any parametric submodel, we have 
\begin{eqnarray*}
0 & =&\E_{\theta}\left[ w_1\left(S_{1},S_{0}\right)\dot{f}_1\left(S_{1},S_{0};\eta_{1,\theta}\right)\left\{m_{1,\theta}\left(S_{1},S_{0}\right)-f_1\left(S_{1},S_{0};\eta_{1,\theta}\right)\right\}\right] \\
 & =&\iint w_1\left(s_{1},s_{0}\right)\dot{f}_1\left(s_{1},s_{0};\eta_{1,\theta}\right)\left\{m_{1,\theta}\left(s_{1},s_{0}\right)-f_1\left(s_{1},s_{0};\eta_{1,\theta}\right)\right\}e_{\theta}\left(s_{1},s_{0}\right)\d s_{1}\d s_{0}\\
 & =&\iint w_1\left(s_{1},s_{0}\right)\dot{f}_1\left(s_{1},s_{0};\eta_{1,\theta}\right)\left\{\lambda_{1,\theta}\left(s_{1},s_{0}\right)-e_{\theta}\left(s_{1},s_{0}\right)f_1\left(s_{1},s_{0};\eta_{1,\theta}\right)\right\}\d s_{1}\d s_{0},
\end{eqnarray*}
where the third equality follows from~\eqref{eqn::lambda-mu}.
Taking the first order derivative with respect to $\theta$ and evaluating
at $\theta=0$, we have 
\begin{eqnarray*}
0 & =&\left.\frac{\partial}{\partial\theta}\iint w_1\left(s_{1},s_{0}\right)\dot{f}_1\left(s_{1},s_{0};\eta_{1,\theta}\right)\left\{\lambda_{1,\theta}\left(s_{1},s_{0}\right)-e_{\theta}\left(s_{1},s_{0}\right)f_1\left(s_{1},s_{0};\eta_{1,\theta}\right)\right\}\d s_{1}\d s_{0}\right|_{\theta=0}\\
 & =&\iint w_1\left(s_{1},s_{0}\right)\ddot{f}_1\left(s_{1},s_{0};\eta_{1}\right)\left.\dot{\eta}_{1,\theta}\right|_{\theta=0}\left\{\lambda_{1}\left(s_{1},s_{0}\right)-e\left(s_{1},s_{0}\right)f_1\left(s_{1},s_{0};\eta_{1}\right)\right\}\d s_{1}\d s_{0}\\
 &&+ \iint w_1\left(s_{1},s_{0}\right)\dot{f}_1\left(s_{1},s_{0};\eta_{1}\right)\left\{\left.\dot{\lambda}_{1,\theta}\left(s_{1},s_{0}\right)\right|_{\theta=0}-\left.\dot{e}_{\theta}\left(s_{1},s_{0}\right)\right|_{\theta=0}f_1\left(s_{1},s_{0};\eta_{1}\right)\right\}\d s_{1}\d s_{0}\\
 &&- \iint w_1\left(s_{1},s_{0}\right)\dot{f}_1\left(s_{1},s_{0};\eta_{1}\right)\dot{f}_1\left(s_{1},s_{0};\eta_{1}\right)^{\T}e\left(s_{1},s_{0}\right)\left.\dot{\eta}_{1,\theta}\right|_{\theta=0}\d s_{1}\d s_{0},
\end{eqnarray*}
where the second equality follows from the chain rule. 
Therefore, we can write
\begin{eqnarray}
 \left.\dot{\eta}_{1,\theta}\right|_{\theta=0} &=& H_{1}^{-1}\iint w_1\left(s_{1},s_{0}\right)\dot{f}_1\left(s_{1},s_{0};\eta_{1}\right)\left\{\left.\dot{\lambda}_{1,\theta}\left(s_{1},s_{0}\right)\right|_{\theta=0}-\left.\dot{e}_{\theta}\left(s_{1},s_{0}\right)\right|_{\theta=0}f_1\left(s_{1},s_{0};\eta_{1}\right)\right\}\d s_{1}\d s_{0},\notag   \\
    \label{equ:eta1_dot}
\end{eqnarray}
where
\begin{eqnarray*}
H_{1} &=& - \iint w_1\left(s_{1},s_{0}\right)\ddot{f}_1\left(s_{1},s_{0};\eta_{1}\right)\left\{ \lambda_1 \left(s_{1},s_{0}\right)-f_1\left(s_{1},s_{0};\eta_{1}\right)e\left(s_{1},s_{0}\right) \right\} \d s_{1}\d s_{0}\\
&&+ \iint w_1\left(s_{1},s_{0}\right)\dot{f}_1\left(s_{1},s_{0};\eta_{1}\right)\dot{f}_1\left(s_{1},s_{0};\eta_{1}\right)^{\T}e\left(s_{1},s_{0}\right)\d s_{1}\d s_{0} \\
&=& \iint w_1(s_1,s_0)\left[\dot{f}_1\left(s_{1},s_{0};\eta_{1}\right)\dot{f}_1\left(s_{1},s_{0};\eta_{1}\right)^{\T} - \left\{m_1(s_1,s_0) - f_1(s_1,s_0;\eta_1)\right\}\ddot{f}_1(s_1,s_0;\eta_1) \right] e\left(s_{1},s_{0}\right)\d s_{1}\d s_{0} \\
&=& \E\left(w_1(S_1,S_0)\left[\dot{f}_1\left(S_{1},S_{0};\eta_{1}\right)\dot{f}_1\left(S_{1},S_{0};\eta_{1}\right)^{\T} - \left\{m_1(S_1,S_0) - f_1(S_1,S_0;\eta_1)\right\}\ddot{f}_1(S_1,S_0;\eta_1) \right] \right), 
\end{eqnarray*}
where the second equality follows from~\eqref{eqn::lambda-mu}. Taking the second-order derivative of the objective function in the optimization problem~\eqref{equ:opt_m1}, we have the Hessian matrix equal to
\begin{equation*}
2\E\left(w_1(S_1,S_0)\left[\dot{f}_1\left(S_{1},S_{0};\eta_{1}\right)\dot{f}_1\left(S_{1},S_{0};\eta_{1}\right)^{\T} - \left\{m_1(S_1,S_0) - f_1(S_1,S_0;\eta_1)\right\}\ddot{f}_1(S_1,S_0;\eta_1) \right] \right).
\end{equation*}
Therefore,  $2H_1$ is equal to the Hessian matrix.

We will calculate the two terms within the integral in~\eqref{equ:eta1_dot} separately.

\noindent \textbf{Part I.} By Lemma \ref{lemma:lambda}, we have
\begin{eqnarray*}
 \iint w_1\left(s_{1},s_{0}\right)\dot{f}_1\left(s_{1},s_{0};\eta_{1}\right)\left.\dot{\lambda}_{1,\theta}\left(s_{1},s_{0}\right)\right|_{\theta=0}\d s_{1}\d s_{0}&=& B_1+B_2+B_3,
\end{eqnarray*}
where
\begin{eqnarray*}
B_1&=&\iint w_1\left(s_{1},s_{0}\right)\dot{f}_1\left(s_{1},s_{0};\eta_{1}\right)\E\left\{ e\left(s_{1},s_{0},X\right)\mu_{1}\left(s_{1},X\right)\score\left(X\right)\right\} \d s_{1}\d s_{0},\\
B_2 &=& \iint w_1\left(s_{1},s_{0}\right)\dot{f}_1\left(s_{1},s_{0};\eta_{1}\right)\E\left\{ \left.\dot{e}_{\theta}\left(s_{1},s_{0},X\right)\right|_{\theta=0}\mu_{1}\left(s_{1},X\right)\right\} \d s_{1}\d s_{0},\\
B_3 &=& \iint w_1\left(s_{1},s_{0}\right)\dot{f}_1\left(s_{1},s_{0};\eta_{1}\right)\E\left\{ e\left(s_{1},s_{0},X\right)\left.\dot{\mu}_{1,\theta}\left(s_{1},X\right)\right|_{\theta=0}\right\} \d s_{1}\d s_{0}.
\end{eqnarray*}
First, consider $B_{1}$. We have
\begin{eqnarray*}
B_1 &=& \E\left\{ \iint w_1\left(s_{1},s_{0}\right)\dot{f}_1\left(s_{1},s_{0};\eta_{1}\right)e\left(s_{1},s_{0},X\right)\mu_{1}\left(s_{1},X\right)\d s_{1}\d s_{0}\cdot \score\left(X\right)\right\}.
\end{eqnarray*}
Because $\E\left\{ \score\left(X\right)\right\} =0$,  we can write
\begin{eqnarray*}
B_1  &=& \E\left\{ h_{1}\left(X\right)\score\left(X\right)\right\} ,
\end{eqnarray*}
where 
\begin{eqnarray*}
h_{1}\left(X\right)&=&\iint w_1\left(s_{1},s_{0}\right)\dot{f}_1\left(s_{1},s_{0};\eta_{1}\right)e\left(s_{1},s_{0},X\right)\mu_{1}\left(s_{1},X\right)\d s_{1}\d s_{0}\\
&&-\E \left\{\iint w_1\left(s_{1},s_{0}\right)\dot{f}_1\left(s_{1},s_{0};\eta_{1}\right)e\left(s_{1},s_{0},X\right)\mu_{1}\left(s_{1},X\right)\d s_{1}\d s_{0}\right\}\\
&=&\iint w_1\left(s_{1},s_{0}\right)\dot{f}_1\left(s_{1},s_{0};\eta_{1}\right)e\left(s_{1},s_{0},X\right)\mu_{1}\left(s_{1},X\right)\d s_{1}\d s_{0}\\
&&-\E \left\{\iint w_1\left(s_{1},s_{0}\right)\dot{f}_1\left(s_{1},s_{0};\eta_{1}\right)e\left(s_{1},s_{0},X\right) f_1\left(s_{1},s_{0};\eta_{1}\right) \d s_{1}\d s_{0}\right\} \quad (\text{Lemma~\ref{lemma:pd+om_ee}})\\
&=&\iint w_1\left(s_{1},s_{0}\right)\dot{f}_1\left(s_{1},s_{0};\eta_{1}\right)e\left(s_{1},s_{0},X\right)\mu_{1}\left(s_{1},X\right)\d s_{1}\d s_{0}\\
&&-\iint w_1\left(s_{1},s_{0}\right)\dot{f}_1\left(s_{1},s_{0};\eta_{1}\right)f_1\left(s_{1},s_{0};\eta_{1}\right)e\left(s_{1},s_{0}\right)\d s_{1}\d s_{0}.
\end{eqnarray*}

Next, consider $B_{2}$. By Lemma \ref{lemma:e},  we can further write
\begin{eqnarray*}
B_2 = B_{21} + B_{22} + B_{23} + B_{24},
\end{eqnarray*}
where
\begin{eqnarray*}
B_{21} &=& \iint w_1\dot{f}_1\E\left\{e(s_1, s_0, X)\score(S=s_1\mid Z=1,X)\mu_{1}\left(s_{1},X\right)\right\}\d s_{1}\d s_{0}, \\
B_{22} &=& \iint w_1\dot{f}_1\E\left\{e(s_1, s_0, X)\score(S=s_0\mid Z=0,X)\mu_{1}\left(s_{1},X\right)\right\}\d s_{1}\d s_{0}, \\
B_{23} &=& \iint w_1\dot{f}_1\E\left[ \dot{c}_{u}\E\left\{ 1(S\leq s_1)s(S\mid Z=1,X)\mid Z=1,X\right\} p_{1}\left(s_{1},X\right)p_{0}\left(s_{0},X\right)\mu_{1}\left(s_{1},X\right)\right]\d s_{1}\d s_{0}, \\
B_{24} &=& \iint w_1\dot{f}_1\E\left[ \dot{c}_{v}\E\left\{ 1(S\leq s_0)s(S\mid Z=0,X)\mid Z=0,X\right\} p_{1}\left(s_{1},X\right)p_{0}\left(s_{0},X\right)\mu_{1}\left(s_{1},X\right)\right]\d s_{1}\d s_{0}.
\end{eqnarray*}
We suppress the dependence of $w_1(s_1,s_0)$, $\dot{f}_1(s_1,s_0;\eta_1)$, $\dot{c}_{u}(\F_1(s_1,X),\F_0(s_0,X))$, and $\dot{c}_{v}(\F_1(s_1,X),\F_0(s_0,X))$ on $(s_1,s_0,X)$ and $\eta_1$ for ease of notation.

Denote
\begin{eqnarray*}
\tilde{\mu}_{c11}(s_1,X) &=& \int w_1\dot{f}_1 \frac{e(s_1,s_0,X)}{p_1(s_1,X)}\mu_{1}\left(s_{1},X\right)\d s_0, \\
\tilde{\mu}_{c01}(s_0,X) &=& \int w_1\dot{f}_1 \frac{e(s_1,s_0,X)}{p_0(s_0,X)}\mu_{1}\left(s_{1},X\right)\d s_1.
\end{eqnarray*}
For $B_{21}$, we have
\begin{eqnarray*}
B_{21} &=& \iint w_1\dot{f}_1\E\left\{e(s_1, s_0, X)\score(S=s_1\mid Z=1,X)\mu_{1}\left(s_{1},X\right)\right\}\d s_{1}\d s_{0} \\
&=& \E\left[ \int \left\{\int w_1\dot{f}_1 \frac{e(s_1,s_0,X)}{p_1(s_1,X)}\mu_{1}\left(s_{1},X\right)\d s_0\right\}p_1(s_1,X) \score(S=s_1\mid Z=1,X)\d s_1\right] \\
&=& \E\left\{\int \tilde{\mu}_{c11}(s_1,X)p_1(s_1,X) \score(S=s_1\mid Z=1,X)\d s_1 \right\} \\
&=& \E\left[\E\left\{\tilde{\mu}_{c11}(S,X)\score(S\mid Z=1,X)\mid Z=1,X\right\}\right] \\
&=& \E\left\{\frac{Z}{\pi(X)}\tilde{\mu}_{c11}(S,X)\score(S\mid Z,X)\right\},
\end{eqnarray*}
where the last equality follows from the law of iterated expectations. Similarly, for $B_{22}$, we have
\begin{eqnarray*}
    B_{22} &=& \E\left\{ \frac{1-Z}{1-\pi\left(X\right)}\tilde{\mu}_{c01}\left(S,X\right)\score\left(S\mid Z,X\right)\right\}.
\end{eqnarray*}
Denote
\begin{eqnarray*}
\tilde{\mu}_{\dot{c}11}\left(t_1,X\right) &=& \iint 1(s_1\geq t_1)w_1\dot{f}_1\dot{c}_{u}p_1(s_1,X)p_0(s_0,X)\mu_{1}\left(s_{1},X\right)\d s_{1}\d s_{0}, \\
\tilde{\mu}_{\dot{c}01}\left(t_0,X\right) &=& \iint 1(s_0\geq t_0)w_1\dot{f}_1\dot{c}_{v}p_1(s_1,X)p_0(s_0,X)\mu_{1}\left(s_{1},X\right)\d s_{1}\d s_{0},
\end{eqnarray*}
For $B_{23}$, we have 
\begin{eqnarray*}
B_{23} &=& \iint w_1\dot{f}_1\E\left[ \dot{c}_{u}\E\left\{ 1(S\leq s_1)s(S\mid Z=1,X)\mid Z=1,X\right\} p_{1}\left(s_{1},X\right)p_{0}\left(s_{0},X\right)\mu_{1}\left(s_{1},X\right)\right]\d s_{1}\d s_{0} \\
&=& \E\left[ \iint w_1\dot{f}_1\dot{c}_{u}\left\{\int1(t_1\leq s_1)\score(t_1\mid Z=1,X)p_1(t_1,X)\d t_1\right\} p_{1}\left(s_{1},X\right)p_{0}\left(s_{0},X\right)\mu_{1}\left(s_{1},X\right)\d s_{1}\d s_{0}  \right] \\
&=& \E\left[ \int\left\{ \iint 1(s_1\geq t_1)w_1\dot{f}_1\dot{c}_{u}p_{1}\left(s_{1},X\right)p_{0}\left(s_{0},X\right)\mu_{1}\left(s_{1},X\right)\d s_{1}\d s_{0} \right\}\score(t_1\mid Z=1,X)p_1(t_1,X)\d t_1 \right] \\
&=& \E\left\{\int \tilde{\mu}_{\dot{c}11}(t_1,X)\score(t_1\mid Z=1,X)p_1(t_1,X)\d t_1 \right\} \\
&=& \E\left[\E\left\{\tilde{\mu}_{\dot{c}11}(S,X)\score(S\mid Z=1,X)\mid Z=1,X \right\} \right] \\
&=& \E\left\{\frac{Z}{\pi(X)}\tilde{\mu}_{\dot{c}11}(S,X)\score(S\mid Z,X) \right\},
\end{eqnarray*}
where $\score(t_1\mid Z=1,X)$ denotes $\score(S=t_1\mid Z=1,X)$, and the last equality follows from the law of iterated expectations. Similarly, for $B_{24}$, we have 
\begin{eqnarray*}
B_{24} &=& \E\left\{ \frac{1-Z}{1-\pi\left(X\right)}\tilde{\mu}_{\dot{c}01}\left(S,X\right)\score\left(S\mid Z,X\right)\right\}.
\end{eqnarray*}
Therefore, combining the four terms of $B_2$, we have
\begin{eqnarray*}
B_{2} &=& \E\left(\left[\frac{Z}{\pi\left(X\right)} \left\{\tilde{\mu}_{c11}\left(S,X\right) + \tilde{\mu}_{\dot{c}11}\left(S,X\right)\right\} + \frac{1-Z}{1-\pi\left(X\right)}\left\{\tilde{\mu}_{c01}\left(S,X\right) + \tilde{\mu}_{\dot{c}01}\left(S,X\right)\right\}\right]\score\left(S\mid Z,X\right)\right) \\
&=& \E\left[\left\{\frac{Z}{\pi\left(X\right)}\tilde{\mu}_{11}\left(S,X\right) + \frac{1-Z}{1-\pi\left(X\right)}\tilde{\mu}_{01}\left(S,X\right)\right\}\score\left(S\mid Z,X\right)\right],
\end{eqnarray*}
where we denote  $\tilde{\mu}_{z1}\left(s_z,X\right)=\tilde{\mu}_{cz1}\left(s_z,X\right) + \tilde{\mu}_{\dot{c}z1}\left(s_z,X\right)$ for $z=0,1$.
As a result, we have 
\begin{eqnarray*}
B_{2}  &=& \E\left\{ h_{2}\left(S,Z,X\right)\score\left(S\mid Z,X\right)\right\},
\end{eqnarray*}
where 
\begin{eqnarray*}
h_{2}\left(S,Z,X\right) &=& \frac{Z}{\pi\left(X\right)}\tilde{\mu}_{11}\left(S,X\right) + \frac{1-Z}{1-\pi\left(X\right)}\tilde{\mu}_{01}\left(S,X\right)\\
 &&- \E\left\{\frac{Z}{\pi\left(X\right)}\tilde{\mu}_{11}\left(S,X\right) + \frac{1-Z}{1-\pi\left(X\right)}\tilde{\mu}_{01}\left(S,X\right)\mid Z,X\right\}\\
 &=&\frac{Z}{\pi\left(X\right)}\left[\tilde{\mu}_{11}\left(S,X\right) - \E\left\{ \tilde{\mu}_{11}\left(S,X\right)\mid Z=1,X\right\} \right]\\
 &&+ \frac{1-Z}{1-\pi\left(X\right)}\left[\tilde{\mu}_{01}\left(S,X\right) - \E\left\{ \tilde{\mu}_{01}\left(S,X\right)\mid Z=0,X\right\} \right].
\end{eqnarray*}

Then, consider $B_{3}$. Denote 
\begin{eqnarray*}
\tilde{e}_{1}\left(s_{1},x\right) &=&\int w_1\left(s_{1},s_{0}\right)\dot{f}_1\left(s_{1},s_{0};\eta_{1}\right)e\left(s_{1},s_{0},x\right)\d s_{0}.
\end{eqnarray*}
From Lemma~\ref{lemma:mu_X}, we have
\begin{eqnarray}
B_{3} &=& \iint w_1\left(s_{1},s_{0}\right)\dot{f}_1\left(s_{1},s_{0};\eta_{1}\right) \notag\\
&& \quad\quad \cdot \E\left\{ e\left(s_{1},s_{0},X\right)\E\left[ \left\{Y-\mu_{1}\left(s_{1},X\right)\right\}\score\left(Y\mid s_{1},Z=1,X\right)\mid s_{1},Z=1,X\right\} \right] \d s_{1}\d s_{0} \notag \\
 &=&\int \E\left( \tilde{e}_{1}\left(s_{1},X\right) \E\left[ \left\{Y-\mu_{1}\left(s_{1},X\right)\right\}\score\left(Y\mid s_{1},Z=1,X\right)\mid s_{1},Z=1,X\right] \right) \d s_{1}\notag \\
 &=&\E\left(\int \tilde{e}_{1}\left(s_{1},X\right) \E\left[ \left\{Y-\mu_{1}\left(s_{1},X\right)\right\}\score\left(Y\mid s_{1},Z=1,X\right)\mid s_{1},Z=1,X\right]\d s_{1} \right).
 \label{eqn::B3}
\end{eqnarray}
By the law of iterated expectation, we have 
\begin{eqnarray*}
&&\E\left[ \frac{Z}{\pi\left(X\right)}\frac{\tilde{e}_{1}\left(S,X\right)}{p_{1}\left(S,X\right)}\left\{Y-\mu_{1}\left(S,X\right)\right\}\score\left(Y\mid S,Z,X\right)\right] \\
 & =&\E\left( \E\left[\frac{\tilde{e}_{1}\left(S,X\right)}{p_{1}\left(S,X\right)}\left\{Y-\mu_{1}\left(S,X\right)\right\}\score\left(Y\mid S,Z,X\right)\mid X,Z=1\right]\right) \\
  & =&\E\left(\int \E\left[\frac{\tilde{e}_{1}\left(s_1,X\right)}{p_{1}\left(s_1,X\right)}\left\{Y-\mu_{1}\left(s_1,X\right)\right\}\score\left(Y\mid s_1,Z,X\right)\mid S=s_1,Z=1,X\right]p(S=s_1\mid Z=1,X)\d s_1\right) \\
    & =&\E\left(\int \tilde{e}_{1}\left(s_1,X\right)\E\left[\left\{Y-\mu_{1}\left(s_1,X\right)\right\}\score\left(Y\mid s_1,Z,X\right)\mid S=s_1,Z=1,X\right]\d s_1\right),
\end{eqnarray*}
which, coupled with~\eqref{eqn::B3}, implies that 
\begin{equation*}
B_{3}\ =\ \E\left\{ h_{3}\left(Y,S,Z,X\right)\score\left(Y\mid S,Z,X\right)\right\},
\end{equation*}
where 
\begin{eqnarray*}
h_{3}\left(Y,S,Z,X\right) & =&\frac{Z}{\pi\left(X\right)}\frac{\int w_1\left(S,s_{0}\right)\dot{f}_1\left(S,s_{0};\eta_{1}\right)e\left(S,s_{0},X\right)\d s_{0}}{p_{1}\left(S,X\right)}\left\{Y-\mu_{1}\left(S,X\right)\right\}.
\end{eqnarray*}

\noindent \textbf{Part II.} Consider the second part of $\left.\dot{\eta}_{1,\theta}\right|_{\theta=0}$ within the integral in~\eqref{equ:eta1_dot}. By Lemma~\ref{lemma:e}, we have
\begin{eqnarray*}
 \iint w_1\left(s_{1},s_{0}\right)\dot{f}_1\left(s_{1},s_{0};\eta_{1}\right)f_1\left(s_{1},s_{0};\eta_{1}\right)\left.\dot{e}_{\theta}\left(s_{1},s_{0}\right)\right|_{\theta=0}\d s_{1}\d s_{0}\ =\ B_4 + B_5,
\end{eqnarray*}
where
\begin{eqnarray*}
B_4 &=& \iint w_1\dot{f}_1f_1\E\left\{e\left(s_{1},s_{0},X\right)\score(X)\right\}\d s_{1}\d s_{0},\\
B_5 &=& \iint w_1\dot{f}_1f_1\E\left[ e\left(s_{1},s_{0},X\right)\left\{\score(S=s_1\mid Z=1,X) + \score(S=s_0\mid Z=0, X)\right\}\right] \d s_{1}\d s_{0} \\
&&+ \iint w_1\dot{f}_1f_1\E\left[\dot{c}_{u}\E\left\{1(S\leq s_1)\score(S\mid Z=1,X)\mid Z=1,X\right\}p_{1}\left(s_{1},X\right)p_{0}\left(s_{0},X\right)\right]\d s_{1}\d s_{0} \\
&&+ \iint w_1\dot{f}_1f_1\E\left[\dot{c}_{v}\E\left\{1(S\leq s_0)\score(S\mid Z=0,X)\mid Z=0,X\right\}p_{1}\left(s_{1},X\right)p_{0}\left(s_{0},X\right)\right]\d s_{1}\d s_{0}.
\end{eqnarray*}
Again, we suppress the dependence of $w_1(s_1,s_0)$, $\dot{f}_1(s_1,s_0;\eta_1)$, $f_1(s_1,s_0;\eta_1)$, $\dot{c}_u(\F_1(s_1,X),\F_0(s_0,X))$ and $\dot{c}_v(\F_1(s_1,X),\F_0(s_0,X))$ on $(s_1,s_0,X)$ and $\eta_1$ for ease of notation.

First, consider $B_4$. We have
\begin{eqnarray*}
    B_4 &=& \iint w_1\dot{f}_1f_1\E\left\{e\left(s_{1},s_{0},X\right)\score(X)\right\}\d s_{1}\d s_{0} \\
    &=& \E\left[\left\{\iint w_1\dot{f}_1f_1 e\left(s_{1},s_{0},X\right)\d s_{1}\d s_{0} \right\}\score(X)\right] \\
    &=& \E\left\{h_4(X)\score(X)\right\},
\end{eqnarray*}
where
\begin{eqnarray*}
h_{4}\left(X\right) & =&\iint w_1\dot{f}_1f_1 e\left(s_{1},s_{0},X\right)\d s_{1}\d s_{0} -\E \left\{\iint w_1\dot{f}_1f_1 e\left(s_{1},s_{0},X\right)\d s_{1}\d s_{0}\right\}\\
 & =&\iint w_1\dot{f}_1f_1\left\{e\left(s_{1},s_{0},X\right)-e\left(s_{1},s_{0}\right)\right\}\d s_{1}\d s_{0}.
\end{eqnarray*}

Next, we consider $B_5$. We can further write $B_5$ as
\begin{eqnarray*}
    B_5 &=& B_{51} + B_{52} + B_{53} + B_{54},
\end{eqnarray*}
where
\begin{eqnarray*}
    B_{51} &=& \iint w_1\dot{f}_1f_1\E\left\{ e\left(s_{1},s_{0},X\right)\score(S=s_1\mid Z=1,X)\right\} \d s_{1}\d s_{0}, \\
    B_{52} &=& \iint w_1\dot{f}_1f_1\E\left\{ e\left(s_{1},s_{0},X\right)\score(S=s_0\mid Z=1,X)\right\} \d s_{1}\d s_{0}, \\
    B_{53} &=& \iint w_1\dot{f}_1f_1\E\left[\dot{c}_{u}\E\left\{1(S\leq s_1)\score(S\mid Z=1,X)\mid Z=1,X\right\}p_{1}\left(s_{1},X\right)p_{0}\left(s_{0},X\right)\right]\d s_{1}\d s_{0}, \\
    B_{54} &=& \iint w_1\dot{f}_1f_1\E\left[\dot{c}_{v}\E\left\{1(S\leq s_0)\score(S\mid Z=0,X)\mid Z=0,X\right\}p_{1}\left(s_{1},X\right)p_{0}\left(s_{0},X\right)\right]\d s_{1}\d s_{0}.
\end{eqnarray*}
Denote
\begin{eqnarray*}
    \tilde{f}_{c11}(s_1,X) &=& \int w_1\dot{f}_1f_1 \frac{e(s_1,s_0,X)}{p_1(s_1,X)}\d s_0, \\
    \tilde{f}_{c01}(s_0,X) &=& \int w_1\dot{f}_1f_1 \frac{e(s_1,s_0,X)}{p_0(s_0,X)}\d s_1.
\end{eqnarray*}
For $B_{51}$, we have
\begin{eqnarray*}
    B_{51} &=& \iint w_1\dot{f}_1f_1\E\left\{ e\left(s_{1},s_{0},X\right)\score(S=s_1\mid Z=1,X)\right\} \d s_{1}\d s_{0} \\
    &=& \E\left[\int\left\{\int w_1\dot{f}_1f_1 \frac{e(s_1,s_0,X)}{p_1(s_1,X)}\d s_0\right\}p_1(s_1,X)\score(S=s_1\mid Z=1,X)\d s_1\right] \\
    &=& \E\left\{\int\tilde{f}_{c11}(s_1,X)p_1(s_1,X)\score(S=s_1\mid Z=1,X)\d s_1 \right\} \\
    &=& \E\left[\E\left\{\tilde{f}_{c11}(S,X)\score(S\mid Z=1,X)\mid Z=1,X\right\}\right] \\
    &=& \E\left\{\frac{Z}{\pi(X)}\tilde{f}_{c11}(S,X)\score(S\mid Z,X)\right\}
\end{eqnarray*}
and similarly,
\begin{eqnarray*}
    B_{52} &=& \E\left\{\frac{1-Z}{1-\pi(X)}\tilde{f}_{c01}(S,X)\score(S\mid Z,X)\right\}.
\end{eqnarray*}

Denote
\begin{eqnarray*}
\tilde{f}_{\dot{c}11}\left(t_1,X\right) &=& \iint 1(s_1\geq t_1)w_1\dot{f}_1f_1\dot{c}_{u}p_1(s_1,X)p_0(s_0,X)\d s_{1}\d s_{0}, \\
\tilde{f}_{\dot{c}01}\left(t_0,X\right) &=& \iint 1(s_0\geq t_0)w_1\dot{f}_1f_1\dot{c}_{v}p_1(s_1,X)p_0(s_0,X)\d s_{1}\d s_{0}.
\end{eqnarray*}
For $B_{53}$, we have
\begin{eqnarray*}
    B_{53} &=& \iint w_1\dot{f}_1f_1\E\left[\dot{c}_{u}\E\left\{1(S\leq s_1)\score(S\mid Z=1,X)\mid Z=1,X\right\}p_{1}\left(s_{1},X\right)p_{0}\left(s_{0},X\right)\right]\d s_{1}\d s_{0} \\
    &=& \E\left[ \iint w_1\dot{f}_1f_1\dot{c}_{u}\left\{\int1(t_1\leq s_1)\score(t_1\mid Z=1,X)p_1(t_1,X)\d t_1\right\} p_{1}\left(s_{1},X\right)p_{0}\left(s_{0},X\right)\d s_{1}\d s_{0}  \right] \\
    &=& \E\left[ \int\left\{ \iint 1(s_1\geq t_1)w_1\dot{f}_1f_1\dot{c}_{u}p_{1}\left(s_{1},X\right)p_{0}\left(s_{0},X\right)\d s_{1}\d s_{0} \right\}\score(t_1\mid Z=1,X)p_1(t_1,X)\d t_1 \right] \\
    &=& \E\left\{\int \tilde{f}_{\dot{c}11}\left(t_1,X\right)p_1(t_1,X)\score(t_1\mid Z=1,X)\d t_1 \right\} \\
    &=& \E\left[\E\left\{\tilde{f}_{\dot{c}11}\left(S,X\right)\score(S\mid Z=1,X)\mid Z=1,X\right\}\right] \\
    &=& \E\left\{\frac{Z}{\pi(X)}\tilde{f}_{\dot{c}11}\left(S,X\right)\score(S\mid Z,X)\right\}
\end{eqnarray*}
and similarly, 
\begin{eqnarray*}
    B_{54} &=& \E\left\{\frac{1-Z}{1-\pi(X)}\tilde{f}_{\dot{c}01}\left(S,X\right)\score(S\mid Z,X)\right\}.
\end{eqnarray*}
Therefore, we obtain
\begin{eqnarray*}
B_{5} &=& \E\left(\left[\frac{Z}{\pi\left(X\right)} \left\{\tilde{f}_{c11}\left(S,X\right) + \tilde{f}_{\dot{c}11}\left(S,X\right)\right\} + \frac{1-Z}{1-\pi\left(X\right)}\left\{\tilde{f}_{c01}\left(S,X\right) + \tilde{f}_{\dot{c}01}\left(S,X\right)\right\}\right]\score\left(S\mid Z,X\right)\right) \\
&=& \E\left[\left\{\frac{Z}{\pi\left(X\right)}\tilde{f}_{11}\left(S,X\right) + \frac{1-Z}{1-\pi\left(X\right)}\tilde{f}_{01}\left(S,X\right)\right\}\score\left(S\mid Z,X\right)\right],
\end{eqnarray*}
where we denote $\tilde{f}_{z1}\left(s_z,X\right)=\tilde{f}_{cz1}\left(s_z,X\right) + \tilde{f}_{\dot{c}z1}\left(s_z,X\right)$ for $z=0,1$. As a result, we have
\begin{eqnarray*}
B_{5} &=& \E\left\{ h_{5}\left(S,Z,X\right)\score\left(S\mid Z,X\right)\right\},
\end{eqnarray*}
where
\begin{eqnarray*}
h_{5}\left(S,Z,X\right) &=& \frac{Z}{\pi\left(X\right)}\tilde{f}_{11}\left(S,X\right) + \frac{1-Z}{1-\pi\left(X\right)}\tilde{f}_{01}\left(S,X\right)\\
 && -\E\left\{\frac{Z}{\pi\left(X\right)}\tilde{f}_{11}\left(S,X\right) + \frac{1-Z}{1-\pi\left(X\right)}\tilde{f}_{01}\left(S,X\right)\mid Z,X\right\}\\
 &=&\frac{Z}{\pi\left(X\right)}\left[\tilde{f}_{11}\left(S,X\right) - \E\left\{ \tilde{f}_{11}\left(S,X\right)\mid Z=1,X\right\} \right]\\
 && +\frac{1-Z}{1-\pi\left(X\right)}\left[\tilde{f}_{01}\left(S,X\right) - \E\left\{ \tilde{f}_{01}\left(S,X\right)\mid Z=0,X\right\} \right].
\end{eqnarray*}

\noindent \textbf{Summary of the proof.}
We have
\begin{eqnarray*}
H_{1}\left.\dot{\eta}_{1,\theta}\right|_{\theta=0} &=&B_1+B_2+B_3+B_4+B_5\\
&=&\E\left\{ h_{1}\left(X\right)\score\left(X\right)\right\} +\E\left\{ h_{2}\left(S,Z,X\right)\score\left(S\mid Z,X\right)\right\} \\
&&+ \E\left\{ h_{3}\left(Y,S,Z,X\right)\score\left(Y\mid S,Z,X\right)\right\} \\
&&- \E\left\{ h_{4}\left(X\right)\score\left(X\right)\right\} -\E\left\{ h_{5}\left(S,Z,X\right)\score\left(S\mid Z,X\right)\right\} \\
&=& \E\left[ \left\{h_{1}\left(X\right)-h_{4}\left(X\right)\right\}\score\left(X\right)\right] \\
&&+ \E\left[ \left\{h_{2}\left(S,Z,X\right)-h_{5}\left(S,Z,X\right)\right\}\score\left(S\mid Z,X\right)\right] \\
&&+ \E\left[ h_{3}\left(Y,S,Z,X\right)\score\left(Y\mid S,Z,X\right)\right] .
\end{eqnarray*}
We can verify that 
\begin{eqnarray*}
H_{1}^{-1}\left\{h_{1}\left(X\right)-h_{4}\left(X\right)\right\}& \in& \H_{1},\\
H_{1}^{-1}\left\{h_{2}\left(S,Z,X\right)-h_{5}\left(S,Z,X\right)\right\}& \in& \H_{3},\\
H_{1}^{-1}h_{3}\left(Y,S,Z,X\right) & \in& \H_{4}.
\end{eqnarray*}
By the fact that $\H_{1}$, $\H_{2}$, $\H_{3}$, and $\H_{4}$ are orthogonal
to each other, we have 
\begin{eqnarray*}
\left.\dot{\eta}_{1,\theta}\right|_{\theta=0} & =&\E\left[ H_1^{-1}\left\{h_{1}\left(X\right)-h_{4}\left(X\right)\right\}\score\left(V\right)\right] \\
 &&+ \E\left[ H_1^{-1}\left\{h_{2}\left(S,Z,X\right)-h_{5}\left(S,Z,X\right)\right\}\score\left(V\right)\right] \\
 &&+ \E\left\{ H_1^{-1}h_{3}\left(Y,S,Z,X\right)\score\left(V\right)\right\}. 
\end{eqnarray*}

Therefore, the EIF for $\eta_1$ is  
\begin{equation*}
    \varphi_{1}\left(V\right)\ =\ H_{1}^{-1} \left\{\ell_{1}\left(X\right)+\ell_{2}\left(S,Z,X\right)+\ell_{3}\left(Y,S,Z,X\right)\right\},
\end{equation*}
where 
\begin{eqnarray*}
    \ell_{1}\left(X\right) &=& h_{1}\left(X\right)-h_{4}\left(X\right)\\
    &=&\iint w_1\left(s_{1},s_{0}\right)\dot{f}_1\left(s_{1},s_{0};\eta_{1}\right)e\left(s_{1},s_{0},X\right)\left\{\mu_{1}\left(s_{1},X\right) - f_1\left(s_{1},s_{0};\eta_{1}\right)\right\}\d s_{1}\d s_{0},\\
    \ell_{2}\left(S,Z,X\right) &=& h_{2}\left(S,Z,X\right)-h_{5}\left(S,Z,X\right)\\
    &=& \frac{Z}{\pi\left(X\right)}\left[m_{11}\left(S,X\right)-\E\left\{m_{11}\left(S,X\right)\mid Z=1,X\right\} \right]\\
    &&+ \frac{1-Z}{1-\pi\left(X\right)}\left[m_{01}\left(S,X\right)-\E\left\{ m_{01}\left(S,X\right)\mid Z=0,X\right\} \right],\\
    \ell_{3}\left(Y,S,Z,X\right) &=&h_{3}\left(Y,S,Z,X\right)\\
    &=& \frac{Z}{\pi\left(X\right)}\frac{ \int w_1\left(S,s_{0}\right)\dot{f}_1\left(S,s_{0};\eta_{1}\right)e\left(S,s_{0},X\right)\d s_{0}}{p_{1}\left(S,X\right)}\left\{Y-\mu_{1}\left(S,X\right)\right\},
\end{eqnarray*}
with $ m_{z1}\left(S,X\right) = \tilde{\mu}_{z1}\left(S,X\right) - \tilde{f}_{z1}\left(S,X\right)$ for $z=0,1$.

Finally, we simplify the expression of $l_2(S,Z,X)$. 
Denote $m_{11}\left(S,X\right) = m_{c11}\left(S,X\right)+m_{\dot{c}11}\left(S,X\right)$ with
\begin{eqnarray*}
m_{c11}\left(S,X\right)&=& \int w_1(S,s_0)\dot{f}_1(S,s_0;\eta_1) \frac{e(S,s_0,X)}{p_1(S,X)}\left\{\mu_{1}\left(S,X\right)-f_1(S,s_0;\eta_1) \right\}\d s_0,\\
m_{\dot{c}11}\left(S,X\right)&=& \iint 1(s_1\geq S)w_1(s_1,s_0)\dot{f}_1(s_1,s_0;\eta_1)\dot{c}_{u}(s_1,s_0,X)p_1(s_1,X)p_0(s_0,X)\\
&&\hspace{1cm}\cdot\left\{\mu_{1}\left(s_{1},X\right)-f_1(s_1,s_0;\eta_1)\right\}\d s_{1}\d s_{0}.
\end{eqnarray*}
%
%
For $m_{c11}\left(S,X\right)$, we have
\begin{eqnarray*}
&&\E \left\{  m_{c11}\left(S,X\right) \mid Z=1,X \right\}\\
&=& \iint w_1\dot{f}_1 \frac{e(s_1,s_0,X)}{p_1(s_1,X)}\left\{\mu_{1}\left(s_{1},X\right)-f_1 \right\} p_1(s_1,X)  \d s_1 \d s_0\\
&=&\iint w_1\dot{f}_1 e(s_1,s_0,X)\left\{\mu_{1}\left(s_{1},X\right)-f_1 \right\}  \d s_1 \d s_0.
\end{eqnarray*}
 For $m_{\dot{c}11}\left(S,X\right)$, we have
\begin{eqnarray*}
&&\E\left\{m_{\dot{c}11}\left(S,X\right)\mid Z=1,X \right\}\\
&=&\iint  1(t_1\geq s_1)w_1(t_1,s_0)\dot{f}_1\dot{c}_{u}(t_1,s_0,X)p_1(s_1,X)p_1(t_1,X)p_0(s_0,X)\left\{\mu_{1}\left(t_{1},X\right)-f_1\right\}\d t_{1}\d s_{0} \d s_1\\
&=&\iint  1(t_1\geq s_1)w_1(t_1,s_0)\dot{f}_1\dot{c}_{u}(t_1,s_0,X)p_1(s_1,X)p_1(t_1,X)p_0(s_0,X)\left\{\mu_{1}\left(t_{1},X\right)-f_1\right\}\d t_{1}\d s_{0} \d s_1\\
&=&\iint  \F_1(t_1,X)w_1(t_1,s_0)\dot{f}_1\dot{c}_{u}(t_1,s_0,X)p_1(t_1,X)p_0(s_0,X)\left\{\mu_{1}\left(t_{1},X\right)-f_1\right\}\d t_{1}\d s_{0}\\
&=&\iint  \F_1(s_1,X)w_1(s_1,s_0)\dot{f}_1\dot{c}_{u}(s_1,s_0,X)p_1(s_1,X)p_0(s_0,X)\left\{\mu_{1}\left(s_{1},X\right)-f_1\right\}\d s_{1}\d s_{0},
\end{eqnarray*}
where we suppress the dependence of $\dot f_1$ and $f_1$ on the parameters and change $t_1$ to $s_1$ in the last equality. Therefore, we obtain
\begin{eqnarray*}
m_{11}\left(S,X\right)-\E\left\{m_{11}\left(S,X\right)\mid Z=1,X\right\} &=& m_{c11}\left(S,X\right) - m_{r11}\left(S,X\right), 
\end{eqnarray*}
where
\begin{eqnarray*}
m_{r11}\left(S,X\right) &=&\E \left\{  m_{c11}\left(S,X\right) \mid Z=1,X \right\}-\left[m_{\dot{c}11}\left(S,X\right)-\E \left\{  m_{\dot{c}11}\left(S,X\right) \mid Z=1,X \right\} \right]\\
  &=&\iint w_1(s_1,s_0)\dot{f}_1(s_1,s_0;\eta_1) e(s_1,s_0,X)\left\{\mu_{1}\left(s_{1},X\right)-f_1(s_1,s_0;\eta_1) \right\}  \d s_1 \d s_0\\
  &&-\iint w_1(s_1,s_0)\dot{f}_1(s_1,s_0;\eta_1)\frac{\dot{c}_{u}(s_1,s_0,X)}{c(s_1,s_0,X)}e(s_1,s_0,X)\\
&&\hspace{1cm}\cdot\left\{1(S\leq s_1)-\F_1(s_1,X)\right\}  \left\{\mu_{1}\left(s_{1},X\right)-f_1(s_1,s_0;\eta_1)\right\}\d s_{1}\d s_{0}\\
&=&\iint w_1(s_1,s_0)\dot{f}_1(s_1,s_0;\eta_1) e(s_1,s_0,X)\left\{\mu_{1}\left(s_{1},X\right)-f_1(s_1,s_0;\eta_1) \right\} \\
&&\hspace{1cm}\cdot\left[1- \frac{\dot{c}_{u}(s_1,s_0,X)}{c(s_1,s_0,X)}\left\{1(S\leq s_1)-\F_1(s_1,X)\right\}  \right] \d s_1 \d s_0
\end{eqnarray*}
Similarly, we can denote $m_{01}\left(S,X\right) = m_{c01}\left(S,X\right)+m_{\dot{c}01}\left(S,X\right)$ with
\begin{eqnarray*}
m_{c01}\left(S,X\right)&=& \int w_1(s_1,S)\dot{f}_1(s_1,S;\eta_1) \frac{e(s_1,S,X)}{p_0(S,X)}\left\{\mu_{1}\left(s_1,X\right)-f_1(s_1,S;\eta_1) \right\}\d s_1,\\
m_{\dot{c}01}\left(S,X\right)&=& \iint 1(s_0\geq S)w_1(s_1,s_0)\dot{f}_1(s_1,s_0;\eta_1)\dot{c}_{v}(s_1,s_0,X)p_1(s_1,X)p_0(s_0,X)\\
&&\hspace{1cm}\cdot\left\{\mu_{1}\left(s_{1},X\right)-f_1(s_1,s_0;\eta_1)\right\}\d s_{1}\d s_{0}
\end{eqnarray*}
and obtain 
\begin{eqnarray*}
m_{01}\left(S,X\right)-\E\left\{m_{11}\left(S,X\right)\mid Z=0,X\right\} &=& m_{c01}\left(S,X\right) - m_{r01}\left(S,X\right), 
\end{eqnarray*}
where
\begin{eqnarray*}
m_{r01}\left(S,X\right) &=&\E \left\{  m_{c11}\left(S,X\right) \mid Z=0,X \right\}-\left[m_{\dot{c}11}\left(S,X\right)-\E \left\{  m_{\dot{c}11}\left(S,X\right) \mid Z=0,X \right\} \right]\\
  &=&\iint w_1(s_1,s_0)\dot{f}_1(s_1,s_0;\eta_1) e(s_1,s_0,X)\left\{\mu_{1}\left(s_{1},X\right)-f_1(s_1,s_0;\eta_1) \right\}  \d s_1 \d s_0\\
  &&-\iint w_1(s_1,s_0)\dot{f}_1(s_1,s_0;\eta_1)\frac{\dot{c}_{v}(s_1,s_0,X)}{c(s_1,s_0,X)}e(s_1,s_0,X)\\
&&\hspace{1cm}\cdot\left\{1(S\leq s_0)-\F_0(s_0,X)\right\}  \left\{\mu_{1}\left(s_{1},X\right)-f_1(s_1,s_0;\eta_1)\right\}\d s_{1}\d s_{0}\\
&=&\iint w_1(s_1,s_0)\dot{f}_1(s_1,s_0;\eta_1) e(s_1,s_0,X)\left\{\mu_{1}\left(s_{1},X\right)-f_1(s_1,s_0;\eta_1) \right\} \\
&&\hspace{1cm}\cdot\left[1- \frac{\dot{c}_{v}(s_1,s_0,X)}{c(s_1,s_0,X)}\left\{1(S\leq s_0)-\F_0(s_0,X)\right\}  \right] \d s_1 \d s_0.
\end{eqnarray*}
\QEDB

\subsection{Proof of Example \ref{exmp:constant}}
First, we have
\begin{eqnarray*}
    \eta_{1} & =&  \arg\min_{\eta}\E \left\{m_{1}\left(S_{1},S_{0}\right)-\eta \right\}^{2} \ =\  \E \left\{m_{1}\left(S_{1},S_{0}\right)\right\} \ = \ \E(Y_{1}).
\end{eqnarray*}
\allowdisplaybreaks
Because $w_1(s_1,s_0)=1$ and $f_1(s_1,s_0;\eta)=\eta$  for all $s_1,s_0$, we have $\dot f_1(s_1,s_0;\eta)=1$ and $\ddot f_1(s_1,s_0;\eta)=0$. Thus, $H_1=1$ and $\varphi_1(V)=\ell_1(X) + \ell_2(S,Z,X) + \ell_3(Y,S,Z,X)$. We will compute these three terms using results in Theorem \ref{thm:eif_m1} and the following properties of the copula density function $c(\cdot,\cdot)$:
\begin{eqnarray}
    \int c(s_1,s_0,X)p_0(s_0,X) \d s_0 &=& \int \frac{e(s_1,s_0,X)}{p_1(s_1,X)} \d s_0 \ =\ 1, \label{equ::copula_d_s_0} \\
    \int \dot{c}_u (s_1,s_0,X)p_0(s_0,X)\d s_0 &=& 0, \label{equ::dot_copula_u}
\end{eqnarray}
where \eqref{equ::copula_d_s_0} follows from \eqref{equ::principal_density} and~\eqref{equ::dot_copula_u} follows from \eqref{equ::copula_d_s_0}  by taking the first order derivative with respective to $\F_1(s_1,X)$ on both sides.

For $\ell_1(X)$, we have
\begin{eqnarray*}
    \ell_{1}\left(X\right) &=& \iint e\left(s_{1},s_{0},X\right)\left\{\mu_{1}\left(s_{1},X\right) - \eta_1 \right\}\d s_{1}\d s_{0} \\
    &=& \E\left(Y\mid Z=1,X\right)-\eta_1.
\end{eqnarray*}

For $\ell_2(S,Z,X)$, we compute $m_{c11}(S,X)$, $m_{r11}(S,X)$, $m_{c01}(S,X)$, and $m_{r01}(S,X)$ separately based on the expressions in the proof of Theorem~\ref{thm:eif_m1}.
We have 
\begin{eqnarray*}
m_{c11}(S,X)&=& \int  \frac{e(S,s_0,X)}{p_1(S,X)}\left\{\mu_{1}\left(S,X\right)-\eta_1\right\}\d s_0 \ = \ \mu_{1}\left(S,X\right)-\eta_1.
\end{eqnarray*}
and 
\begin{eqnarray*}
m_{\dot{c}11}(S,X)
&=& \iint  \dot{c}_{u}(s_1,s_0,X)p_1(s_1,X)p_0(s_0,X)1(S\leq s_1) \left\{\mu_{1}\left(s_{1},X\right)-\eta_1\right\}\d s_{1}\d s_{0}\\
&=& \int   \left\{\int \dot{c}_{u}(s_1,s_0,X)p_0(s_0,X)\d s_{0}\right\} p_1(s_1,X)1(S\leq s_1) \left\{\mu_{1}\left(s_{1},X\right)-\eta_1\right\}\d s_{1}\\
&=& 0,
\end{eqnarray*}
where the last equality follows from~\eqref{equ::dot_copula_u}. 
Therefore,
\begin{eqnarray*}
m_{r01}(S,X)&=& \E\{m_{c11}(S,X)\mid Z=1,X\} \ = \ \E(Y\mid Z=1,X)-\eta_1.
\end{eqnarray*}

For $m_{c01}(S,X)$, we consider
\begin{eqnarray}
\nonumber m_{c01}(s_0,X)&=& \int \frac{e(s_1,s_0,X)}{p_0(s_0,X)}\left\{\mu_{1}\left(s_{1},X\right)-\eta_1 \right\}\d s_1\\
\label{eqn::ex1-md01}&=& \int c(s_1,s_0,X)p_1(s_1,X)\left\{\mu_{1}\left(s_{1},X\right)-\eta_1 \right\}\d s_1.
\end{eqnarray}
From the chain rule, we have 
\begin{eqnarray*}
    \frac{\partial}{\partial t_0}c(s_1,t_0,X) &=& \dot{c}_v(s_1,t_0,X)p_0(t_0,X).
\end{eqnarray*}
Integrating from $-\infty$ to $s_0$ on both sides, we obtain
\begin{eqnarray}
 \label{eqn::ex1-partial}   c(s_1,s_0,X) &=&\int_{-\infty}^{s_0} \dot{c}_v(s_1,t_0,X)p_0(t_0,X)\d t_0 ,
\end{eqnarray}
where we use $c(s_1,-\infty,X)=0$.
Plugging~\eqref{eqn::ex1-partial} into~\eqref{eqn::ex1-md01} yields
\begin{eqnarray*}
m_{c01}(s_0,X)&=&  \int  \left\{\int_{-\infty}^{s_0} \dot{c}_v(s_1,t_0,X)p_0(t_0,X)\d t_0 \right\}p_1(s_1,X)\left\{\mu_{1}\left(s_{1},X\right)-\eta_1 \right\}\d s_1\\
 &=&\iint 1(t_0<s_0) \dot{c}_v(s_1,t_0,X)p_0(t_0,X)p_1(s_1,X)\left\{\mu_{1}\left(s_{1},X\right)-\eta_1 \right\}\d s_1 \d t_0.
\end{eqnarray*}
Therefore,
\begin{eqnarray*}
m_{c01}(S,X)&=& \iint 1(t_0<S) \dot{c}_v(s_1,t_0,X)p_0(t_0,X)p_1(s_1,X)\left\{\mu_{1}\left(s_{1},X\right)-\eta_1 \right\}\d s_1 \d t_0\\
 &=& \iint 1(s_0<S) \dot{c}_v(s_1,s_0,X)p_1(s_1,X)p_0(s_0,X)\left\{\mu_{1}\left(s_{1},X\right)-\eta_1 \right\}\d s_1 \d s_0,
\end{eqnarray*}
where we change $t_0$ to $s_0$ in the last equality.
For  $m_{\dot{c}01}(S,X)$, we have
\begin{eqnarray*}
m_{\dot{c}01}(S,X)
&=&\iint 1(S\leq s_0)\dot{c}_{v}(s_1,s_0,X)p_1(s_1,X)p_0(s_0,X) \left\{\mu_{1}\left(s_{1},X\right)-\eta_1\right\}\d s_{1}\d s_{0}.
\end{eqnarray*}
Therefore,
\begin{eqnarray*}
m_{c01}(S,X)+m_{\dot{c}01}(S,X)&=& \iint \dot{c}_{v}(s_1,s_0,X)p_1(s_1,X)p_0(s_0,X) \left\{\mu_{1}\left(s_{1},X\right)-\eta_1\right\}\d s_{1}\d s_{0},
\end{eqnarray*}
which does not depend on $S$. As a result,
\begin{eqnarray*}
m_{c01}(S,X)-m_{r01}(S,X)&=&m_{c01}(S,X)+m_{\dot{c}01}(S,X)-\E\left\{m_{c01}(S,X)+m_{\dot{c}01}(S,X)\mid Z=0,X\right\}\\
&=&0.
\end{eqnarray*}
Combining results on  $m_{c11}(S,X)$, $m_{r11}(S,X)$, $m_{c01}(S,X)$, and $m_{r01}(S,X)$, we obtain
\begin{eqnarray*}
  \ell_2(S,Z,X) = \frac{Z}{\pi(X)}\left\{\mu_{1}\left(S,X\right)-\E(Y\mid Z=1,X)\right\}.
\end{eqnarray*}

For $\ell_3(Y,S,Z,X)$, we have
\begin{eqnarray*}
    \ell_{3}\left(Y,S,Z,X\right) &=& \frac{Z}{\pi\left(X\right)}\frac{ \int e\left(S,s_{0},X\right)\d s_{0}}{p_{1}\left(S,X\right)}\left\{Y-\mu_{1}\left(S,X\right)\right\} \\
    &=& \frac{Z}{\pi\left(X\right)}\left\{Y-\mu_{1}\left(S,X\right)\right\}.
\end{eqnarray*}

Combining these three terms gives the result in Example \ref{exmp:constant}. \QEDB

\subsection{Proof of Example \ref{exmp:linear_eta}}
Under the linear working model $f_1\left(s_{1},s_{0};\eta_1 \right)=\eta^{\T}_1g_1\left(s_{1},s_{0}\right)$,  we have $\dot{f}_1\left(s_{1},s_{0};\eta_1 \right) = g_1\left(s_{1},s_{0}\right)$ and $\ddot{f}_1\left(s_{1},s_{0};\eta_1 \right) = 0$. We can then obtain the EIF by plugging the expressions of $\dot{f}_1\left(s_{1},s_{0};\eta_1 \right) $ and $\ddot{f}_1\left(s_{1},s_{0};\eta_1 \right) $ in the general EIF given in Theorem~\ref{thm:eif_m1}. \QEDB

\subsection{Proof of Theorem \ref{thm:consistency}}
We first introduce the following lemma to simplify the proof.
\begin{lemma}[Valid estimating equation]
\label{lemma:valid_ee}
    $\E\left\{ D_{1,\textup{eif}}\left(V;\eta_1,\bar{\pi},\bar{e},\bar{\mu}_{1}\right)\right\}=0$ if either $(\bar \pi, \bar e) = (\pi,e)$ or $(\bar e, \bar \mu_1) = (e,\mu_1)$.
\end{lemma}

\noindent {\it Proof of Lemma \ref{lemma:valid_ee}.} Denote
\begin{eqnarray*}
\bar \nu_1(s_1,s_0,X)&=&  w_1\left(s_{1},s_{0}\right)\dot{f}_1\left(s_{1},s_{0};\eta_{1}\right)\bar e\left(s_{1},s_{0},X\right)\left\{\bar \mu_{1}\left(s_{1},X\right) - f_1\left(s_{1},s_{0};\eta_{1}\right)\right\},\\
\bar r_u(s_1,s_0,S,X)&=& 1- \frac{\bar{\dot{c}}_{u}(s_1,s_0,X)}{\bar c(s_1,s_0,X)}\left\{1(S\leq s_1)-\bar \F_1(s_1,X)\right\},\\
\bar r_v(s_1,s_0,S,X)&=& 1- \frac{\bar{\dot{c}}_{v}(s_1,s_0,X)}{\bar c(s_1,s_0,X)}\left\{1(S\leq s_0)-\bar \F_0(s_0,X)\right\},
\end{eqnarray*}
where $\bar \F_z(s_z,X)$ is the probability limit of the cumulative distribution function of $S_z$ given $X$ for $z=0,1$ and $\bar{\dot{c}}_{*}(s_1,s_0,X_i)/\bar c(s_1,s_0,X_i)$ equal to $\partial \log c(u,v) / \partial *$ for $*=u,v$ evaluated at the $\bar \F_1(s_1,X_i)$ and $\bar \F_0(s_0,X_i)$. 
We have 
\begin{eqnarray}
\nonumber  &&  \E\left\{ D_{1,\textup{eif}}\left(V;\eta_1,\bar{\pi},\bar{e},\bar{\mu}_{1}\right)\right\} \\
 \label{eqn::robust-eif}   &=& \E\left\{ \ell_{1}\left(X;\eta_1,\bar{e},\bar{\mu}_{1}\right)\right\}+ \E\left\{ \ell_{2}\left(S,Z,X;\eta_1,\bar{\pi},\bar{e},\bar{\mu}_{1}\right)\right\} 
    + \E\left\{ \ell_{3}\left(Y,S,Z,X;\eta_1,\bar{\pi},\bar{e},\bar{\mu}_{1}\right)\right\},
\end{eqnarray}
where $V=(Y,S,Z,X)$ and
\begin{eqnarray*}
\ell_{1}\left(X;\eta_1,\bar{e},\bar{\mu}_{1}\right) &=& \iint  \bar \nu_1(s_1,s_0,X) \d s_{1}\d s_{0},\\
\ell_{2}\left(S,Z,X;\eta_1,\bar{\pi},\bar{e},\bar{\mu}_{1}\right) &=& 
\frac{Z}{\bar\pi\left(X\right)}\left\{ \int \frac{\bar \nu_1(S,s_0,X)}{\bar p_1(S,X)}\d s_0 - \iint \bar \nu_1(s_1,s_0,X) \bar r_u(s_1,s_0,S,X) \d s_1 \d s_0\right\}\\
&&+ \frac{1-Z}{1-\bar \pi\left(X\right)}\left\{\int  \frac{\bar \nu_1(s_1,S,X)}{\bar p_0(S,X)} \d s_1 - \iint \bar \nu_1(s_1,s_0,X) \bar r_v(s_1,s_0,S,X) \d s_1 \d s_0\right\},\\
\ell_{3}\left(Y,S,Z,X;\eta_1,\bar{\pi},\bar{e},\bar{\mu}_{1}\right) &=& \frac{Z}{\bar \pi\left(X\right)}\frac{ \int w_1\left(S,s_{0}\right)\dot{f}_1\left(S,s_{0};\eta_{1}\right)\bar e\left(S,s_{0},X\right)\d s_{0}}{\bar p_{1}\left(S,X\right)}\left\{Y-\bar \mu_{1}\left(S,X\right)\right\}.
\end{eqnarray*}
For ease of notation, we suppress the dependence of the functions on $(s_1,s_0,S,X)$ and $\eta_1$ when no confusion arises. We have
\begin{eqnarray}
\label{eqn::robust-m1}\bar{\nu}_1 &=&w_1\dot{f}_1(\bar{e}-e)\left(\mu_{1}-f_1\right)+w_1\dot{f}_1\bar{e}\left(\bar{\mu}_{1}-\mu_1\right)+w_1\dot{f}_1e\left(\mu_{1}-f_1\right),\\
\nonumber \bar r_u &=&  r_u  + \left(\frac{\bar{\dot c}_u}{\bar c} - \frac{\dot c_u}{c} \right)\left\{\bar \F_1-1(S\leq s_1)\right\} + \frac{\dot{c}_{u}}{c}\left(\bar \F_1-\F_1\right), \\
\nonumber \bar r_v &=&  r_v  + \left(\frac{\bar{\dot c}_v}{\bar c} - \frac{\dot c_v}{c} \right)\left\{\bar \F_0-1(S\leq s_0)\right\} + \frac{\dot{c}_{v}}{c}\left(\bar \F_0-\F_0\right).
\end{eqnarray}
We then compute the three terms in~\eqref{eqn::robust-eif} separately. 
For the first term, we have
\begin{eqnarray}
\nonumber    \E\left\{ \ell_{1}\left(X;\eta_1,\bar{e},\bar{\mu}_{1}\right)\right\} &=& \E\left\{ \iint  \bar \nu_1 \d s_{1}\d s_{0}\right\} \\
  \label{eqn::robust-l1}    &=& \E\left\{ \iint w_1\dot{f}_1\left(\bar{e}-e\right)\left(\mu_{1}-f_1\right)\d s_{1}\d s_{0}\right\} + \E\left\{ \iint w_1\dot{f}_1\bar{e}\left(\bar{\mu}_{1}-\mu_1\right)\d s_{1}\d s_{0}\right\},
\end{eqnarray}
where the second equality follows from $$\E\left\{\iint w_1\dot{f}_1e\left(\mu_{1}-f_1\right)\d s_1 \d s_0\right\} =0.$$ 
For the second term, we have
\begin{eqnarray*}
\E\left\{ \frac{Z}{\bar \pi\left(X\right)} \int  \frac{\bar  \nu_1(S,s_0,X)}{\bar  p_1(S,X)}\d s_0\right\}
&=&\E\left[ \E\left\{ \frac{Z}{\bar \pi\left(X\right)} \int  \frac{\bar  \nu_1(S,s_0,X)}{\bar  p_1(S,X)}\d s_0 \right\}\mid X \right]\\
&=&\E\left[ \frac{\pi(X)}{\bar \pi\left(X\right)}  \E\left\{ \left. \int  \frac{\bar \nu_1(S,s_0,X)}{\bar p_1(S,X)}\d s_0\ \right | \ Z=1, X  \right\}\right]\\
&=&\E\left( \frac{\pi}{\bar \pi }    \iint  \frac{\bar \nu_1 p_1 }{\bar p_1 }\d s_1 \d s_0\right)\\
&=&\E\left( \frac{\pi}{\bar \pi }    \iint  \frac{\bar \nu_1 (p_1-\bar p_1) }{\bar p_1 }\d s_1 \d s_0\right)+\E\left( \frac{\pi}{\bar \pi }    \iint  \bar \nu_1 \d s_1 \d s_0\right)
\end{eqnarray*}
and 
\begin{eqnarray*}
&&\E\left\{ \frac{Z}{\bar \pi\left(X\right)} \iint  \bar  \nu_1 \bar r_u\d s_1 \d s_0\right\} \\
&=&\E\left\{ \frac{Z}{\bar \pi\left(X\right)} \iint  \bar  \nu_1 (\bar r_u-r_u)\d s_1 \d s_0\right\} +\E\left\{ \frac{Z}{\bar \pi\left(X\right)} \iint  \bar  \nu_1  r_u\d s_1 \d s_0\right\} \\
&=&\E\left[ \frac{Z}{\bar \pi\left(X\right)} \iint  \bar  \nu_1 \left(\frac{\bar{\dot c}_u}{\bar c} - \frac{\dot c_u}{c} \right)\left\{\bar \F_1-1(S\leq s_1)\right\} \d s_1 \d s_0\right]\\
&&+ \E\left\{ \frac{Z}{\bar \pi\left(X\right)} \iint \bar \nu_1 \left(\bar \F_1-\F_1\right)\frac{\dot c_u}{c} \d s_1 \d s_0 \right\} + \E\left\{ \frac{Z}{\bar \pi\left(X\right)} \iint  \bar  \nu_1  r_u\d s_1 \d s_0\right\} \\
&=&\E\left\{ \frac{\pi}{\bar \pi } \iint  \bar  \nu_1 \left(\frac{\bar{\dot c}_u}{\bar c} - \frac{\dot c_u}{c} \right) \left(\bar \F_1-\F_1\right) \d s_1 \d s_0\right\}\\
&&+ \E\left\{ \frac{\pi}{\bar \pi } \iint \bar \nu_1 \left(\bar \F_1-\F_1\right)\frac{\dot c_u}{c} \d s_1 \d s_0 \right\} + \E\left\{ \frac{Z}{\bar \pi\left(X\right)} \iint  \bar  \nu_1  r_u\d s_1 \d s_0\right\} \\
&=&\E\left\{ \frac{\pi}{\bar \pi } \iint  \bar  \nu_1 \frac{\bar{\dot c}_u}{\bar c} (\bar \F_1-\F_1)\d s_1 \d s_0\right\} +\E\left\{ \frac{Z}{\bar \pi\left(X\right)} \iint  \bar  \nu_1  r_u\d s_1 \d s_0\right\}.
\end{eqnarray*}
From the law of iterated expectations,
\begin{eqnarray*}
\E\left\{ \frac{Z}{\bar \pi\left(X\right)} \iint  \bar  \nu_1  r_u\d s_1 \d s_0\right\}&=& \E\left\{ \frac{\pi}{\bar \pi} \iint  \bar  \nu_1 \E( r_u\mid Z=1,X )\d s_1 \d s_0\right\}\ = \ \E\left( \frac{\pi}{\bar \pi }    \iint  \bar \nu_1 \d s_1 \d s_0\right),
\end{eqnarray*}
where the second equality follows from $\E( r_u\mid Z=1,X )=1$.
Therefore, we obtain
\begin{eqnarray}
\nonumber &&\E\left[ \frac{Z}{\bar \pi\left(X\right)} \left\{ \int  \frac{\bar  \nu_1(S,s_0,X)}{\bar  p_1(S,X)}\d s_0 - \iint \bar \nu_1(s_1,s_0,X) \bar r_u(s_1,s_0,S,X) \d s_1 \d s_0 \right\} \right]\\
\label{eqn::robust-l2-1}   &=&-\E\left[ \frac{\pi}{\bar \pi } \iint  \bar  \nu_1 \left\{ \frac{\bar p_1 - p_1}{\bar p_1}+ \frac{\bar{\dot c}_u}{\bar c} (\bar \F_1-\F_1) \right\}\d s_1 \d s_0\right].
\end{eqnarray}
Similarly, we have 
\begin{eqnarray}
\nonumber &&\E\left[ \frac{1-Z}{1-\bar \pi\left(X\right)} \left\{ \int  \frac{\bar  \nu_1(s_1,S,X)}{\bar  p_0(S,X)}\d s_0 - \iint  \bar  \nu_1(s_1,s_0,X) \bar r_v(s_1,s_0,S,X) \d s_1 \d s_0 \right\} \right]\\
\label{eqn::robust-l2-2} &=& -\E\left[ \frac{1-\pi}{1-\bar \pi } \iint  \bar  \nu_1 \left\{ \frac{\bar p_0 - p_0}{\bar p_0}+ \frac{\bar{\dot c}_v}{\bar c} (\bar \F_0-\F_0) \right\}\d s_1 \d s_0\right].
\end{eqnarray}
For the third term, we have
\begin{eqnarray}
\nonumber      && \E\left[\frac{Z}{\bar{\pi}(X)}\frac{\int w_1(S,s_0)\dot{f}_1(S,s_0)\bar{e}(S,s_0,X)\d s_{0}}{\bar{p}_1(S,X)}\left\{Y-\bar{\mu}_{1}(S,X)\right\}\right] \\
\nonumber      &=& \E\left(\frac{\pi(X)}{\bar{\pi}(X)} \E\left[\frac{\int w_1(S,s_0)\dot{f}_1(S,s_0)\bar{e}(S,s_0,X)\d s_{0}}{\bar{p}_1(S,X)}\left\{Y-\bar{\mu}_{1}(S,X)\right\}\mid X,Z=1\right]\right) \\
\nonumber           &=& \E\left(\frac{\pi(X)}{\bar{\pi}(X)} \E\left[\frac{\int w_1(S,s_0)\dot{f}_1(S,s_0)\bar{e}(S,s_0,X)\d s_{0}}{\bar{p}_1(S,X)}\left\{\mu_{1}(S,X)-\bar{\mu}_{1}(S,X)\right\}\mid X,Z=1\right]\right) \\
\nonumber      &=& \E\left[\frac{\pi(X)}{\bar{\pi}(X)}\int  \frac{\int w_1(s_1,s_0)\dot{f}_1(s_1,s_0)\bar{e}(s_1,s_0,X)\d s_{0}}{\bar{p}_1(s_1,X)}\left\{\mu_{1}(s_1,X)-\bar{\mu}_{1}(s_1,X)\right\} p_1(s_1,X)\d s_1\right] \\
\label{eqn::robust-l3}      &=& -\E\left\{\frac{\pi}{\bar{\pi}}\iint w_1\dot{f}_1\bar{e}\left(\bar{\mu}_{1}-\mu_{1}\right)\frac{p_{1}}{\bar{p}_{1}}\d s_{1}\d s_{0}\right\},
\end{eqnarray}
where the first two equalities follow from the law of iterated expectations.

Combining~\eqref{eqn::robust-l1}~to~\eqref{eqn::robust-l3}, we have
\begin{eqnarray*}
&&\E\left\{ D_{1,\textup{eif}}\left(V;\eta_1,\bar{\pi},\bar{e},\bar{\mu}_{1}\right)\right\} \\
&=&  \E\left\{ \iint w_1\dot{f}_1\left(\bar{e}-e\right)\left(\mu_{1}-f_1\right)\d s_{1}\d s_{0}\right\} + \E\left\{ \iint w_1\dot{f}_1\bar{e}\left(\bar{\mu}_{1}-\mu_1\right)\d s_{1}\d s_{0}\right\}\\
&&- \E\left[ \frac{\pi}{\bar \pi } \iint  \bar  \nu_1 \left\{ \frac{\bar p_1 - p_1}{\bar p_1}+ \frac{\bar{\dot c}_u}{\bar c} (\bar \F_1-\F_1) \right\}\d s_1 \d s_0\right]\\
&&- \E\left[ \frac{1-\pi}{1-\bar \pi } \iint  \bar  \nu_1 \left\{ \frac{\bar p_0 - p_0}{\bar p_0}+ \frac{\bar{\dot c}_v}{\bar c} (\bar \F_0-\F_0) \right\}\d s_1 \d s_0\right]-\E\left\{\frac{\pi}{\bar{\pi}}\iint w_1\dot{f}_1\bar{e}\left(\bar{\mu}_{1}-\mu_{1}\right)\frac{p_{1}}{\bar{p}_{1}}\d s_{1}\d s_{0}\right\}.
\end{eqnarray*}
Plugging~\eqref{eqn::robust-m1} into the above equation, we obtain
\begin{eqnarray*}
    \E\left\{ D_{1,\textup{eif}}\left(V;\eta_1,\bar{\pi},\bar{e},\bar{\mu}_{1}\right)\right\}&=& \E\left\{\iint w_1\dot{f}_1D_{\textup{integrand}}\left(V;\eta_1,\bar{\pi},\bar{e},\bar{\mu}_{1}\right)\d s_{1}\d s_{0}\right\},
\end{eqnarray*}
where
\begin{eqnarray*}
    && D_{\textup{integrand}}\left(V;\eta_1,\bar{\pi},\bar{e},\bar{\mu}_{1}\right) \notag \\
    &=&\left(\bar{e}-e\right)\left(\mu_{1}-f_1\right)+ \bar{e}\left(\bar{\mu}_{1}-\mu_1\right)-\frac{\pi}{\bar \pi }  \bar{e}\left(\bar{\mu}_{1}-\mu_{1}\right)\frac{p_{1}}{\bar{p}_{1}} \\
    &&- \frac{\pi}{\bar \pi } \left\{\bar e (\mu_1-f_1)+\bar e (\bar \mu_1 - \mu_1) \right\}    \left\{ \frac{\bar p_1 - p_1}{\bar p_1}+ \frac{\bar{\dot c}_u}{\bar c} (\bar \F_1-\F_1) \right\}\\
    &&- \frac{1-\pi}{1-\bar \pi } \left\{\bar e (\mu_1-f_1)+\bar e (\bar \mu_1 - \mu_1) \right\}   \left\{ \frac{\bar p_0 - p_0}{\bar p_0}+ \frac{\bar{\dot c}_v}{\bar c} (\bar \F_0-\F_0) \right\}\\
    &=&\left(\bar{e}-e\right)\left(\mu_{1}-f_1\right)+ \bar{e}\left(\bar{\mu}_{1}-\mu_1\right)\left(1- \frac{\pi p_1}{\bar \pi \bar p_1} \right)\\
    &&+ \frac{\bar \pi-\pi}{\bar \pi } \left\{\bar e (\mu_1-f_1)+\bar e (\bar \mu_1 - \mu_1) \right\}    \left\{ \frac{\bar p_1 - p_1}{\bar p_1}+ \frac{\bar{\dot c}_u}{\bar c} (\bar \F_1-\F_1) \right\}\\
    &&-\left\{\bar e (\mu_1-f_1)+\bar e (\bar \mu_1 - \mu_1) \right\}    \left\{ \frac{\bar p_1 - p_1}{\bar p_1}+ \frac{\bar{\dot c}_u}{\bar c} (\bar \F_1-\F_1) \right\}\\
    &&-\frac{\bar \pi-\pi}{1-\bar \pi } \left\{\bar e (\mu_1-f_1)+\bar e (\bar \mu_1 - \mu_1) \right\}   \left\{ \frac{\bar p_0 - p_0}{\bar p_0}+ \frac{\bar{\dot c}_v}{\bar c} (\bar \F_0-\F_0) \right\}\\
    &&-\left\{\bar e (\mu_1-f_1)+\bar e (\bar \mu_1 - \mu_1) \right\}   \left\{ \frac{\bar p_0 - p_0}{\bar p_0}+ \frac{\bar{\dot c}_v}{\bar c} (\bar \F_0-\F_0) \right\}.
\end{eqnarray*}
Rearranging the terms, we obtain
\begin{eqnarray*}
&& D_{\textup{integrand}}\left(V;\eta_1,\bar{\pi},\bar{e},\bar{\mu}_{1}\right) \\
&=&\left(\bar{\pi}-\pi\right)\left\{\frac{\bar{p}_1-p_1}{\bar{p}_1} + \frac{\bar{\dot c}_u}{\bar c}  \left(\bar{\F}_1-\F_1\right)\right\}\frac{\bar{e}\left(\mu_1-f_1\right)}{\bar{\pi}} - \left(\bar{\pi}-\pi\right)\left\{\frac{\bar{p}_0-p_0}{\bar{p}_0} + \frac{\bar{\dot c}_v}{\bar c}  \left(\bar{\F}_0-\F_0\right)\right\}\frac{\bar{e}\left(\mu_1-f_1\right)}{1-\bar{\pi}} \label{eqn::pi_e} \\
&&+ \left(\bar{\pi}-\pi\right) \frac{\bar{\dot c}_u}{\bar c}\left(\bar{\F}_1-\F_1\right)\left(\bar{\mu}_1-\mu_1\right)\frac{\bar{e}}{\bar{\pi}} - \left(\bar{\pi}-\pi\right)\left\{\frac{\bar{p}_0-p_0}{\bar{p}_0} +  \frac{\bar{\dot c}_v}{\bar c} \left(\bar{\F}_0-\F_0\right)\right\}\left(\bar{\mu}_1-\mu_1\right)\frac{\bar{e}}{1-\bar{\pi}} \label{eqn::pi_e_mu1} \\
&&- \bar{e}\left(\mu_1 - f_1\right)\left\{\frac{\bar{p}_1-p_1}{\bar{p}_1} + \frac{\bar{\dot c}_u}{\bar c}  \left(\bar{\F}_1-\F_1\right) + \frac{\bar{p}_0-p_0}{\bar{p}_0} +\frac{\bar{\dot c}_v}{\bar c} \left(\bar{\F}_0-\F_0\right)\right\} \label{eqn::e}\\
&&- \left\{\frac{\bar{\dot c}_u}{\bar c} \left(\bar{\F}_1-\F_1\right) + \frac{\bar{p}_0-p_0}{\bar{p}_0} + \frac{\bar{\dot c}_v}{\bar c}\left(\bar{\F}_0-\F_0\right)\right\}\left(\bar{\mu}_{1}-\mu_1\right)\bar{e} \label{eqn::e_mu1} \\
&&+ \left(\bar{e}-e\right)\left(\mu_{1}-f_1\right)+  \left(\bar{\pi}-\pi\right)\left(\bar{\mu}_{1}-\mu_1\right)\frac{\bar{e}}{\bar{\pi}}.
\end{eqnarray*}

When the principal density model is correctly specified,  $\left(\bar{p}_{1},\bar{p}_{0},\bar{\F}_1,\bar{\F}_0, \bar{e}\right)=\left(p_{1},p_{0},\F_1,\F_0,e\right)$, and thus
$$
D_{\textup{integrand}}\left(V;\eta_1,\bar{\pi},\bar{e},\bar{\mu}_{1}\right)\ = \ \left(\bar{\pi}-\pi\right)\left(\bar{\mu}_{1}-\mu_1\right)\frac{\bar{e}}{\bar{\pi}},
$$
which is equal to zero if either  $\bar \pi = \pi$ or $ \bar \mu_1 = \mu_1$.
As a result, we have $ \E\left\{ D_{1,\textup{eif}}\left(V;\eta_1,\bar{\pi},\bar{e},\bar{\mu}_{1}\right)\right\}=0 $ if either $(\bar \pi, \bar e) = (\pi,e)$ or $(\bar e, \bar \mu_1) = (e,\mu_1)$.
\QEDB

\subsubsection*{Proof of Theorem~\ref{thm:consistency}(a)}
Our proposed estimator $\hat{\eta}_{1,\textup{eif}}$ is a Z-estimator with the system of estimating equations $D_{1,\textup{eif}}$. By Lemma~\ref{lemma:valid_ee}, the Glivenko-Cantelli Theorem, and the existence and uniqueness of $\eta_1$, consistency of the Z-estimator $\hat{\eta}_{1,\textup{eif}}$ to $\eta_1$ follows from Theorem 5.9 in \citet{van2000asymptotic}.
\subsubsection*{Proof of Theorem~\ref{thm:consistency}(b)}
From Theorem 5.31 in \citet{van2000asymptotic}, under the conditions stated in the theorem, the convergence rate of $\hat{\eta}_{1,\textup{eif}}$ is
\[
 O_{\P}\left\{ n^{-1/2} + \textup{Rem} \right\},
\]
where $\textup{Rem}=\norm{\E\left\{D_{1,\textup{eif}}\left(V;\eta,\hat{\pi},\hat{e},\hat{\mu}_{1}\right)\right\}}$. Following similar arguments as in the proof of Lemma \ref{lemma:valid_ee}, we have
\begin{eqnarray}
\label{eqn::Dintergrand}\E\left\{ D_{1,\textup{eif}}\left(V;\eta,\hat{\pi},\hat{e},\hat{\mu}_{1}\right)\right\}=\E\left\{ \iint w_1\dot{f}_1D_{\textup{integrand}}\left(V;\eta,\hat{\pi},\hat{e},\hat{\mu}_{1}\right)\d s_{1}\d s_{0}\right\},
\end{eqnarray}
where
\begin{eqnarray}
    && D_{\textup{integrand}}\left(V;\eta,\hat{\pi},\hat{e},\hat{\mu}_{1}\right) \notag \\
    &=& \left(\hat{\pi}-\pi\right)\left(\hat{\mu}_{1}-\mu_1\right)\frac{\hat{e}}{\hat{\pi}} \label{eqn::pi_mu1_est} \\
   \notag &&+ \left(\hat{\pi}-\pi\right)\left[ \frac{1}{\hat \pi}\left\{\frac{\hat{p}_1-p_1}{\hat{p}_1} + \frac{\hat{\dot c}_u}{\hat c}  \left(\hat{\F}_1-\F_1\right)\right\} - \frac{1}{1-\hat \pi}\left\{\frac{\hat{p}_0-p_0}{\hat{p}_0} + \frac{\hat{\dot c}_v}{\hat c} \left(\hat{\F}_0-\F_0\right)\right\}\right]\hat{e}\left(\mu_1-f_1\right)\\
    \label{eqn::pi_e_est} \\
    &&- \left\{\frac{\hat{\dot c}_u}{\hat c} \left(\hat{\F}_1-\F_1\right) + \frac{\hat{p}_0-p_0}{\hat{p}_0} + \frac{\hat{\dot c}_v}{\hat c} \left(\hat{\F}_0-\F_0\right)\right\}\left(\hat{\mu}_{1}-\mu_1\right)\hat{e} \label{eqn::e_mu1_est} \\
  \notag  &&+ \left(\hat{\pi}-\pi\right)\frac{\hat{\dot c}_u}{\hat c} \left(\hat{\F}_1-\F_1\right)\left(\hat{\mu}_1-\mu_1\right)\frac{\hat{e}}{\hat{\pi}} - \left(\hat{\pi}-\pi\right)\left\{\frac{\hat{p}_0-p_0}{\hat{p}_0} + \frac{\hat{\dot c}_v}{\hat c}  \left(\hat{\F}_0-\F_0\right)\right\}\left(\hat{\mu}_1-\mu_1\right)\frac{\hat{e}}{1-\hat{\pi}} \\
    \label{eqn::pi_e_mu1_est} \\
    &&+ \left(\mu_1 - f_1\right)\left[ \left(\hat{e}-e\right)- \hat{e}\left\{\frac{\hat{p}_1-p_1}{\hat{p}_1} + \frac{\hat{\dot c}_u}{\hat c}  \left(\hat{\F}_1-\F_1\right) + \frac{\hat{p}_0-p_0}{\hat{p}_0} + \frac{\hat{\dot c}_v}{\hat c} \left(\hat{\F}_0-\F_0\right)\right\}\right]. \label{eqn::e_est}
\end{eqnarray}
Plugging the expression of  $ D_{\textup{integrand}}\left(V;\eta,\hat{\pi},\hat{e},\hat{\mu}_{1}\right) $ into~\eqref{eqn::Dintergrand}, the corresponding first term
\begin{eqnarray*}
   \left\Vert\E\left\{ \iint w_1\dot{f}_1\left(\hat{\pi}-\pi\right)\left(\hat{\mu}_{1}-\mu_1\right)\frac{\hat{e}}{\hat{\pi}} \d s_{1}\d s_{0}\right\}\right\Vert
    &\leq& C\norm{\hat{\pi}-\pi}_{2}\norm{\hat{\mu}_{1}-\mu_{1}}_{2}
\end{eqnarray*}
by the Cauchy-Schwarz inequality and the fact that $w_1 \hat{e}/\hat{\pi} \norm{\dot f_1 }$  is bounded. Similarly, we can bound the terms in~\eqref{eqn::pi_e_est}--\eqref{eqn::pi_e_mu1_est} and simplify the bound using the fact that the convergence rate of $\hat{\F}_z$ is no slower than $\hat{p}_z$ for $z=0,1$ in both nonparametric and parametric estimation of the conditional distribution function and conditional density function \citep{van2000asymptotic,hall1999methods,fan2003nonlinear,li2023nonparametric}. For~\eqref{eqn::e_est}, we can write the term in the square brackets as
\begin{eqnarray*}
    && c\left(\hat{\F}_1,\hat{\F}_0\right)\hat{p}_1\hat{p}_0 - c\left(\F_1,\F_0\right)p_1p_0 - c\left(\hat{\F}_1,\hat{\F}_0\right)\left(\hat{p}_1-p_1\right)\hat{p}_0 - \dot{c}_u\left(\hat{\F}_1,\hat{\F}_0\right)\hat{p}_1\hat{p}_0\left(\hat{\F}_1-\F_1\right) \\
    &&- c\left(\hat{\F}_1,\hat{\F}_0\right)\left(\hat{p}_0-p_0\right)\hat{p}_1 - \dot{c}_v\left(\hat{\F}_1,\hat{\F}_0\right)\hat{p}_1\hat{p}_0\left(\hat{\F}_0-\F_0\right) \\
    &=& \hat{p}_1\hat{p}_0\left\{ -c\left(\F_1,\F_0\right)+ c\left(\hat{\F}_1,\hat{\F}_0\right) + \dot{c}_u\left(\hat{\F}_1,\hat{\F}_0\right)\left(\F_1-\hat{\F}_1\right)  +  \dot{c}_v\left(\hat{\F}_1,\hat{\F}_0\right)\left(\F_0-\hat{\F}_0\right)\right\}\\
    &&+ \left\{ c\left(\F_1,\F_0\right) -c\left(\hat{\F}_1,\hat{\F}_0\right) \right\} \left\{\left(\hat{p}_1-p_1\right)\hat{p}_0+\left(\hat{p}_0-p_0\right)\hat{p}_1 \right\}\\
    &&-c\left(\F_1,\F_0\right) \left\{ p_1p_0+\left(\hat{p}_1-p_1\right)\hat{p}_0+\left(\hat{p}_0-p_0\right)\hat{p}_1-\hat p_1\hat p_0 \right\}\\  
    &=&  -\hat{p}_1\hat{p}_0\left\{\frac{1}{2}\ddot{c}_{uu}\left(\hat{\F}_1,\hat{\F}_0\right)\left(\hat{\F}_1-\F_1\right)^2+\frac{1}{2}\ddot{c}_{vv}\left(\hat{\F}_1,\hat{\F}_0\right)\left(\hat{\F}_0-\F_0\right)^2+\ddot{c}_{uv}\left(\hat{\F}_1,\hat{\F}_0\right)\left(\hat{\F}_1-\F_1\right)\left(\hat{\F}_0-\F_0\right)\right\}\\
    && + \left\{\dot{c}_u\left(\hat{\F}_1,\hat{\F}_0\right)\left(\F_1-\hat{\F}_1\right) + \dot{c}_v\left(\hat{\F}_1,\hat{\F}_0\right)\left(\F_0-\hat{\F}_0\right)\right\} \left\{\left(\hat{p}_1-p_1\right)\hat{p}_0 + \left(\hat{p}_0-p_0\right)\hat{p}_1\right\} \\
    && - c\left(\F_1,\F_0\right)\left(\hat{p}_1-p_1\right)\left(\hat{p}_0-p_0\right) \\
    && + o\left\{\left(\hat{\F}_1-\F_1\right)^2 + \left(\hat{\F}_0-\F_0\right)^2 \right\} \\
    && + o\left\{\left(\hat{\F}_1-\F_1\right)\left(\hat{p}_1-p_1\right) + \left(\hat{\F}_1-\F_1\right)\left(\hat{p}_0-p_0\right) + \left(\hat{\F}_0-\F_0\right)\left(\hat{p}_1-p_1\right) + \left(\hat{\F}_0-\F_0\right)\left(\hat{p}_0-p_0\right)\right\},
\end{eqnarray*}
where $\ddot{c}_{uu}(u,v)=\partial^2c(u,v)/\partial u^2$, $\ddot{c}_{vv}(u,v)=\partial^2c(u,v)/\partial v^2$, $\ddot{c}_{uv}(u,v)=\partial^2c(u,v)/\partial u\partial v$, and the equality follows from Taylor expansion.

Therefore,~\eqref{eqn::e_est} can be bounded by $\norm{\hat{p}_{1}-p_{1}}_{2}\norm{\hat{p}_{0}-p_{0}}_{2}$ again due to the fact that $\hat{\F}_z$ has a faster convergence rate than $\hat{p}_z$ for $z=0,1$. Combining all of these terms gives the results in Theorem Theorem~\ref{thm:consistency}(b).  \QEDB

\subsection{Proof of Theorem~\ref{thm:asymptotic_distribution}}
By Theorem 5.31 in \citet{van2000asymptotic}, under the conditions stated in the theorem, we have
\begin{eqnarray}
    \nonumber \sqrt{n}\left(\hat{\eta}_{1,\textup{eif}}-\eta_1\right) &=& -C_1^{-1} \sqrt{n} \E\left\{ D_{1,\textup{eif}}\left(V;\eta_1,\hat{\pi},\hat{e},\hat{\mu}_{1}\right)\right\} \\
    && -C_1^{-1} \left\{\frac{1}{\sqrt{n}}\sumn D_{1,\textup{eif}}\left(V_{i};\eta_1,\pi, e, \mu_{1}\right) \right\} +o_{\P}\left(1+\sqrt{n}\textup{Rem}\right), \label{eqn::theorem531_dist}
\end{eqnarray}
where $C_{1} = C(\eta_1, \bar{\xi})$ denotes the derivative in Assumption \ref{assump:differentiablility}.
When Assumption \ref{assump:nuisance_rate} is satisfied, the first term in~\eqref{eqn::theorem531_dist}, also called the drift term, $\E\left\{ D_{1,\textup{eif}}\left(V;\eta_1,\hat{\pi},\hat{e},\hat{\mu}_{1}\right)\right\}=o_{\P}\left(n^{-1/2}\right)$, and the remainder term is of higher order. Thus $\sqrt{n}\left(\hat{\eta}_{1,\textup{eif}}-\eta_1\right)$ reduces to 
\[
-C_{1}^{-1}\left\{\frac{1}{\sqrt{n}}\sumn D_{1,\textup{eif}}\left(V_{i};\eta_1,\pi, e, \mu_{1}\right) \right\},
\]
and
\[
\sqrt{n}\left(\hat{\eta}_{1,\textup{eif}}-\eta_1\right) \overset{d}{\rightarrow}\mathcal{N}\left(0,C_{1}^{-1}\E\left\{ D_{1,\textup{eif}}\left(V;\eta_1,\xi\right)D_{1,\textup{eif}}\left(V;\eta_1,\xi\right)^{\T}\right\} C_{1}^{-1}\right),
\]
where $C_{1} = C(\eta_1, \bar{\xi})$ denotes the derivative in Assumption \ref{assump:differentiablility} \citep{van2000asymptotic}. 

By the fact that $\E\left\{\varphi_{1}\left(V;\eta_1,\xi\right)\varphi_{1}\left(V;\eta_1,\xi\right)^{\T}\right\} = H_{1}^{-1}\E\left\{ D_{1,\textup{eif}}\left(V;\eta_1,\xi\right)D_{1,\textup{eif}}\left(V;\eta_1,\xi\right)^{\T}\right\} H_{1}^{-1}$, it remains to show $H_{1}=C_{1}$. Under Assumption \ref{assump:nuisance_rate}, $\bar{\xi}=\xi$ and 
\begin{align*}
    C_{1}&=\frac{\partial}{\partial \eta_1}\E\left\{ D_{1,\textup{eif}}\left(V;\eta_1,\bar{\xi}\right)\right\} \\&=\frac{\partial}{\partial \eta_1}\E\left\{ \ell_{1}\left(X;\eta_1,\bar{\xi}\right)\right\} +\frac{\partial}{\partial \eta_1}\E\left\{ \ell_{2}\left(S,Z,X;\eta_1,\bar{\xi}\right)\right\} +\frac{\partial}{\partial \eta_1}\E\left\{ \ell_{3}\left(Y,S,Z,X;\eta_1,\bar{\xi}\right)\right\}.
\end{align*}
For the first term, we have
\begin{eqnarray*}
&& \frac{\partial}{\partial \eta_1}\E\left\{ \ell_{1}\left(X;\eta_1,\xi\right)\right\} \\
 &=&\E\left[ \iint w_1\left\{\ddot{f}_1e\left(s_{1},s_{0},X\right)\mu_{1}\left(s_{1},X\right)-\ddot{f}_1e\left(s_{1},s_{0}\right)f_1-\dot{f}_1\dot{f}_1^{\T}e\left(s_{1},s_{0}\right)\right\}\d s_{1}\d s_{0}\right] \\
    & =&\iint w_1\ddot{f}_1\E\left\{ e\left(s_{1},s_{0},X\right)\mu_{1}\left(s_{1},X\right)\right\} \d s_{1}\d s_{0}-\iint w_1e\left(s_{1},s_{0}\right)\left(\ddot{f}_1f_1-\dot{f}_1\dot{f}_1^{\T}\right)\d s_{1}\d s_{0}\\
    &=&H_{1}.
\end{eqnarray*}

For the second term, we have 
\begin{eqnarray*}
    &&\E\left[\frac{Z}{\pi\left(X\right)}\left\{  m_{11}\left(S,X;\eta_1\right)-\E\left( m_{11}\left(S,X;\eta_1\right) \mid Z=1,X\right)\right\} \right]\\
    &=&\E\left[\E\left\{ \frac{Z}{\pi\left(X\right)}\left\{  m_{11}\left(S,X;\eta_1\right)-\E\left( m_{11}\left(S,X;\eta_1\right)\mid Z=1,X\right)\right\} \mid X\right\} \right]\\
    &=&\E\left[\frac{\E\left(Z\mid X\right)}{\pi\left(X\right)}\E \left\{  m_{11}\left(S,X;\eta_1\right)-\E\left( m_{11}\left(S,X;\eta_1\right)\mid Z=1,X\right) \mid X,Z=1\right\} \right]\\
    &=&\E\left[\E\left\{  m_{11}\left(S,X;\eta_1\right)-\E\left( m_{11}\left(S,X;\eta_1\right)\mid Z=1,X\right) \mid X,Z=1\right\} \right]\\
    &=&0,
\end{eqnarray*}
where the first equality is by the law of iterated expectation and the second equality is by treatment ignorability.
Similarly, we have
\begin{eqnarray*}
 \E\left[\frac{1-Z}{1-\pi\left(X\right)}\left\{m_{01}\left(S,X;\eta_1\right) -\E\left(m_{01}\left(S,X;\eta_1\right) \mid Z=0,X\right) \right\}\right]\ =\ 0.
\end{eqnarray*}
As a result, $\E\left\{ \ell_{2}\left(S,Z,X;\eta_1,\bar{\xi}\right)\right\} =0$ for all $\eta_1$ and hence $\partial \E\left\{ \ell_{2}\left(S,Z,X;\eta_1,\xi\right)\right\} / \partial \eta_1 = 0$.

For the third term, we have
\begin{eqnarray*}
    && \frac{\partial}{\partial \eta_1}\E\left\{ \ell_{3}\left(Y,S,Z,X;\eta_1,\xi\right)\right\} \\
    &=& \E\left[ \frac{Z}{\pi\left(X\right)} \frac{\partial \int w_1(S,s_0)\dot{f}_1(S,s_0;\eta_1)e(S,s_0,X)\d s_{0}/\partial \eta_1}{p_{1}\left(S,X\right)}\left\{Y-\mu_{1}\left(S,X\right)\right\}\right] \\
    &=&\E\left[\E\left\{ \frac{\partial \int w_1(S,s_0)\dot{f}_1(S,s_0;\eta_1)e(S,s_0,X)\d s_{0}/\partial \eta_1}{p_{1}\left(S,X\right)}\left(Y-\mu_{1}\left(S,X\right)\right)\mid X,Z=1\right\} \right]\\
    &=&\E\left[\E\left\{ \frac{\partial \int w_1(S,s_0)\dot{f}_1(S,s_0;\eta_1)e(S,s_0,X)\d s_{0}/\partial \eta_1}{p_{1}\left(S,X\right)}\left(\E\left(Y\mid S,X,Z=1\right)-\mu_{1}\left(S,X\right)\right)\mid X,Z=1\right\} \right]\\
     &=&0,
\end{eqnarray*}
where the second and the third equalities follow from the law of total expectation.
Therefore, 
\begin{align*}
    C_{1}&=\frac{\partial}{\partial \eta_1}\E\left\{ \ell_{1}\left(X;\eta_1,\bar{\xi}\right)\right\} +\frac{\partial}{\partial \eta_1}\E\left\{ \ell_{2}\left(S,Z,X;\eta_1,\bar{\xi}\right)\right\} +\frac{\partial}{\partial \eta_1}\E\left\{ \ell_{3}\left(Y,S,Z,X;\eta_1,\bar{\xi}\right)\right\}=H_{1}.
\end{align*}
As a result, $\sqrt{n}\left(\hat{\eta}_{\textup{eif}}-\eta\right) \overset{d}{\rightarrow}\mathcal{N}\left(0,\E\left\{ \varphi_{1}\left(V;\eta_1,\xi\right)\varphi_{1}\left(V;\eta_1,\xi\right)^{\T}\right\}\right)$. \QEDB

\section{More examples: projection onto space of polynomial functions}
\label{app::example}
\label{sec::more_examples_linear_form}
This section presents two examples of the working model: a linear function of $s_1$ and polynomial functions of $(s_1,s_0)$.
For any function $r\left(Y,S,X\right)$, define
\begin{equation*}
    \psi_{r\left(Y_{z},S_{z},X\right)} = \frac{1(Z=z)}{\P(Z=z \mid X)}\left[ r\left(Y,S,X\right) - \E\left\{r\left(Y,S,X\right) \mid X, Z=z\right\} \right] + \E\left\{r\left(Y,S,X\right) \mid X, Z=z\right\}.
\end{equation*}
Under Assumption~\ref{assump:treatment_ignorability}, $\E\{\psi_{r\left(Y_{z},S_{z},X\right)}\}=\E\left\{r\left(Y_{z},S_{z},X\right)\right\}$. Below, we provide the EIFs under both models.

\begin{example}[Projection onto space of linear functions of $s_{1}$]
Suppose the weight $w_1\left(s_{1},s_{0}\right)\equiv1$ is constant, $f_1\left(s_{1},s_{0};\eta\right)=\eta^{1} s_{1}+\alpha$ is a linear function of $s_{1}$, and distributions of $S_{1}$ and $S_{0}$ are independent conditional on $X$. The EIF for $\left(\eta^{1},\alpha\right)^{\T}$ is
\[
\varphi_{1}\left(V\right)=H_{1}^{-1}\left(\phi_{1}^{1}\left(V\right),\phi^{\alpha}\left(V\right)\right)^{\T},
\]
where
\begin{eqnarray*}
H_{1} &=& \left(\begin{array}{cc}
\E\left\{\E\left(S^{2}\mid Z=1,X\right)\right\} & \E\left\{\E\left(S\mid Z=1,X\right)\right\}\\
\E\left\{\E\left(S\mid Z=1,X\right)\right\} & 1
\end{array}\right),\\
\phi_{1}^{1}\left(V\right) &=& \psi_{Y_{1}S_{1}} -\eta^{1}\psi_{S^{2}_{1}} - \alpha \psi_{S_{1}}, \\
\phi^{\alpha}\left(V\right) &=& \psi_{Y_{1}} -\eta^{1}\psi_{S_{1}} - \alpha.
\end{eqnarray*}
\label{exmp:linear_s1}
\end{example}

\begin{example}[Projection onto space of polynomial functions of $\left(s_{1},s_{0}\right)$]
Suppose the weight $w_1\left(s_{1},s_{0}\right)\equiv1$ is constant, $f_1\left(s_{1},s_{0};\eta\right)=\sum_{j=1}^{p}\eta_{j}^{1}s_{1}^{j}+\sum_{j=1}^{p}\eta_{j}^{0}s_{0}^{j}+\alpha$ is a polynomial function of $s_{1}$ and $s_{0}$, and distributions of $S_{1}$ and $S_{0}$ are independent conditional on $X$. Let $\bar{S}$ denote the $(2p+1)-$dimensional vector $(S_{1}, \ldots,S_{1}^{p}, S_{0},\ldots,S_{0}^{p}, 1)^{\T}$. The EIF for $\eta = (\eta_{1}^{1},\ldots,\eta_{p}^{1},\eta_{1}^{0},\ldots,\eta_{p}^{0},\alpha)^{\T}$ is
\[
\varphi_{1}\left(V\right)=H_{1}^{-1}\left(\phi_{1}^{1}\left(V\right),\ldots,\phi_{p}^{1}\left(V\right), \phi_{1}^{0}\left(V\right),\ldots,\phi_{p}^{0}\left(V\right),\phi_{\alpha}\left(V\right) \right)^{\T}
\]
where
$H_{1} = \E(\bar{S}\bar{S}^{\T})$
with $\E(S_1^j S_0^k) = \E\{\E(S^j\mid Z=1,X)\E(S^k\mid Z=0,X)\}$ for $j,k=1,\ldots,p$, and 
\begin{eqnarray*}
\phi_{k}^{1}\left(V\right) &=& \psi_{Y_{1}S_{1}^{k}} - \sum_{j=1}^{p}\eta_{j}^{1}\psi_{S_{1}^{j+k}} - \sum_{j=1}^{p}\eta_{j}^{0}\psi_{S_{1}^{k}}\psi_{S_{0}^{j}} - \alpha \psi_{S_{1}^{k}}, \\
\phi_{k}^{0}\left(V\right) &=& \psi_{Y_1}\psi_{S_0^k} - \sum_{j=1}^{p}\eta_{j}^{1}\psi_{S_1^j}\psi_{S_0^k} - \sum_{j=1}^{p}\eta_{j}^{0}\psi_{S_0^{j+k}} - \alpha\psi_{S_0^k},\\
\phi^{\alpha}\left(V\right) &=& \psi_{Y_{1}} - \sum_{j=1}^{p}\eta_{j}^{1}\psi_{S_1^j} - \sum_{j=1}^{p}\eta_{j}^{0}\psi_{S_0^j} -\alpha.
\end{eqnarray*}
for $k=1,\ldots,p$.
\label{exmp:poly}
\end{example}
Because working models are linear in the unknown parameter $\eta_1$, we can construct a closed-form estimator based on the EIFs.

\section{EIF for $\eta_{0}$}
\label{app::eif-eta0}
In this section, we provide the EIF for $\eta_0$ and the closed-form solution of $\eta_0$ under a linear working model.
Denote 
\begin{eqnarray*}
\nu_0(s_1,s_0,X)&=&  w_0\left(s_{1},s_{0}\right)\dot{f}_0\left(s_{1},s_{0};\eta_{0}\right)e\left(s_{1},s_{0},X\right)\left\{\mu_{0}\left(s_{0},X\right) - f_0\left(s_{1},s_{0};\eta_{0}\right)\right\}.
\end{eqnarray*}
 The following theorem presents the EIF for $\eta_0$.

\begin{theorem}
\label{thm:eif_m0}
Under the nonparametric model, the EIF for $\eta_{0}$ is
\begin{equation*}
    \varphi_{0}\left(V\right) = H_{0}^{-1} \left\{k_{1}\left(X\right)+k_{2}\left(S,Z,X\right)+k_{3}\left(Y,S,Z,X\right)\right\},
\end{equation*}
where
\begin{eqnarray*}
H_{0} &=& \E\left(w_0(S_1,S_0)\left[\dot{f}_0\left(S_{1},S_{0};\eta_{0}\right)\dot{f}_0\left(S_{1},S_{0};\eta_{0}\right)^{\T} - \left\{m_0(S_1,S_0) - f_0(S_1,S_0;\eta_0)\right\}\ddot{f}_0(S_1,S_0;\eta_0) \right] \right)
\end{eqnarray*}
and
\begin{eqnarray*}
    k_{1}\left(X\right) &=& \iint  \nu_0(s_1,s_0,X) \d s_{1}\d s_{0},\\
    k_{2}\left(S,Z,X\right) &=& \frac{Z}{\pi\left(X\right)}\left\{\frac{\int \nu_0(S,s_0,X)\d s_0}{p_1(S,X)} - \iint \nu_0(s_1,s_0,X) r_u(s_1,s_0,S,X) \d s_1 \d s_0\right\}\\
    &&+ \frac{1-Z}{1-\pi\left(X\right)}\left\{\frac{\int \nu_0(s_1,S,X)\d s_1 }{p_0(S,X)} - \iint \nu_0(s_1,s_0,X) r_v(s_1,s_0,S,X) \d s_1 \d s_0\right\},\\
    k_{3}\left(Y,S,Z,X\right) &=& \frac{1-Z}{1-\pi\left(X\right)}\frac{\int w_0\left(s_1,S\right)\dot{f}_0\left(s_1,S;\eta_{0}\right)e\left(s_1,S,X\right)\d s_{1}}{p_{0}\left(S,X\right)}\left\{Y-\mu_{0}\left(S,X\right)\right\}.
\end{eqnarray*}
\end{theorem}
The following example provides the EIF under a linear working model.
\begin{example}[Working model is linear in $\eta_{0}$]
Suppose that the working model $f_{0}\left(s_{1},s_{0};\eta_{0} \right)$ is linear in $\eta_{0}$, i.e., $f_{0}\left(s_{1},s_{0};\eta_{0} \right)=\eta_{0}^{\T}g_{0}\left(s_{1},s_{0}\right)$ for some function $g_{0}$. 
Denote 
\begin{eqnarray*}
G_0(s_1,s_0,X; h) &=& w_0(s_1,s_0)g_0(s_1,s_0)e(s_1,s_0,X)h(s_1,s_0,X),
\end{eqnarray*}
where $h(s_1,s_0,X)$ can be a scalar function or vector function whose dimension is compatible with $g_0(s_1,s_0)$.
The EIF for $\eta_0$ is 
\begin{equation*}
    \varphi_{0}\left(V\right)=H_{0}^{-1}\left\{A_3(Y,S,Z,X) - A_4(Y,S,Z,X) \eta_0\right\},
\end{equation*}
where
\begin{eqnarray*}
H_{0} &=& \iint w_0\left(s_{1},s_{0}\right)g_0\left(s_{1},s_{0}\right)g_0\left(s_{1},s_{0}\right)^{\T}e\left(s_{1},s_{0}\right)\d s_{1}\d s_{0}, \\
A_3(Y,S,Z,X) &=& \frac{Z}{\pi\left(X\right)}\left[\frac{\int G_0(S,s_0,X;\mu_0)\d s_0}{p_1(S,X)} - \iint G_0(s_1,s_0,X;\mu_0) r_u(s_1,s_0,S,X) \d s_1 \d s_0\right]\\
&&+\frac{1-Z}{1-\pi\left(X\right)}\left[\frac{\int G_0(s_1,S,X;\mu_0)d s_1}{p_0(S,X)} \ - \iint G_0(s_1,s_0,X;\mu_0)  r_v(s_1,s_0,S,X) \d s_1 \d s_0\right]\\
&&+\iint G_0(s_1,s_0,X; \mu_0)\d s_1 \d s_0 + \frac{1-Z}{1-\pi\left(X\right)}\frac{ \int G_0(s_1,S,X;1)\d s_{1}}{p_{0}\left(S,X\right)}\left\{Y-\mu_{0}\left(S,X\right)\right\}, \\
A_4(Y,S,Z,X) &=&\frac{Z}{\pi\left(X\right)}\left[\frac{\int G_0(S,s_0,X;g_0)\d s_0}{p_1(S,X)} - \iint G_0(s_1,s_0,X;g_0) r_u(s_1,s_0,S,X) \d s_1 \d s_0\right]\\
&&+\frac{1-Z}{1-\pi\left(X\right)}\left[\frac{\int G_0(s_1,S,X;g_0)\d s_1}{p_0(S,X)}  - \iint G_0(s_1,s_0,X;g_0)  r_v(s_1,s_0,S,X) \d s_1 \d s_0\right]\\
&&+ \iint G_0(s_1,s_0,X; g_0)\d s_1 \d s_0.
\end{eqnarray*}
We can  solve for the closed-form solution of $\eta_0$ and construct the estimator, $\hat{\eta}_{0,\textup{eif}}$, from the EIF.
\label{exmp:linear_eta_0}
\end{example}

\section{Projection of $\tau\left(s_{1}, s_{0}\right)$ and EIF for $\eta_{\tau}$}
\label{app::eif-tau}
We can project the PCE surface $\tau\left(s_{1},s_{0};\eta\right)$ directly without the two separate projections for the average potential outcome means. This section  presents the results for the projection parameter $\eta_{\tau}$, defined as
\begin{equation}
\eta_{\tau}=\arg\min_{\eta\in\mathbb{R}^{q}}\E\left[ w_{\tau}\left(S_{1},S_{0}\right)\left\{\tau\left(S_{1},S_{0}\right)-f_{\tau}\left(S_{1},S_{0};\eta\right)\right\}^{2}\right].
\label{equ:opt_tau}
\end{equation}
The identification of $\eta_{\tau}$ also follows from the identification of $\tau\left(s_{1},s_{0}\right)$. Denote 
\begin{eqnarray*}
\nu_{\tau}(s_1,s_0,X)&=&  w_{\tau}\left(s_{1},s_{0}\right)\dot{f}_{\tau}\left(s_{1},s_{0};\eta_{\tau}\right)e\left(s_{1},s_{0},X\right)\left\{\mu_{1}\left(s_{1},X\right) - \mu_{0}\left(s_{0},X\right) - f_{\tau}\left(s_{1},s_{0};\eta_{\tau}\right)\right\},\\
r_u(s_1,s_0,S,X)&=& 1- \frac{\dot{c}_{u}(s_1,s_0,X)}{c(s_1,s_0,X)}\left\{1(S\leq s_1)-\F_1(s_1,X)\right\},\\
r_v(s_1,s_0,S,X)&=& 1- \frac{\dot{c}_{v}(s_1,s_0,X)}{c(s_1,s_0,X)}\left\{1(S\leq s_0)-\F_0(s_0,X)\right\}.
\end{eqnarray*}
The following theorem presents the EIF for $\eta_{\tau}$.
\begin{theorem}
\label{thm:eif_mtau}
Under the nonparametric model, the EIF for $\eta_{\tau}$ is
\begin{equation*}
    \varphi_{\tau}\left(V\right) = H_{\tau}^{-1} \left\{t_{1}\left(X\right)+t_{2}\left(S,Z,X\right)+t_{3}\left(Y,S,Z,X\right)\right\},
\end{equation*}
where
\begin{eqnarray*}
H_{\tau} &=& \E\left(w_{\tau}(S_1,S_0)\left[\dot{f}_{\tau}\left(S_{1},S_{0};\eta_{\tau}\right)\dot{f}_{\tau}\left(S_{1},S_{0};\eta_{\tau}\right)^{\T} - \left\{m_{\tau}(S_1,S_0) - f_{\tau}(S_1,S_0;\eta_{\tau})\right\}\ddot{f}_{\tau}(S_1,S_0;\eta_{\tau}) \right] \right)
\end{eqnarray*}
and
\begin{eqnarray*}
    t_{1}\left(X\right) &=& \iint  \nu_{\tau}(s_1,s_0,X) \d s_{1}\d s_{0},\\
    t_{2}\left(S,Z,X\right) &=& \frac{Z}{\pi\left(X\right)}\left\{\frac{\int_{\mathcal{S}_0} \nu_{\tau}(S,s_0,X)\d s_0}{p_1(S,X)} - \iint \nu_{\tau}(s_1,s_0,X) r_u(s_1,s_0,S,X) \d s_1 \d s_0\right\}\\
    &&+ \frac{1-Z}{1-\pi\left(X\right)}\left\{\frac{\int_{\mathcal{S}_1} \nu_{\tau}(s_1,S,X)\d s_1}{p_0(S,X)} - \iint \nu_{\tau}(s_1,s_0,X) r_v(s_1,s_0,S,X) \d s_1 \d s_0\right\},\\
    t_{3}\left(Y,S,Z,X\right) &=& \frac{Z}{\pi\left(X\right)}\frac{\int_{\mathcal{S}_0}w_{\tau}(S,s_0)\dot{f}_{\tau}(S,s_0;\eta_\tau)e(S,s_0,X)\d s_0}{p_{1}\left(S,X\right)}\left\{Y-\mu_{1}\left(S,X\right)\right\} \\
     &&- \frac{1-Z}{1-\pi\left(X\right)}\frac{\int_{\mathcal{S}_1}w_{\tau}(s_1,S)\dot{f}_{\tau}(s_1,S;\eta_\tau)e(s_1,S,X)\d s_1}{p_{0}\left(S,X\right)}\left\{Y-\mu_{0}\left(S,X\right)\right\}.
\end{eqnarray*}
\end{theorem}

We can also construct an estimator based on the EIF.

\subsection{Proof of Theorem \ref{thm:eif_mtau}}
To find the $\varphi_{\tau}(V)$ satisfies
\[
\left.\dot{\eta}_{\tau,\theta}\right|_{\theta=0}\equiv\left.\frac{\partial\eta_{\tau,\theta}}{\partial\theta}\right|_{\theta=0}=\mathbb{E}\left\{ \varphi_{\tau}(V)\score(V)\right\},
\]
we calculate the derivative $\left.\dot{\eta}_{\tau,\theta}\right|_{\theta=0}$.
Taking the first order derivative with respect to $\eta_\tau$ on the objective function in~\eqref{equ:opt_tau}.
\begin{equation}
0=\E\left[ w_{\tau}\left(S_{1},S_{0}\right)\dot{f}_{\tau}\left(S_{1},S_{0};\eta_{\tau}\right)\left\{\tau\left(S_{1},S_{0}\right)-f_{\tau}\left(S_{1},S_{0};\eta_{\tau}\right)\right\}\right].
\label{equ:foc}
\end{equation}
 By the fact that~\eqref{equ:foc} holds for any parametric
submodel, we have 
\begin{align*}
0 & =\E_{\theta}\left[ w_{\tau}\left(S_{1},S_{0}\right)\dot{f}_{\tau}\left(S_{1},S_{0};\eta_{\tau,\theta}\right)\left\{\tau_{\theta}\left(S_{1},S_{0}\right)-f_{\tau}\left(S_{1},S_{0};\eta_{\tau,\theta}\right)\right\}\right] \\
 & =\iint w_{\tau}\left(s_{1},s_{0}\right)\dot{f}_{\tau}\left(s_{1},s_{0};\eta_{\tau,\theta}\right)\left\{\tau_{\theta}\left(s_{1},s_{0}\right)-f_{\tau}\left(s_{1},s_{0};\eta_{\tau,\theta}\right)\right\}e_{\theta}\left(s_{1},s_{0}\right)\d s_{1}\d s_{0}\\
 & =\iint w_{\tau}\left(s_{1},s_{0}\right)\dot{f}_{\tau}\left(s_{1},s_{0};\eta_{\tau,\theta}\right)\left\{\frac{\lambda_{\theta}\left(s_{1},s_{0}\right)}{e_{\theta}\left(s_{1},s_{0}\right)}-f_{\tau}\left(s_{1},s_{0};\eta_{\tau,\theta}\right)\right\}e_{\theta}\left(s_{1},s_{0}\right)\d s_{1}\d s_{0}\\
 & =\iint w_{\tau}\left(s_{1},s_{0}\right)\dot{f}_{\tau}\left(s_{1},s_{0};\eta_{\tau,\theta}\right)\left\{\lambda_{\theta}\left(s_{1},s_{0}\right)-e_{\theta}\left(s_{1},s_{0}\right)f_{\tau}\left(s_{1},s_{0};\eta_{\tau,\theta}\right)\right\}\d s_{1}\d s_{0},
\end{align*}
where $\lambda_{\theta}\left(s_{1},s_{0}\right)=\E_{\theta}\left[e_{\theta}\left(s_{1},s_{0},X\right)\left\{\mu_{1,\theta}\left(s_{1},X\right)-\mu_{0,\theta}\left(s_{0},X\right)\right\}\right] $.
Taking the first order derivative with respect to $\theta$ and evaluating
at $\theta=0$, we have 
\begin{eqnarray*}
    0 &=& \left.\frac{\partial}{\partial\theta}\iint w_{\tau}\left(s_{1},s_{0}\right)\dot{f}_{\tau}\left(s_{1},s_{0};\eta_{\tau,\theta}\right)\left\{\lambda_{\theta}\left(s_{1},s_{0}\right)-e_{\theta}\left(s_{1},s_{0}\right)f_{\tau}\left(s_{1},s_{0};\eta_{\tau,\theta}\right)\right\}\d s_{1}\d s_{0}\right|_{\theta=0}\\
    &=& \iint w_{\tau}\left(s_{1},s_{0}\right)\ddot{f}_{\tau}\left(s_{1},s_{0};\eta_{\tau}\right)\left.\dot{\eta}_{\tau,\theta}\right|_{\theta=0}\left\{\lambda\left(s_{1},s_{0}\right)-e\left(s_{1},s_{0}\right)f_{\tau}\left(s_{1},s_{0};\eta_{\tau}\right)\right\}\d s_{1}\d s_{0}\\
    &&+ \iint w_{\tau}\left(s_{1},s_{0}\right)\dot{f}_{\tau}\left(s_{1},s_{0};\eta_{\tau}\right)\left\{\left.\dot{\lambda}_{\theta}\left(s_{1},s_{0}\right)\right|_{\theta=0}-\left.\dot{e}_{\theta}\left(s_{1},s_{0}\right)\right|_{\theta=0}f_{\tau}\left(s_{1},s_{0};\eta_{\tau}\right)\right\}\d s_{1}\d s_{0}\\
    &&- \iint w_{\tau}\left(s_{1},s_{0}\right)\dot{f}_{\tau}\left(s_{1},s_{0};\eta_{\tau}\right)\dot{f}_{\tau}\left(s_{1},s_{0};\eta_{\tau}\right)^{\T}e\left(s_{1},s_{0}\right)\left.\dot{\eta}_{\tau,\theta}\right|_{\theta=0}\d s_{1}\d s_{0}.
\end{eqnarray*}
Therefore, we can write
\begin{equation}
    \left.\dot{\eta}_{\tau,\theta}\right|_{\theta=0}\ =\ H_{\tau}^{-1}\iint w_{\tau}\left(s_{1},s_{0}\right)\dot{f}_{\tau}\left(s_{1},s_{0};\eta_{\tau}\right)\left\{\left.\dot{\lambda}_{\theta}\left(s_{1},s_{0}\right)\right|_{\theta=0}-\left.\dot{e}_{\theta}\left(s_{1},s_{0}\right)\right|_{\theta=0}f_{\tau}\left(s_{1},s_{0};\eta_{\tau}\right)\right\}\d s_{1}\d s_{0},
    \label{equ:eta_dot}
\end{equation}
where
\begin{eqnarray*}
    H_{\tau} &=& -\iint w_{\tau}\left(s_{1},s_{0}\right)\ddot{f}_{\tau}\left(s_{1},s_{0};\eta_{\tau}\right)\E\left[e(s_1,s_0,X)\left\{\mu_1(s_1,X)-\mu_0(s_0,X)\right\}\right] \d s_{1}\d s_{0} \\
    &&+ \iint w_{\tau}\left(s_{1},s_{0}\right)\left\{\ddot{f}_{\tau}\left(s_{1},s_{0};\eta_{\tau}\right)f_{\tau}\left(s_{1},s_{0};\eta_{\tau}\right)+\dot{f}_{\tau}\left(s_{1},s_{0};\eta_{\tau}\right)\dot{f}_{\tau}\left(s_{1},s_{0};\eta_{\tau}\right)^{\T}\right\}e\left(s_{1},s_{0}\right) \d s_{1}\d s_{0}.
\end{eqnarray*}
Again, we calculate the two parts separately.

\noindent \textbf{Part I.} By Lemma \ref{lemma:lambda}, we have
\begin{eqnarray*}
    \iint w_{\tau}\left(s_{1},s_{0}\right)\dot{f}_{\tau}\left(s_{1},s_{0};\eta_{\tau}\right)\left.\dot{\lambda}_{\theta}\left(s_{1},s_{0}\right)\right|_{\theta=0}\d s_{1}\d s_{0} &=& B_1 + B_2 + B_3
\end{eqnarray*}
where
\begin{eqnarray*}
    B_1 &=& \iint w_{\tau}\left(s_{1},s_{0}\right)\dot{f}_{\tau}\left(s_{1},s_{0};\eta_{\tau}\right)\E\left[ e\left(s_{1},s_{0},X\right)\left\{\mu_{1}\left(s_{1},X\right)-\mu_{0}\left(s_{0},X\right)\right\}\score\left(X\right)\right] \d s_{1}\d s_{0} \\
    &=& \E\left\{h_1(X)\score(X)\right\}, \\
    B_2 &=& \iint w_{\tau}\left(s_{1},s_{0}\right)\dot{f}_{\tau}\left(s_{1},s_{0};\eta_{\tau}\right)\E\left[ \left.\dot{e}_{\theta}\left(s_{1},s_{0},X\right)\right|_{\theta=0}\left\{\mu_{1}\left(s_{1},X\right)-\mu_{0}\left(s_{0},X\right)\right\}\right] \d s_{1}\d s_{0} \\
    &=& \E\left\{h_2(S,Z,X)\score(S\mid Z,X)\right\}, \\
    B_3 &=& \iint w_{\tau}\left(s_{1},s_{0}\right)\dot{f}_{\tau}\left(s_{1},s_{0};\eta_{\tau}\right)\E\left[ e\left(s_{1},s_{0},X\right)\left\{\left.\dot{\mu}_{1,\theta}\left(s_{1},X\right)\right|_{\theta=0}-\left.\dot{\mu}_{0,\theta}\left(s_{0},X\right)\right|_{\theta=0}\right\}\right] \d s_{1}\d s_{0} \\
    &=& \E\left\{h_3(Y,S,Z,X)\score(Y\mid S,Z,X)\right\}
\end{eqnarray*}
with a little abuse of notation. Through similar arguments, we have
\begin{eqnarray*}
    h_1(X) &=& \iint w_{\tau}\left(s_{1},s_{0}\right)\dot{f}_{\tau}\left(s_{1},s_{0};\eta_{\tau}\right)e\left(s_{1},s_{0},X\right)\left\{\mu_{1}\left(s_{1},X\right)-\mu_{0}\left(s_{0},X\right)\right\}\d s_{1}\d s_{0} \\
    &&- \iint w_{\tau}\left(s_{1},s_{0}\right)\dot{f}_{\tau}\left(s_{1},s_{0};\eta_{\tau}\right)f_{\tau}\left(s_{1},s_{0};\eta_{\tau}\right)e\left(s_{1},s_{0}\right) \d s_{1}\d s_{0}, \\
    h_2(S,Z,X) &=& \frac{Z}{\pi\left(X\right)}\left[\tilde{\mu}_{11}\left(S,X\right)-\E\left\{ \tilde{\mu}_{11}\left(S,X\right)\mid Z=1,X\right\} \right] \\
    &&+ \frac{1-Z}{1-\pi\left(X\right)}\left[\tilde{\mu}_{01}\left(S,X\right)-\E\left\{ \tilde{\mu}_{01}\left(S,X\right)\mid Z=0,X\right\} \right] \\
    &&- \frac{Z}{\pi\left(X\right)}\left[\tilde{\mu}_{10}\left(S,X\right)-\E\left\{ \tilde{\mu}_{10}\left(S,X\right)\mid Z=1,X\right\} \right] \\
    &&- \frac{1-Z}{1-\pi\left(X\right)}\left[\tilde{\mu}_{00}\left(S,X\right)-\E\left\{ \tilde{\mu}_{00}\left(S,X\right)\mid Z=0,X\right\} \right], \\
    h_3(Y,S,Z,X) &=& \frac{Z}{\pi\left(X\right)}\frac{\int_{\mathcal{S}_0}w_{\tau}(S,s_0)\dot{f}_{\tau}(S,s_0;\eta_\tau)e(S,s_0,X)\d s_0}{p_{1}\left(S,X\right)}\left\{Y-\mu_{1}\left(S,X\right)\right\} \\
    &&- \frac{1-Z}{1-\pi\left(X\right)}\frac{\int_{\mathcal{S}_1}w_{\tau}(s_1,S)\dot{f}_{\tau}(s_1,S;\eta_\tau)e(s_1,S,X)\d s_1}{p_{0}\left(S,X\right)}\left\{Y-\mu_{0}\left(S,X\right)\right\},
\end{eqnarray*}
where $\tilde{\mu}_{*}$ are the same as previously defined for $*=11,01,10,00$ except that the weighting function is changed to $w_{\tau}(s_1,s_0)$ and the working model is changed to $f_{\tau}(s_1,s_0;\eta_\tau)$.

\noindent \textbf{Part II.} The second part of $\left.\dot{\eta}_{\theta}\right|_{\theta=0}$ as in Equation \eqref{equ:eta_dot} is exactly the same as in the proof of Theorem \ref{thm:eif_m1}, so we omit the details here. 

\noindent \textbf{Summary of the proof.} We have 
\begin{eqnarray*}
H_{\tau}\left.\dot{\eta}_{\theta}\right|_{\theta=0} &=& \iint w_{\tau}\left(s_{1},s_{0}\right)\dot{f}_{\tau}\left(s_{1},s_{0};\eta_{\tau}\right)\left.\dot{\lambda}_{\theta}\left(s_{1},s_{0}\right)\right|_{\theta=0}\d s_{1}\d s_{0} \\
&&- \iint w_{\tau}\left(s_{1},s_{0}\right)\dot{f}_{\tau}\left(s_{1},s_{0};\eta_{\tau}\right)f_{\tau}\left(s_{1},s_{0};\eta_{\tau}\right)\left.\dot{e}_{\theta}\left(s_{1},s_{0}\right)\right|_{\theta=0}\d s_{1}\d s_{0}\\
&=& \E\left\{ \left(h_{1}\left(X\right)-h_{4}\left(X\right)\right)\score\left(X\right)\right\} \\
&&+ \E\left\{ \left(h_{2}\left(S,Z,X\right)-h_{5}\left(S,Z,X\right)\right)\score\left(S\mid Z,X\right)\right\} \\
&&+ \E\left\{ h_{3}\left(Y,S,Z,X\right)\score\left(Y\mid S,Z,X\right)\right\}
\end{eqnarray*}
where 
\begin{align*}
H_{\tau}^{-1}\left(h_{1}\left(X\right)-h_{4}\left(X\right)\right) & \in \H_{1},\\
H_{\tau}^{-1}\left(h_{2}\left(S,Z,X\right)-h_{5}\left(S,Z,X\right)\right) & \in \H_{3},\\
H_{\tau}^{-1}h_{3}\left(Y,S,Z,X\right) & \in \H_{4}.
\end{align*}
By similar arguments, the EIF is  
\begin{equation*}
    \varphi_{\tau}\left(V\right) = H_{\tau}^{-1} \left\{t_{1}\left(X\right)+t_{2}\left(S,Z,X\right)+t_{3}\left(Y,S,Z,X\right)\right\},
\end{equation*}
where
\begin{eqnarray*}
    t_{1}\left(X\right) &=& \iint w_{\tau}\left(s_{1},s_{0}\right)\dot{f}_{\tau}\left(s_{1},s_{0};\eta_{\tau}\right)e\left(s_{1},s_{0},X\right)\left\{\mu_{1}\left(s_{1},X\right)-\mu_{0}\left(s_{0},X\right) - f_{\tau}\left(s_{1},s_{0};\eta_{\tau}\right) \right\}\d s_{1}\d s_{0}, \\
    t_{2}\left(S,Z,X\right) &=& \frac{Z}{\pi\left(X\right)}\left[\tilde{\mu}_{11}\left(S,X\right)-\E\left\{ \tilde{\mu}_{11}\left(S,X\right)\mid Z=1,X\right\} \right] \\
    &&+ \frac{1-Z}{1-\pi\left(X\right)}\left[\tilde{\mu}_{01}\left(S,X\right)-\E\left\{ \tilde{\mu}_{01}\left(S,X\right)\mid Z=0,X\right\} \right] \\
    &&- \frac{Z}{\pi\left(X\right)}\left[\tilde{\mu}_{10}\left(S,X\right)-\E\left\{ \tilde{\mu}_{10}\left(S,X\right)\mid Z=1,X\right\} \right] \\
    &&- \frac{1-Z}{1-\pi\left(X\right)}\left[\tilde{\mu}_{00}\left(S,X\right)-\E\left\{ \tilde{\mu}_{00}\left(S,X\right)\mid Z=0,X\right\} \right] \\
    &&- \frac{Z}{\pi\left(X\right)}\left[\tilde{f}_{1}\left(S,X\right)-\E\left\{ \tilde{f}_{1}\left(S,X\right)\mid Z=1,X\right\} \right] \\
    &&- \frac{1-Z}{1-\pi\left(X\right)}\left[\tilde{f}_{0}\left(S,X\right)-\E\left\{ \tilde{f}_{0}\left(S,X\right)\mid Z=0,X\right\} \right], \\
    t_{3}\left(Y,S,Z,X\right) &=& \frac{Z}{\pi\left(X\right)}\frac{\int_{\mathcal{S}_0}w_{\tau}(S,s_0)\dot{f}_{\tau}(S,s_0;\eta_\tau)e(S,s_0,X)\d s_0}{p_{1}\left(S,X\right)}\left\{Y-\mu_{1}\left(S,X\right)\right\} \\
    &&- \frac{1-Z}{1-\pi\left(X\right)}\frac{\int_{\mathcal{S}_1}w_{\tau}(s_1,S)\dot{f}_{\tau}(s_1,S;\eta_\tau)e(s_1,S,X)\d s_1}{p_{0}\left(S,X\right)}\left\{Y-\mu_{0}\left(S,X\right)\right\},
\end{eqnarray*}
where $\tilde{\mu}_{*z}$ denotes the weighted integration of the conditional outcome model $\mu_z(s_z,X)$,
\begin{eqnarray*}
    \tilde{\mu}_{11}(s_1,X) &=& \int w_{\tau}(s_1,s_0)\dot{f}_{\tau}(s_1,s_0;\eta_\tau)\frac{e(s_1,s_0,X)}{p_1(s_1,X)}\mu_1(s_1,X)\d s_0 \\
    && + \iint 1(t_1\geq s_1)w_{\tau}(t_1,s_0)\dot{f}_{\tau}(t_1,s_0;\eta_\tau)\dot{c}_u(t_1,s_0,X)p_1(t_1,X)p_0(s_0,X)\mu_1(t_1,X)\d t_1 \d s_0, \\
    \tilde{\mu}_{01}(s_0,X) &=& \int w_{\tau}(s_1,s_0)\dot{f}_{\tau}(s_1,s_0;\eta_\tau)\frac{e(s_1,s_0,X)}{p_0(s_0,X)}\mu_1(s_1,X)\d s_1 \\
    && + \iint 1(t_0\geq s_0)w_{\tau}(s_1,t_0)\dot{f}_{\tau}(s_1,t_0;\eta_\tau)\dot{c}_v(s_1,t_0,X)p_1(s_1,X)p_0(t_0,X)\mu_1(s_1,X)\d s_1 \d t_0, \\
    \tilde{\mu}_{10}(s_1,X) &=& \int w_{\tau}(s_1,s_0)\dot{f}_{\tau}(s_1,s_0;\eta_\tau)\frac{e(s_1,s_0,X)}{p_1(s_1,X)}\mu_0(s_0,X)\d s_0 \\
    && + \iint 1(t_1\geq s_1)w_{\tau}(t_1,s_0)\dot{f}_{\tau}(t_1,s_0;\eta_\tau)\dot{c}_u(t_1,s_0,X)p_1(t_1,X)p_0(s_0,X)\mu_0(s_0,X)\d t_1 \d s_0, \\
    \tilde{\mu}_{00}(s_0,X) &=& \int w_{\tau}(s_1,s_0)\dot{f}_{\tau}(s_1,s_0;\eta_\tau)\frac{e(s_1,s_0,X)}{p_0(s_0,X)}\mu_0(s_0,X)\d s_1 \\
    && + \iint 1(t_0\geq s_0)w_{\tau}(s_1,t_0)\dot{f}_{\tau}(s_1,t_0;\eta_\tau)\dot{c}_v(s_1,t_0,X)p_1(s_1,X)p_0(t_0,X)\mu_0(t_0,X)\d s_1 \d t_0, \\
    \tilde{f}_1(s_1,X) &=& \int w_{\tau}(s_1,s_0)\dot{f}_{\tau}(s_1,s_0;\eta_\tau)f_{\tau}(s_1,s_0;\eta_\tau)\frac{e(s_1,s_0,X)}{p_1(s_1,X)}\d s_0 \\
    && + \iint 1(t_1\geq s_1)w_{\tau}(t_1,s_0)\dot{f}_{\tau}(t_1,s_0;\eta_\tau)f_{\tau}(t_1,s_0;\eta_\tau)\dot{c}_u(t_1,s_0,X)p_1(t_1,X)p_0(s_0,X)\d t_1 \d s_0, \\
    \tilde{f}_0(s_0,X) &=& \int w_{\tau}(s_1,s_0)\dot{f}_{\tau}(s_1,s_0;\eta_\tau)f_{\tau}(s_1,s_0;\eta_\tau)\frac{e(s_1,s_0,X)}{p_0(s_0,X)}\d s_1 \\
    && + \iint 1(t_0\geq s_0)w_{\tau}(s_1,t_0)\dot{f}_{\tau}(s_1,t_0;\eta_\tau)f_{\tau}(s_1,t_0;\eta_\tau)\dot{c}_v(s_1,t_0,X)p_1(s_1,X)p_0(t_0,X)\d s_1 \d t_0.
\end{eqnarray*}
Using similar notation and derivation as in the proof of Theorem~\ref{thm:eif_m1} gives the results stated in the theorem. \QEDB

\begin{corollary}
\label{cor::linear-proj}
    Suppose that
    \begin{enumerate}
        \item the working models have the same linear form in the parameters of interest, i.e., $f_{*}(s_1,s_0;\eta_{*})=\eta_{*}^\top g(s_1,s_0)$;
        \item the specified weights are the same, i.e., $w_1(s_1,s_0)=w_0(s_1,s_0)=w_{\tau}(s_1,s_0)$ for all values of $s_1$ and $s_0$ on the support of $S_1$ and $S_0$.
    \end{enumerate}
    The EIF for $\eta_1 - \eta_0$ is the same as $\eta_\tau$, which indicates that parameters defined by projecting $m_1(s_1,s_0)$ and $m_0(s_1,s_0)$ separately and computing the difference have the same semiparametric efficiency bound as those by projecting $\tau(s_1,s_0)$ directly.
\end{corollary}
The proof follows directly from the algebraic facts that $H_1=H_0=H_{\tau}$ and $t_{j} = \ell_j - k_j$ for $j=1,2,3$ in the EIF formulas.

 From Corollary~\ref{cor::linear-proj}, if $\hat{\eta}_{1,\textup{eif}}$ and $\hat{\eta}_{0,\textup{eif}}$ are the EIF estimators under linear working models
$f_{1}\left(s_{1},s_{0};\eta_{1}\right) = \eta_{1}^{\T}g\left(s_{1},s_{0}\right)$ and $f_{0}\left(s_{1},s_{0};\eta_{0}\right) = \eta_{0}^{\T}g\left(s_{1},s_{0}\right)$, respectively, then the EIF estimator for $\tau(s_1,s_0)$ under the linear working model $f_{\tau}\left(s_{1},s_{0}\right) = \eta^{\T}g\left(s_{1},s_{0}\right)$  is $\hat{\eta}_{\textup{eif}}=\hat{\eta}_{1,\textup{eif}}-\hat{\eta}_{0,\textup{eif}}$. 
\end{document}